\documentstyle[psfig,12pt]{article}
\def\beq{\begin{equation}}
\def\eeq{\end{equation}}
\def\bea{\begin{eqnarray}}
\def\eea{\end{eqnarray}}
\def\bed{\begin{displaymath}}
\def\eed{\end{displaymath}}
\def\omk{\omega _k}
\def\dk{\frac {d^3k}{(2\pi )^3}}
\def\d4k{\frac{d^4K}{(2\pi )^4}}

\def\ag{\buildrel >\over \sim}
\def\al{\buildrel <\over \sim}

\def\sm{\smallskip}
\def\me{\medskip}

\def\bu{$\bullet$}
\newcommand{\sla}{\!\!\!\!/ \,}

\begin{document}
\centerline{\large \bf New Developments and Applications of 
Thermal Field Theory\footnote{Lectures given at the Jyv\"askyl\"a 
Summer School 2000}}

\bigskip

\centerline{\large \bf by Markus H. Thoma\footnote{Email: 
markus.thoma@cern.ch}$^,$\footnote{Heisenberg Fellow}}

\medskip

\centerline{\bf Theory Division, CERN, CH-1211 Geneva 23, Switzerland}

\bigskip
\bigskip
\bigskip
\bigskip
\bigskip
\bigskip

\centerline{\bf Abstract}

\bigskip

The lecture provides an introduction to thermal field theory and its 
applications to the physics of the quark-gluon plasma, possibly created 
in relativistic heavy ion collisions. In particular the Hard Thermal Loop
resummation technique, providing a consistent perturbative description of
relativistic, high-temperature plasmas is introduced. Using this
method interesting quantities of the quark-gluon plasma (damping rates,
energy loss, photon and dilepton production) are discussed. Furthermore 
recent developments on non-equilibrium field theory, which are relevant 
for high-energy heavy ion physics, are presented.

\bigskip
\bigskip
\bigskip

\centerline{{\bf Contents}}

\bigskip

\noindent
1. Introduction and Motivation (Quark-gluon plasma)

\medskip

\noindent
2. Introduction to Thermal Field Theory (Imaginary and real time formalism)

\medskip

\noindent
3. Hard Thermal Loop Resummation (Self energies, propagators, 
dispersion relations, effective perturbation theory)

\medskip

\noindent
4. Applications (Damping rates, energy loss of energetic partons, 
photon and dilepton production)

\medskip

\noindent
5. Non-equilibrium field theory  (Hard Thermal Loop method for a 
non-equilibrated plasma)

\medskip

\noindent
6. Problems

\newpage

\section{Introduction and Motivation}

The aim of relativistic heavy ion collision experiments is the discovery of
a new state of matter, the so-called quark-gluon plasma (QGP). Quarks are 
substructure particles of hadrons. The observed hadron spectrum can be 
described by 6 different quark flavors (up, down, strange, charm, bottom, and 
top). Baryons contain 3 quarks, e.g. the proton 2 up and 1 down quarks. 
Meson consists of a quark and an antiquark, e.g. $\pi^+ \hat = u\; \bar d$.

The strong interaction between quarks is described within QCD, where
the interaction is caused by the exchange of gauge bosons, the so-called
gluons, analogously to photons in QED mediating the electromagnetic 
interaction. In contrast to QED quarks have three ``charges'' (and 
``anticharges'') of the strong interaction, called color. As a consequence
QCD is a non-abelian gauge theory based on the $SU(3)$ group. Also in contrast
to QED the $SU(3)$ group leads to 8 gauge bosons, which can interact directly 
with themselves. The gluon self-interaction causes asymptotic freedom, i.e.,
quarks and gluons interact weakly at small distances or large momentum 
transfers. At large distances, however, the potential between partons
(quarks and gluons) seems to increase linearly, which explains confinement,
i.e., the absence of free quarks and gluons in nature. In this picture
nucleons and other hadrons can be regarded as quark bags, containing 
besides the valence quarks also virtual quarks (sea quarks) and gluons,
which force the partons to stay inside of the bag.

A nucleus can be pictured as a dense system of quark bags. If one increases 
the baryon density, e.g. by compressing the nucleus, or adds further hadrons,
e.g. by increasing the temperature leading to thermal pion production,
these bags will overlap. Then the quarks and gluons are not restricted 
to individual bags anymore but can move around in the entire system,
which is now in the QGP phase. This deconfinement transition is expected
to occur at a critical baryon density of the order of 10 times nuclear
density $\rho_0$, where $\rho _0=0.125 $ GeV/fm$^3$ 
= $2.2 \times 10^{17}$ kg/m$^3$. The critical temperature is estimated by
lattice calculations (see below) to be in the range of $T_c=150-200$ MeV 
= $(1.8-2.4)\times 10^{12}$ K. These considerations give rise to the 
phase diagram shown in Fig.1. In nature the deconfinement phase transition
occurred during the expansion of the early Universe about 2 microseconds
after the Big Bang at zero baryon density (almost equal number of quarks 
and antiquarks). It might be realized also in the core of neutron stars 
at high baryon density. In the laboratory the fireball created in
relativistic heavy ion collisions could be in the QGP phase for 
a short time period. 

\begin{figure}

\centerline{\psfig{figure=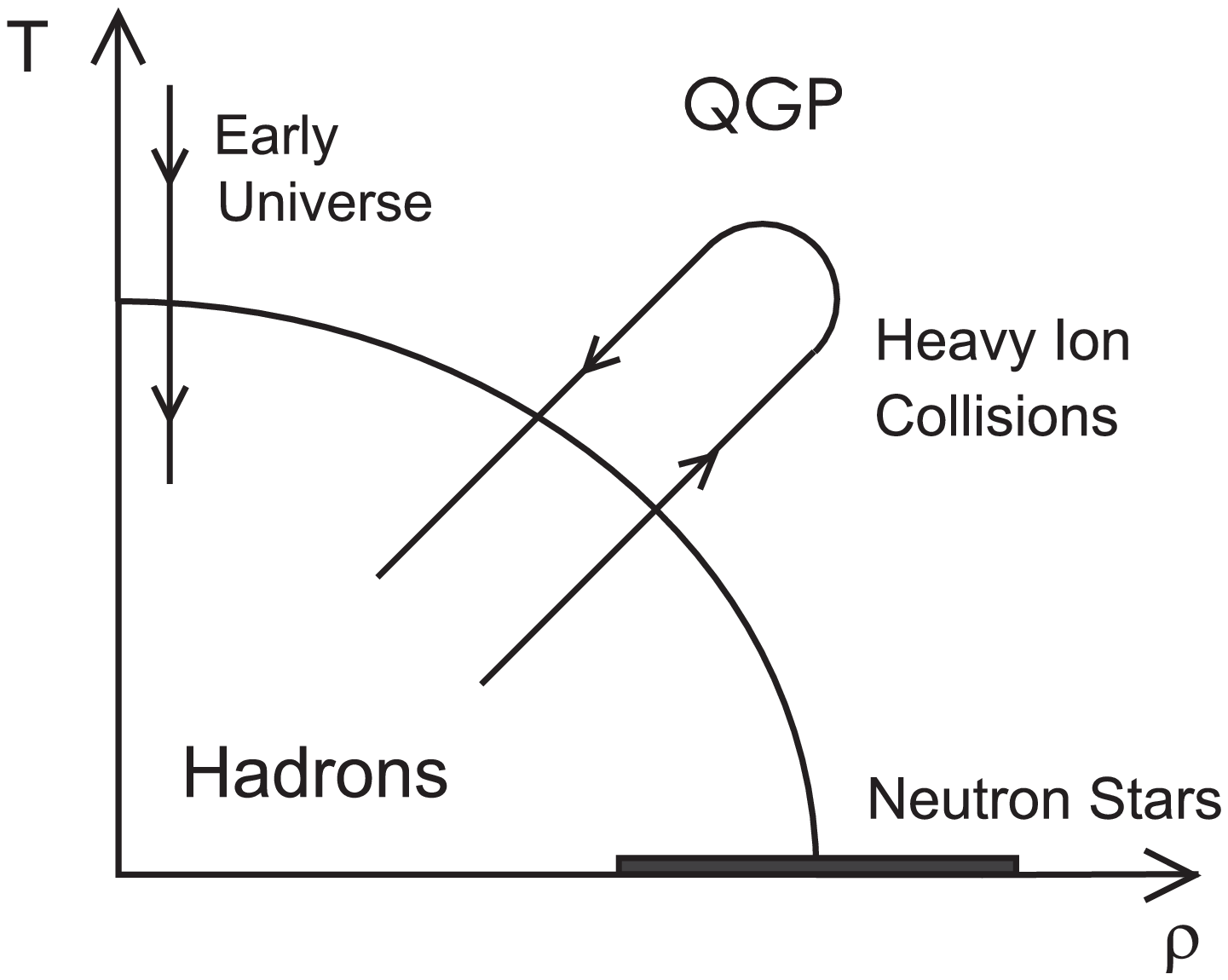,width=10cm}}
\centerline{\it Fig.1}

\end{figure}

At this point let me make a few remarks:

1. The QGP is a relativistic plasma, i.e., the thermal velocity of the
up-, down-quarks, and gluons is relativistic, since their masses are much 
smaller than the temperature of the plasma.

2. The QGP has the astonishing behaviour that it approaches the ideal gas
limit for large densities and temperatures, because the parton interaction
becomes weak due to asymptotic freedom.

3. Are there alternatives to the QGP at high temperature? Could it be that 
the deconfinement transition is never complete even at arbitrary high 
temperatures? Maybe new non-perturbative configurations like QCD monopoles
exist even in the high temperature limit? In order to answer this question 
the first step must be to identify the QGP state in relativistic heavy ion 
collisions.

There are two possible scenarios for producing a QGP fireball in relativistic
heavy ion collisions. At relatively small center of mass energies,
i.e. $\sqrt{s} \ll 100 $ A$\cdot $GeV, one expects that the Lorentz contracted  
nuclei stop each other in the collision and a hot, highly compressed 
fireball with a finite 
baryon or quark density (finite chemical potential), according to the 
surplus of quarks over anti-quarks from the nuclei, is created. 
At much higher energies, however, $\sqrt{s}{\buildrel >\over \sim} 
100$ A$\cdot $GeV,
there is not enough time for stopping and the nuclei penetrate each other
transparently. However, the vacuum in the
space-time volume between the nuclei after the 
collision is highly excited leading to a violent production of
gluons and quark-antiquark pairs. Hence we expect the formation of 
a hot parton gas at zero baryon density (zero chemical potential). The both 
scenarios are sketched in Fig.2.

\begin{figure}

\centerline{\psfig{figure=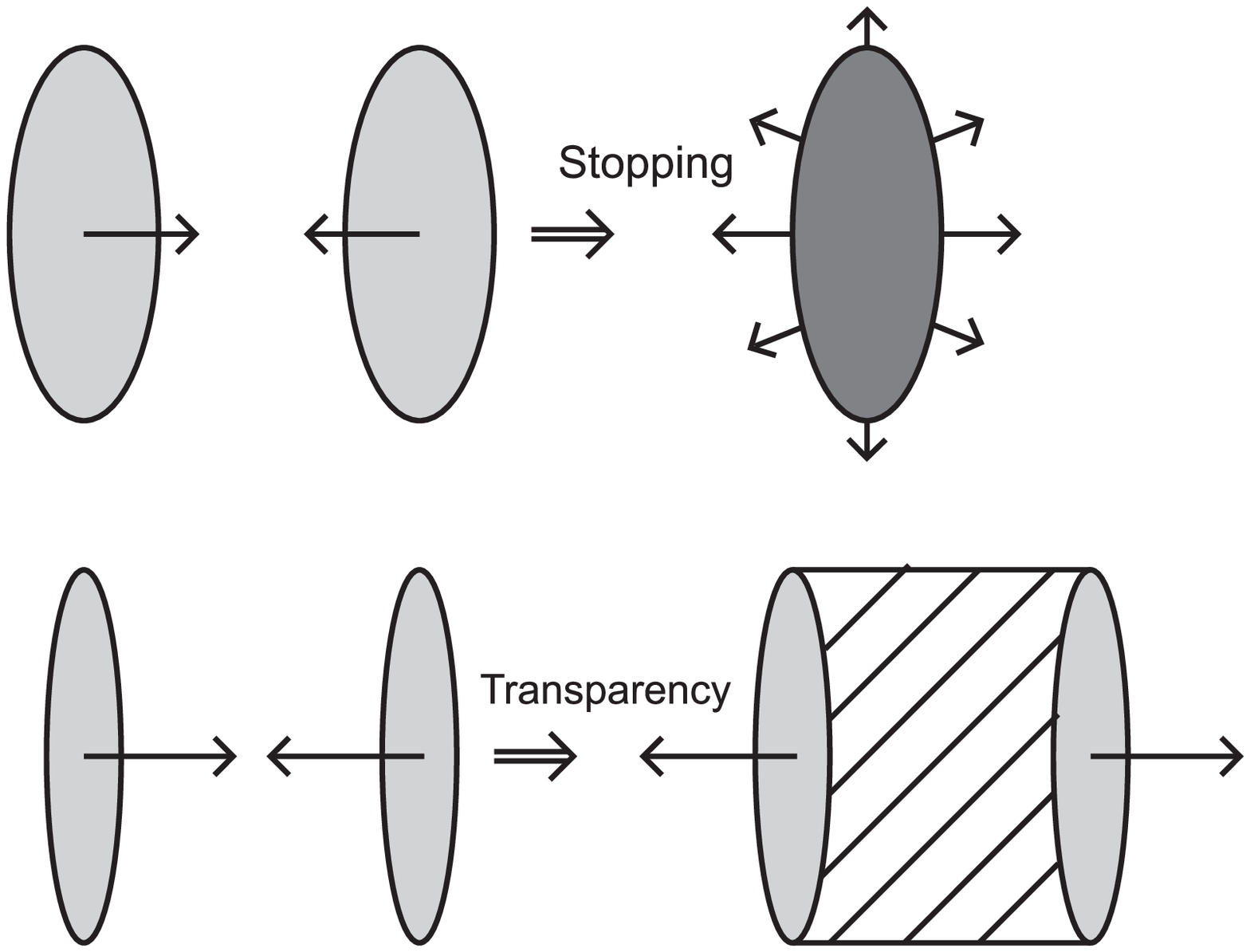,width=10cm}}
\centerline{\it Fig.2}

\end{figure}

In the table below former, present, and future experiments with relativistic
heavy ions are listed. Also a rough estimate for the maximum temperature 
reachable in these collisions is given.
 
\begin{tabbing}
Accelerator\hspace*{1cm} \= Projectile-Target\= $\; \; \sqrt{s}$ [A$\cdot$GeV]\= 
$\; \; T_{max} $[MeV] \\
$\,$ \> $\,$ \> $\,$ \> $\,$ \\ 
AGS (BNL) \> {\it Si-Au}, {\it Au-Au} \> $\; \; $4-5 \> $\; \; $150 \\
SPS (CERN)\> {\it S-U}, {\it Pb-Pb} \> $\; \; $17-20 \> $\; \; $190 \\
RHIC (BNL)\> {\it Au-Au} \> $\; \; $200 \> $\; \; $230 \\
LHC (CERN)\> {\it Pb-Pb} \> $\; \; $5500 \> $\; \; $260 
\end{tabbing}

How does a high energy nucleus-nucleus collision proceed in space and time?
The space-time evolution of the fireball in the ultrarelativistic 
case, $\sqrt{s}{\buildrel >\over \sim} 100 $ A$\cdot $GeV, is shown in Fig.3.
The $x$-axis shows the beam direction and the $y$-axis the time,
with $t=0$ at the maximum overlap of the nuclei. The produced particles
in this diagram lie only above the light-cone due to causality. The hyperbolas
denote curves of constant proper time $\tau=\sqrt{t^2-z^2}$. 
At the beginning the parton gas 
will be not in equilibrium. Only secondary collisions between the partons 
will lead to equilibration. Afterwards the temperature of the fireball
will decrease in the course of its expansion until the critical temperature 
has been reached. Then the system will hadronize, maybe showing a 
substantial mixed phase, depending on the order of the phase transition.
Finally the system will be so dilute that the interactions between the 
hadrons cease (freeze-out).

\begin{figure}

\centerline{\psfig{figure=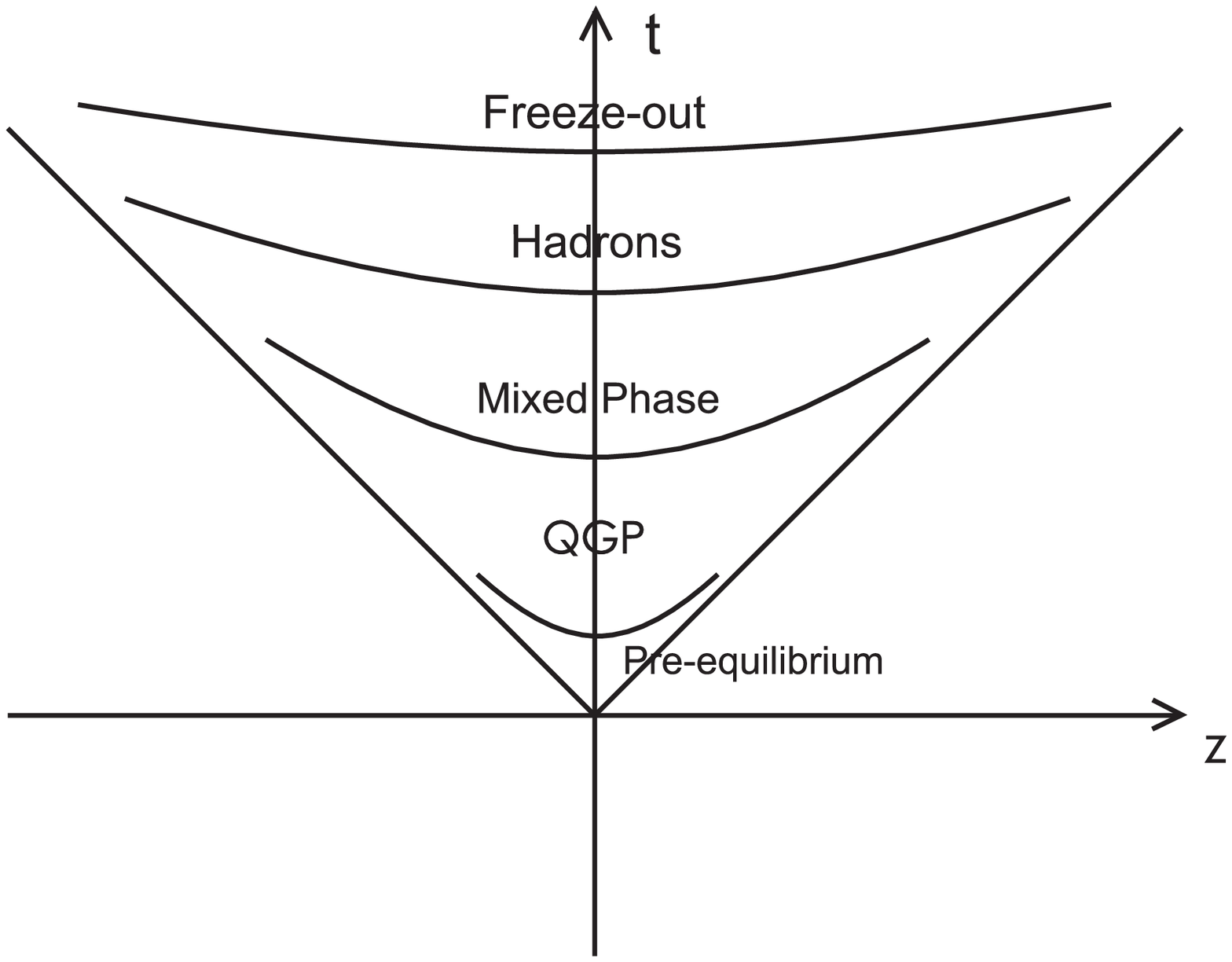,width=12cm}}
\centerline{\it Fig.3}

\end{figure}

In order to speak of the QGP as a thermal system, we need a large
volume and particle number, and a sufficient life time of the 
equilibrated system. The following rough estimates from simple models
support the possibility of a QGP formation in heavy ion collisions: 
maximum volume ({\it U-U}) $\sim$ 3000 fm$^3$; parton number 
$\sim $ 10000; pre-equilibrium period
$\sim $ 1 fm/c = $3\times 10^{-24}$ s; lifetime of QGP $\sim $ 5 - 10 fm/c.

The main problem is the identification of the QGP phase during the early 
stage of the fireball. Hadronic signatures (e.g. $J/\psi$-suppression or 
strangeness enhancement) are often strongly affected by final state 
interactions in the hadronic phase, while electromagnetic signals (photons, 
lepton pairs) are covered up by a huge background of hadronic decays.

In order to understand the properties of a QGP and to make unambiguous 
predictions about signatures for the QGP formation, we need a profound
description of the QGP. For this purpose we have to use QCD at finite 
temperature and chemical potential. There are two different approaches.

1. Lattice QCD is a non-perturbative method for solving the QCD equations
numerically on a 4-dimensional space-time lattice. 
In this way all temperatures from below to above the phase transition are 
accessible. Static quantities as the critical temperature, the order of the
phase transition, or the equation of state of the QGP have been investigated
successfully in this way.
However, it is not possible (or at least very difficult) to address
dynamical quantities, such as the most signatures for the QGP formation,
a finite chemical potential, and non-equilibrium situations. 

2. Perturbative QCD at finite temperature is based on the fact that the
temperature dependent running coupling constant is small at high
temperatures due to asymptotic freedom, $T\rightarrow \infty 
\Rightarrow \alpha _s(T)=g^2/4\pi \rightarrow 0$. At a typical temperature
of $T=250$ MeV we expect $\alpha _s= 0.3$ -- 0.5. This suggest that
perturbation theory could work at least qualitatively. It corresponds to an 
expansion in $\alpha_s$, which can be performed conveniently by using 
Feynman diagrams as the ones in Fig.4 for elastic quark-quark scattering.
In this way  cross sections, life times, production and decay rates etc.
can be calculated. The advantages compared to lattice calculations are,
that one can compute static {\it and} dynamical quantities at $T>0$ {\it and} 
$\mu >0$ and that an extension to non-equilibrium is possible (see chapter 5).
The drawbacks are that it is reliable only at high temperatures ($T\gg T_c$)
and that one encounters infrared singularities, as we will discuss in the 
following in detail.

Finally let me note in this motivation of thermal field theory that there are
more applications besides the QGP in relativistic heavy ion collisions,
namely interactions of neutrinos and other particles in Supernovae
plasmas, a possible quark matter core in neutron stars,
the origin of the baryon asymmetry in the early Universe, and Bose condensates 
in condensed matter physics.

\begin{figure}

\centerline{\psfig{figure=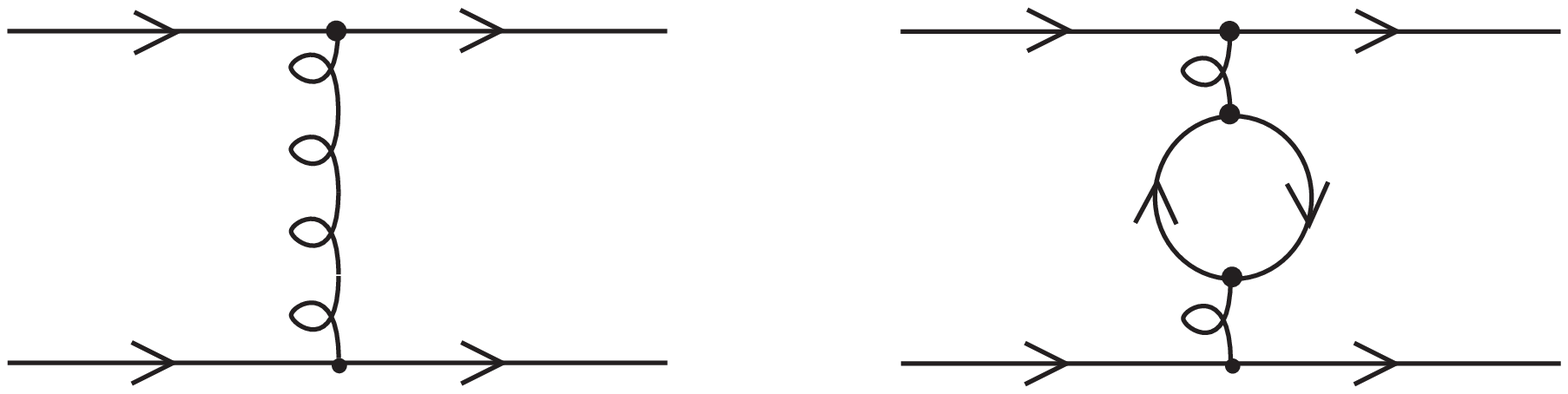,width=10cm}}
\centerline{\it Fig.4}

\end{figure}

\vspace*{15cm}

\newpage

\section{Introduction to Thermal Field Theory}

\subsection{Green functions at $T>0$}

The aim of this lecture is the introduction to thermal field theory (TFT) 
for allowing perturbative calculation of cross sections, life times, 
rates, etc. of particles in an equilibrated, relativistic medium.  
Thermal field theory is a combination of all three basic branches of 
modern physics, namely quantum  mechanics, theory of relativity, and
statistical physics. Therefore one could also call it 
relativistic quantum statistics.

Our aim is to derive Feynman diagrams and rules at $T>0$ ($\mu \neq 0$).
The most important quantity in perturbative field theory is the
2-point Green function or propagator. Therefore we have to ask first: 
How do propagators look like at $T>0$? (In the non-relativistic case Green 
functions at finite temperature are discussed e.g. in {\it Fetter, Walecka, 
Quantum Theory of Many-Particle Systems}.) Here we want to discuss 
the scalar field theory first, which we will use in the following
as a simple toy model to study the different techniques in TFT.

In order to set the notations and for comparing with the vacuum case, we 
repeat some basic facts about quantum field theory at zero temperature.
The Feynman propagator is defined as
\beq
i\> \Delta _F(x-y)\equiv \langle 0|T\{\phi (x)\phi (y)\}|0\rangle, 
\label{2.1}
\eeq
where the scalar field can be expressed by the Fourier transform 
\beq
\phi (x)=\int \frac{d^3k}{(2\pi)^{3/2}}\> \frac{1}{(2\omega _k)^{1/2}}\>
[a({\bf k}) e^{-iK\cdot x}+a^\dagger ({\bf k}) e^{iK\cdot x}]\biggl |_{k_0=\omega_k}.
\label{2.2}
\eeq
Here $x\equiv (x_0, {\bf x})$, $x_0\equiv t$ using natural units
$\hbar =c=k_B=1$. We use the Minkowski metric,
$x^2 = x_\mu x^\mu = g_{\mu \nu}x^\nu x^\mu = x_0^2-{\bf x}^2$.
Four momenta are denoted by
$K\equiv(k_0,{\bf k})$, $k\equiv |{\bf k}|$ and the energy of the field
is given by $\omega_k=\sqrt{k^2+m^2}$. The time ordered product of two fields
is defined as
\bed
T\{\phi (x)\phi (y)\}=\left \{ \begin{array}{c} \phi (x)\phi (y)\; \; \; \; \; x_0>y_0\\
\phi (y)\phi (x)\; \; \; \; \; x_0<y_0 \end{array} \right. .\nonumber 
\eed                                     
The Fourier coefficients in (\ref{2.2}) represent creation $a^\dagger ({\bf k})$ 
and destruction operators $a({\bf k})$, which create or destroy a boson with 
momentum ${\bf k}$ in the state of the system. In particular the vacuum state
is given by $a({\bf k})|0\rangle =0$ for all ${\bf k}$. 

Using (\ref{2.1}) and (\ref{2.2}) the Feynman propagator can be written as
\beq
\Delta _F (x-y)=\int \frac{d^4K}{(2\pi )^4}\> \frac{e^{-iK\cdot (x-y)}}{K^2-m^2+i\epsilon}.
\label{2.3}
\eeq
From this expression we can derive the Feynman rule for the propagator 
in momentum space. $\Delta_F$ describes the free propagation of a free
scalar particle from $y$ to $x$ for $x_0>y_0$
(creation at $y$, destruction at $x$) and from $x$ to $y$ for $x_0<y_0$.
Integrating over $k_0$ in the complex $k_0$-plane
we find in the case $x_0>y_0$
\beq
\Delta_F(x-y)=-i\int \frac{d^3k}{(2\pi )^3}\> \frac{1}{2\omega _k}\>e^{-iK\cdot (x-y)}\biggl 
|_{k_0=\omega _k}.
\label{2.4}
\eeq

Now let us turn to $T>0$.
Vacuum expectation values have now to be replaced by
quantum statistic expectation values, i.e.
\beq
\langle A \rangle \equiv {\rm Tr}\, (\rho A),
\label{2.5}
\eeq
where $A$ is an arbitrary quantum operator and $\rho $ the
density operator or matrix. Choosing the canonical ensemble
it is given by
\beq
\rho = \frac{1}{Z}\> e^{-\beta H},
\label{2.6}
\eeq
where $\beta \equiv 1/T$, $H$ is the Hamiltonian of the system 
with the eigenvalues and eigenstates $H|n\rangle = E_n|n\rangle$.
(For $\mu \neq 0$ $H$ is replaced by $H-\mu N$ with the number operator
$N$.) The partition function is $Z={\rm Tr}\, (e^{-\beta H})$. Then we can
write 
\beq
\langle A \rangle =\frac{1}{Z}\> {\rm Tr}\, (Ae^{-\beta H})
=\frac{1}{Z}\> \sum_n \> \langle n| A|n\rangle \> e^{-\beta E_n},
\label{2.7}
\eeq
where the sum goes over all thermally exited states weighted with the
Boltzmann factor $|n\rangle $ and $\exp (-\beta E_n)$.

(It is possible to formulate the Boltzmann factor in a Lorentz invariant way
by introducing the four velocity $u_\mu$: $\exp{(-\beta u_\mu P^\mu)}$ 
with $p_0=E$. In the following we will consider only the Lorentz frame of 
heat bath: $u_\mu = (1,0,0,0)$.)

Now we will apply (\ref{2.7}) to the scalar propagator, yielding
\beq
i\> \Delta_F^{T>0} (x-y) = \frac {1}{Z}\> \sum _n \> \langle n|T\{\phi (x)
\phi (y)\}|n\rangle \> e^{-\beta E_n}\; .
\label{2.8}
\eeq
Using (\ref{2.2}) in (\ref{2.8}) for $x_0>y_0$ we obtain 
\bea
&&i\> \Delta _F^{T>0} (x-y) = \frac{1}{Z} \> \int \frac{d^3k}{(2\pi)^{3/2}}\frac{d^3k'}{(2\pi)^{3/2}}
\frac{1}{(2\omega _k)^{1/2}} \frac{1}{(2\omega _k')^{1/2}} \sum _n e^{-\beta E_n}\nonumber \\
&&\langle n| [a({\bf k}) e^{-iK\cdot x}+a^\dagger ({\bf k}) e^{iK\cdot x}]\>
[a({\bf k'}) e^{-iK'\cdot y}+a^\dagger ({\bf k'}) e^{iK'\cdot y}]|n\rangle .
\label{2.8a}
\eea
The multi-boson states are given by acting repeatedly with the creation
operator on the vacuum state according to
\beq
|n\rangle = |n_1({\bf k}_1), n_2({\bf k}_2), ...\rangle = \prod _i \frac
{[a^\dagger ({\bf k}_i)]^{n_i({\bf k}_i)}}{\sqrt {n_i({\bf k}_i)!}}\>
|0\rangle.
\label{2.9}
\eeq
The states $|n\rangle $ are orthonormalized. In order to evaluate 
(\ref{2.8a}) we need
\bea
a({\bf k}_i)|n\rangle & = & \sqrt{n({\bf k}_i)}\> |n_1({\bf k}_1), n_2({\bf k}_2), ..., 
n_i({\bf k}_i)-1, ...\rangle ,\nonumber \\
a^\dagger({\bf k}_i)|n\rangle & = & \sqrt{n({\bf k}_i)+1}\> |n_1({\bf k}_1), n_2({\bf k}_2), ..., 
n_i({\bf k}_i)+1, ...\rangle ,
\label{2.10}
\eea
which leads to
\bea
&&i\> \Delta _F^{T>0} (x-y) = \frac{1}{Z} \> \int \frac{d^3k}{(2\pi)^{3/2}}\frac{d^3k'}{(2\pi)^{3/2}}
\frac{1}{(2\omega _k)^{1/2}} \frac{1}{(2\omega _k')^{1/2}} \sum _n e^{-\beta E_n}\nonumber \\
&&\{ [n({\bf k})+1] \delta^3({\bf k}-{\bf k'}) e^{-iK\cdot x+iK'\cdot y}+
n({\bf k}) \delta^3({\bf k}-{\bf k'}) e^{iK\cdot x-iK'\cdot y}\}\nonumber \\
&&=\frac{1}{Z} \> \int \frac{d^3k}{(2\pi)^{3}} \frac{1}{2\omega _k}\sum _n e^{-\beta E_n}
\{ [n({\bf k})+1] e^{-iK\cdot (x-y)}+n({\bf k}) e^{iK\cdot (x-y)}\}\biggl |_{k_0=\omega_k}.
\label{2.11}
\eea
Now we use 
\beq
\frac{1}{Z}\sum _n n({\bf k}) e^{-\beta E_n}=\frac{1}{\exp{(\beta \omega_k)}-1}\equiv n_B(\omega _k)
\label{2.12}
\eeq
where $E_n=\sum_{\bf k} \omega_k n({\bf k})$ (see e.g. {\it Reif, 
Fundamentals of Statistical and Thermal Physics}) and
$n_B(\omega_k)$ is the Bose-Einstein distribution. 
(In the case of fermions, we have according to the Pauli exclusion principle 
$n({\bf k})\epsilon \{0,1\}$, from which the Fermi-Dirac distribution
$n_F(\omega _k)=1/[\exp(\beta(\omega_k-\mu))+1]$ follows. Owing to
particle number conservation, e.g. charge or baryon number conservation,
the average number of particles is fixed, which is taken into account 
by introducing the chemical potential $\mu$.)
Combining (\ref{2.11}) and (\ref{2.12}) we find
\beq
i\> \Delta_F^{T>0} (x-y) = \int \dk \> \frac {1}{2\omk }\>
\left \{ [1+n_B(\omk )]\> e^{-iK(x-y)} + n_B(\omk )\>
e^{iK(x-y)}\right \}.
\label{2.13}
\eeq
Indeed for $T=0$, i.e. $n_B=0$, we recover the vacuum result (\ref{2.4}). 

The physical interpretation of this expression is the following: 
As at zero temperature the finite temperature propagator describes the 
propagation of a scalar particle from $y$ to $x$ ($x_0>y_0$). However,
besides spontaneous creation at $y$ there is also induced
creation ($\sim n_B$) and absorption ($\sim n_B$) at $x$ due to the 
presence of the thermal particles in the heat bath.

Next we want to give a 4-dimensional $K$-integral representation of 
$\Delta_F^{T>0}$ from which we can derive Feynman rules in momentum space.

\subsection{Imaginary Time Formalism (ITF)}

We start with the following statement: 
Going to imaginary times $t$ with $0\leq \tau \equiv it <\beta$
and summing over discrete energies $k_0=2\pi i n T$ (Matsubara frequencies)
instead of integrating, i.e.,
\bed
\int \frac{dk_0}{2\pi} \rightarrow iT\> \sum_{n=-\infty}^\infty, \nonumber
\eed
the propagator (\ref{2.13}) can be written as
\beq
i\> \Delta _F^{T>0} (x)=iT\> \sum_n \int \frac{d^3k}{(2\pi )^3}\> 
\frac{i}{K^2-m^2}\>e^{-iK\cdot x}.
\label{2.14}
\eeq

\medskip

Proof: (\ref{2.14}) can be written as
\bed
i\> \Delta _F^{T>0} (x)=-T\> \int \frac{d^3k}{(2\pi )^3}\> \sum_n 
\frac{1}{k_0^2-\omega_k^2}\> e^{-k_0\cdot \tau}\> e^{i{\bf k}\cdot {\bf x}}.
\nonumber
\eed

Now we use the formula
\bea
T \sum _{n=-\infty}^{\infty} f(k_0=2\pi inT) & = & \frac {1}{2\pi i}\> \int
_{-i\infty}^{i\infty} dk_0 \frac {1}{2}\> [f(k_0)+f(-k_0)]\nonumber \\
                                             & +  & \frac {1}{2\pi i}\>
\int_{-i\infty +\epsilon}^{i\infty +\epsilon} dk_0\> [f(k_0)+f(-k_0)]\>
n_B(k_0),
\nonumber
\eea
which holds if $f(k_0)$ has no poles on the imaginary axis (see problem \#2).

 Choosing
\bed
f(k_0)=\frac{-1}{k_0^2-\omega_k^2}\> e^{-k_0\tau}
\nonumber
\eed
we get
\bed
-T\> \sum_{n=-\infty}^\infty \frac{1}{k_0^2-\omega_k^2}\> 
e^{-k_0\cdot \tau}=I_1+I_2+I_3+I_4,
\nonumber
\eed
with
\bea
I_1 &=& -\frac{1}{2\pi i}\> \frac{1}{2}\> \int_{-i\infty}^{i\infty} dk_0\>
\frac{e^{-k_0\tau}}{k_0^2-\omega_k^2}\nonumber \\
&=&\frac{1}{2}\> {\rm Res}\, \left [\frac{e^{-k_0\tau}}{k_0^2-\omega_k^2}
\right ]_{k_0=\omega_k}\nonumber \\
&=& \frac{1}{2}\> \frac{e^{-\omega_k\tau}}{2\omega_k},\nonumber
\eea
where we have closed the contour in the half plane ${\rm Re}\, k_0>0$
(see Fig.5).
Analogously, closing the contour for  ${\rm Re}\, k_0<0$ we find 
$I_2=I_1$.

Furthermore we have
\bea
I_3 &=& -\frac{1}{2\pi i}\> \int_{-i\infty+\epsilon}^{i\infty+\epsilon} 
dk_0\> \frac{1}{k_0^2-\omega_k^2} \frac{e^{-k_0\tau}}{e^{\beta k_0}-1}\nonumber \\
&=& n_B(\omega_k)\> \frac{e^{-\omega_k\tau}}{2\omega_k},\nonumber
\eea
where we have used the contour shown in Fig.6. For
$I_4$ we have to use the same contour as in Fig.6, since $\beta >\tau$,
yielding
\bed
I_4=n_B(\omega_k)\> \frac{e^{+\omega_k\tau}}{2\omega_k}.\nonumber
\eed

\begin{figure}

\centerline{\psfig{figure=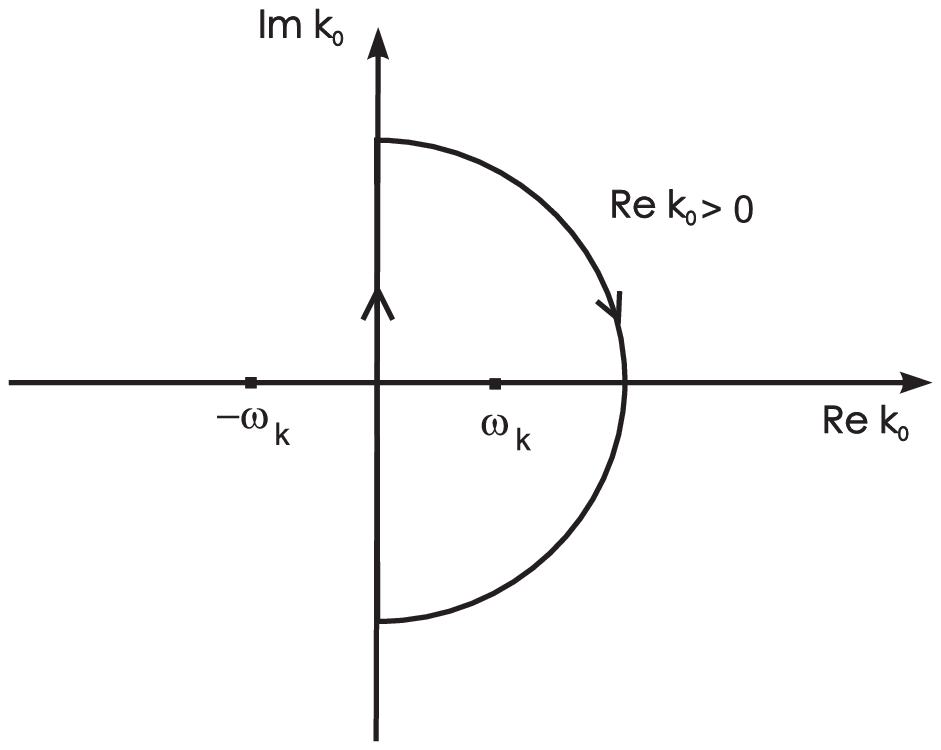,width=6cm}}
\centerline{\it Fig.5}

\end{figure}

Combining these expressions we find
\bea
&& i\> \Delta_F^{T>0} (x) = \int \dk \> e^{i{\bf k}\cdot{\bf x}}\> \left 
[\frac{e^{-\omk \tau}}{2\omk }+\frac{n_B(\omk )}{2\omk }\> \left (e^{-\omk \tau}+
e^{\omk \tau}\right )\right ]\nonumber \\
&& =\int \dk \> \frac {1}{2\omk }\> \left \{ [1+n_B(\omk )]\> e^{-i\omk t}\> 
e^{i{\bf k}\cdot {\bf x}} + n_B(\omk )\> e^{i\omk t}\> e^{i{\bf k}\cdot {\bf x}}
\right \}.\nonumber
\eea
Replacing ${\bf k}$ by $-{\bf k}$ in the second part of the last expression
we recover (\ref{2.13}).\hfill $\Box $

\begin{figure}

\centerline{\psfig{figure=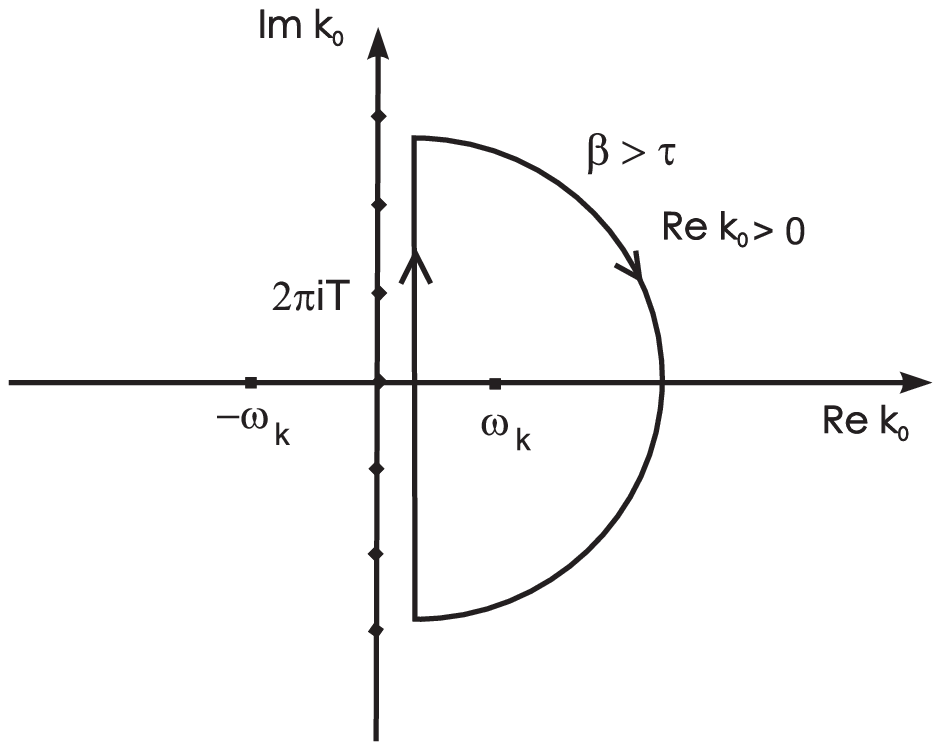,width=6cm}}
\centerline{\it Fig.6}

\end{figure}

\medskip

Another motivation for introducing an imaginary time is the following:
for $\tau=it \rightarrow \beta$ the Boltzmann factor $\exp (-\beta H)$ 
assumes the form of the time evolution operator $\exp (-iHt)$.
As a consequence thermal propagators (more general, all Green functions) 
become periodic with $\beta$, since
\bea
&&\Delta_F^{T>0} (x-y)=\Delta_F^{T>0} ({\bf x},{\bf y},\tau ,0)\qquad (\tau_x=\tau ,
\tau_y=0)\nonumber \\
&&=\frac{1}{Z}\> {\rm Tr}\, \left [e^{-\beta H}\> T\{\phi ({\bf x},\tau)\phi ({\bf y},0)
\}\right ]\qquad
(\beta>\tau >0)\nonumber \\
&&=\frac{1}{Z}\> {\rm Tr}\, \left [e^{-\beta H}\> \phi ({\bf x},\tau)\> \phi ({\bf y},0)
\right ]\nonumber \\
&&=\frac{1}{Z}\> {\rm Tr}\, \left [\phi ({\bf y},0)\> e^{-\beta H}\> \phi ({\bf x},\tau)
\right ]\nonumber \\
&&=\frac{1}{Z}\> {\rm Tr}\, \left [e^{-\beta H}\> e^{\beta H}\>\phi ({\bf y},0)\> 
e^{-\beta H}\> \phi ({\bf x},\tau) \right ]\nonumber \\
&&=\frac{1}{Z}\> {\rm Tr}\, \left [e^{-\beta H}\> \phi ({\bf y},\beta)\> \phi ({\bf x},\tau)
\right ]\nonumber \\
&&=\frac{1}{Z}\> {\rm Tr}\, \left [e^{-\beta H}\> T\{\phi ({\bf x},\tau)\phi ({\bf y},\beta)
\}\right ]\nonumber \\
&&=\Delta_F^{T>0} ({\bf x},{\bf y},\tau ,\beta).\nonumber 
\eea
In general 
\beq
\Delta_F^{T>0} (\tau )=\Delta_F^{T>0}(\tau +n\beta)\qquad n\; {\rm integer}
\label{2.15}
\eeq
holds. This has two consequences

1. The time $\tau $ is restricted to the interval $[0,\beta [$, 
known as Kubo-Martin-Schwinger- or KMS-condition.

2. The Fourier integral over $k_0$ at $T=0$ goes over to a Fourier series over
the Matsubara frequencies $k_0=2\pi inT$.

(For fermions we have $S_F^{T>0}(\tau)=(-1)^n\, S_F^{T>0}(\tau+n\beta)$
due to a minus sign in the definition of time ordering corresponding
to anti-commuting fields, from which we get $k_0=(2n+1)i\pi T$.)

The Feynman rules in the ITF for example for the $\phi^4$-theory
now read:

1. The propagator is given by $i\> \Delta_F^{T>0}(K) =i/(K^2-m^2)$ with 
$k_0=2\pi i nT$.

2. In loop integrals we have to make the replacement $\int {d^4K}/{(2\pi )^4} 
\rightarrow iT\> \sum_{k_0} \int {d^3k}/{(2\pi )^3}$.

3. The vertex reads as in vacuum $-i\> 4!\> g^2$. (In order to compare 
with gauge theories we denote the coupling constant as $g^2$.)

4. Symmetry factors, e.g. $1/2$ for tadpole, are the same as in vacuum. 

\begin{figure}

\centerline{\psfig{figure=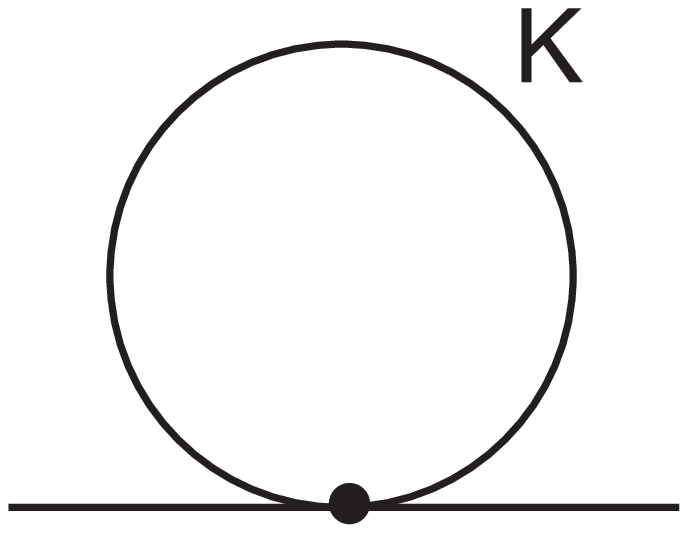,width=4cm}}
\centerline{\it Fig.7}

\end{figure}

As the simplest example for a loop diagram we consider the tadpole of
the $\phi^4$-theory shown in Fig.7. According to the above Feynman rules 
it is given by
\bea
\Pi & = & \frac {i}{2}\> (-i\, 4!\, g^2)\> iT\> \sum _{k_0} \int \dk \>
i\> \frac{1}{K^2-m^2}\nonumber \\
& = & -12\> g^2\> T\> \int \dk \> \sum _{n=-\infty}^{\infty} \frac
{1}{(2\pi inT)^2-\omk ^2}\nonumber \\
& = & 12\> g^2\> T\> \frac{1}{(2\pi T)^2}\> \int \dk \> \sum _{n=-\infty}^{\infty} \frac
{1}{n^2+\Omega^2}\nonumber 
\eea
with $\Omega\equiv \omk/(2\pi T)$. 
Using (see {\it Gradshteyn, Ryzhik, Tables of Integrals})
\bed
\coth(\pi \Omega)=\frac{1}{\pi \Omega}+\frac{2\Omega}{\pi} \sum_{n=1}^\infty
\frac{1}{n^2+\Omega^2} \nonumber
\eed
we obtain
\bea 
&&\sum_{n=-\infty}^\infty \frac{1}{n^2+\Omega^2}=2 \sum_{n=1}^\infty
\frac{1}{n^2+\Omega^2}+\frac{1}{\Omega^2}=\frac{\pi}{\Omega}\> \coth (\pi \Omega)\nonumber \\
&&=\frac{\pi}{\Omega}\> \frac{e^{\pi \Omega}+e^{-\pi \Omega}}{e^{\pi \Omega}-e^{-\pi \Omega}}
=\frac{\pi}{\Omega}\> \frac{e^{2\pi \Omega}+1}{e^{2\pi \Omega}-1}\nonumber \\
&&=\frac{\pi}{\Omega}\> \left (1+\frac{2}{e^{2\pi \Omega}-1}\right )=
\frac{2\pi^2T}{\omk}\> [1+2n_B(\omk )]\nonumber
\eea
leading to 
\beq
\Pi = 6\> g^2\> \int \dk \> \frac{1}{\omk}\> [1+2n_B(\omk)].
\label{2.16}
\eeq

Note that only the vacuum part ($n_B=0$) is ultraviolet divergent,
since the distribution functions falls off exponentially for large momenta.
Hence we can use the same renormalization as at $T=0$. Using 
dimensional regularization the tadpole at $T=0$ vanishes,
resulting in
\bed
\Pi = 12\> g^2\> \frac{4\pi}{(2\pi )^3}\> \int_0^\infty dk \> 
\frac{k^2}{\omk}\> n_B(\omk).
\eed
An analytic expression for this integral is only possible for $m=0$,
for which we get
\bed
\Pi = \frac{6}{\pi^2}\> g^2\> \int_0^\infty dk \> \frac{k}{e^{k/T}-1}
{\buildrel{x\equiv k/T}\over =} \frac{6}{\pi^2}\> g^2\> T^2\> 
\int_0^\infty dx \> \frac{x}{e^{x}-1}.\nonumber
\eed
Using
\bed
\int_0^\infty dx\> \frac{x^{n-1}}{e^{x}-1} = (n-1)!\> \zeta(n), \qquad \qquad  
\zeta(2) =\frac{\pi^2}{6}\nonumber
\eed
we end up with the simple result
\beq
\Pi = g^2T^2.
\label{2.17}
\eeq

\begin{figure}

\centerline{\psfig{figure=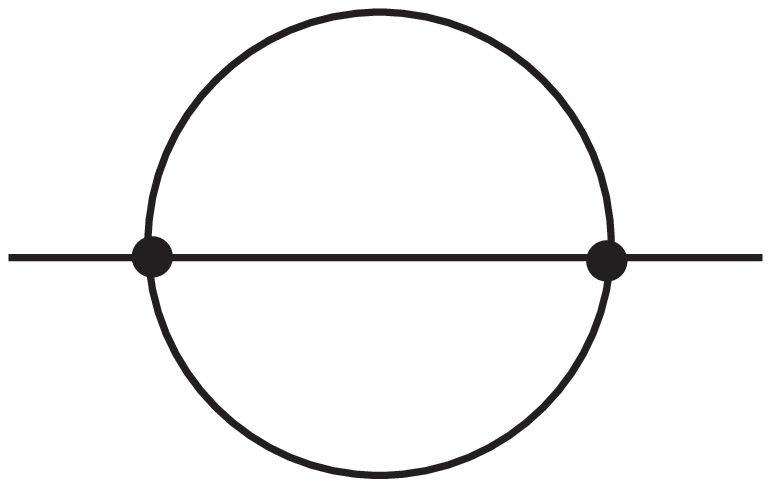,width=4cm}}
\centerline{\it Fig.8}

\end{figure}

The problem is now that for diagrams with more than one propagator, e.g.
the one shown in Fig.8, the summation over $k_0$ becomes difficult.
A convenient way out of this problem is the so-called
Saclay method which is a mixed representation performing the Fourier 
transformation in time only. (In the following we use $\Delta \equiv 
\Delta_F^{T>0}$.) This leads to
\beq
\Delta (\tau, \omk )=-T\> \sum _{k_0} e^{-k_0\tau }\> \Delta (K),
\label{2.18}
\eeq
where the Fourier coefficients are given as
\beq
\Delta (K) = -\int _0^\beta d\tau \> e^{k_0\tau }\> \Delta (\tau, \omk ).
\label{2.19}
\eeq
Following the proof of (\ref{2.14}) we can perform the sum (\ref{2.18}), 
\bed
\Delta (\tau, \omk )=\frac{1}{2\omk }\> \left \{ [1+n_B(\omk )]\>
e^{-\omk \tau } + n_B(\omk )\> e^{\omk \tau }\right \}.
\nonumber
\eed
Now the Matsubara frequency $k_0$, over which we have to sum, appears in 
the propagator (\ref{2.19}) only in the exponent, which makes the summation 
simple at the expense of introducing another integral over $\tau $.
For example for the tadpole we get now 
\bea
\Pi & = & \frac {i}{2}\> (-i\, 4!\, g^2)\> iT\> \sum _{k_0} \int \dk \>
i\> (-1)\int _0^\beta d\tau \> e^{k_0\tau }\> \Delta (\tau, \omk ) \nonumber \\
& = & 12\> g^2\> \int \dk \int _0^\beta\> d\tau 
\Delta (\tau, \omk )\> T\> \sum _{n=-\infty}^{\infty} e^{k_0\tau}. \nonumber 
\eea
The sum over $k_0$ reduces to a $\delta$-function 
\bed
T\> \sum _{n=-\infty}^{\infty} e^{k_0(\tau -\tau ')}=T\> \sum_n
e^{2\pi inT(\tau -\tau')}= T \delta(T(\tau -\tau')) =\delta(\tau -\tau').
\nonumber
\eed
This formula makes the Saclay method very convenient in the case of 
two or more propagators. In the case of the tadpole it yields 
\bea
\Pi &=& 12\> g^2\> \int \dk \Delta (0,\omk)\nonumber \\
&=& 6\> g^2\> \int \dk \> \frac{1}{\omk}\> [1+2n_B(\omk)],
\nonumber
\eea
which is identical to our result (\ref{2.16}).

\subsection{Real Time Formalism (RTF)}

Here we do not aim at a formal derivation, but will present only a
motivation and plausibility arguments. For a more detailed derivation
the interested reader is referred e.g. to Ref.[2].

The starting point is again our basic equation (\ref{2.13}) for the
propagator,
\bed
i\> \Delta (x-y) = \int \dk \> \frac {1}{2\omk }\>
\left \{ [1+n_B(\omk )]\> e^{-iK(x-y)} + n_B(\omk )\>
e^{iK(x-y)}\right \}\biggl |_{k_0=\omk}.
\nonumber
\eed
Instead of introducing Matsubara frequencies we will give an
alternative expression for writing (\ref{2.13}) as a 4-dimensional 
integral, namely
\beq
i\> \Delta (x-y) = \int \frac {d^4K}{(2\pi )^4}\> \left [\frac {i}
{K^2-m^2+i\epsilon } + 2\pi \> n_B(|k_0|)\> \delta (K^2-m^2)\right ]\>
e^{-iK(x-y)}.
\label{2.20}
\eeq

The complex $k_0$-integration over the first term gives the $T=0$
term (\ref{2.4}), i.e., the first term of (\ref{2.13}) following 
from setting $n_B=0$. The second term gives
\bea
&& \int \frac {d^4K}{(2\pi )^4}\> 2\pi \> n_B(|k_0|)\> \delta (K^2-m^2)\>
e^{-iK(x-y)}\nonumber \\
&& =\int \dk \> dk_0\> n_B(|k_0|)\> \frac{1}{2\omk }\> [\delta (k_0-\omk )
+\delta (k_0+\omk)]\> e^{-ik_0(x_0-y_0)}\> e^{i{\bf k}\cdot ({\bf x}-{\bf y})}
\nonumber \\
&& =\int \dk \> \frac{n_B(\omk )}{2\omk}\> \left [e^{-i\omk (x_0-y_0)}+
e^{i\omk (x_0-y_0)}\right ] \> e^{i{\bf k}\cdot ({\bf x}-{\bf y})}
\nonumber \\
&& =\int \dk \> \frac {1}{2\omk }\>
\left [ n_B(\omk )\> e^{-iK(x-y)} + n_B(\omk )\> e^{iK(x-y)}\right 
]\biggl |_{k_0=\omk},
\eea
which proves the equivalence of (\ref{2.20}) and (\ref{2.13}).

The advantages of this representation are the following.

1. There is no Matsubara sum, but an integration over $k_0$ as at $T=0$.

2. We do not need an imaginary time, which is restricted to finite 
temperatures and cannot be generalized to non-equilibrium. Actually, as we 
will see in chapter 5, the RTF can be used as a starting point for
non-equilibrium field theory.

3. The distribution functions appear from the beginning. Hence  the
$T=0$ and $T>0$ contributions are disentangled.

The Feynman rule in momentum space in the RTF for the scalar propagator reads
now:
\beq
\frac {i}{K^2-m^2+i\epsilon } + 2\pi \> n_B(|k_0|)\> 
\delta (K^2-m^2).
\label{2.21}
\eeq

Otherwise we have the same rules as at $T=0$.

The fermion propagator is given by
\beq
S(P)=(P\sla+M)\> \left [\frac {i}{P^2-M^2+i\epsilon } - 2\pi \> 
n_F(|p_0|)\> \delta (P^2-M^2)\right ].
\label{2.22}
\eeq

Again we will consider the tadpole in $\phi^4$-theory as an example. 
In the RTF it is given as
\bea
\Pi & = & \frac {i}{2}\> (-i\, 4!\, g^2)\> \int \d4k \>
i\> \Delta (K) \nonumber \\
& = & 12\> g^2\> \int \d4k \> \left [\frac {i}{K^2-m^2+i\epsilon } + 2\pi \> 
n_B(|k_0|)\> \delta (K^2-m^2)\right ]\nonumber \\
& = & 12\> g^2\> \int \dk \> dk_0\> n_B(|k_0|)\> \frac{1}{2\omk }\> [\delta (k_0-\omk )
+\delta (k_0+\omk)]\nonumber \\
& = & 12\> g^2\> \int \dk \> \frac{n_B(\omk)}{\omk },\nonumber
\eea
where we have neglected the vacuum part, which vanishes after renormalization,
again. This result coincides with the one, (\ref{2.16}), found in the ITF.

\begin{figure}

\centerline{\psfig{figure=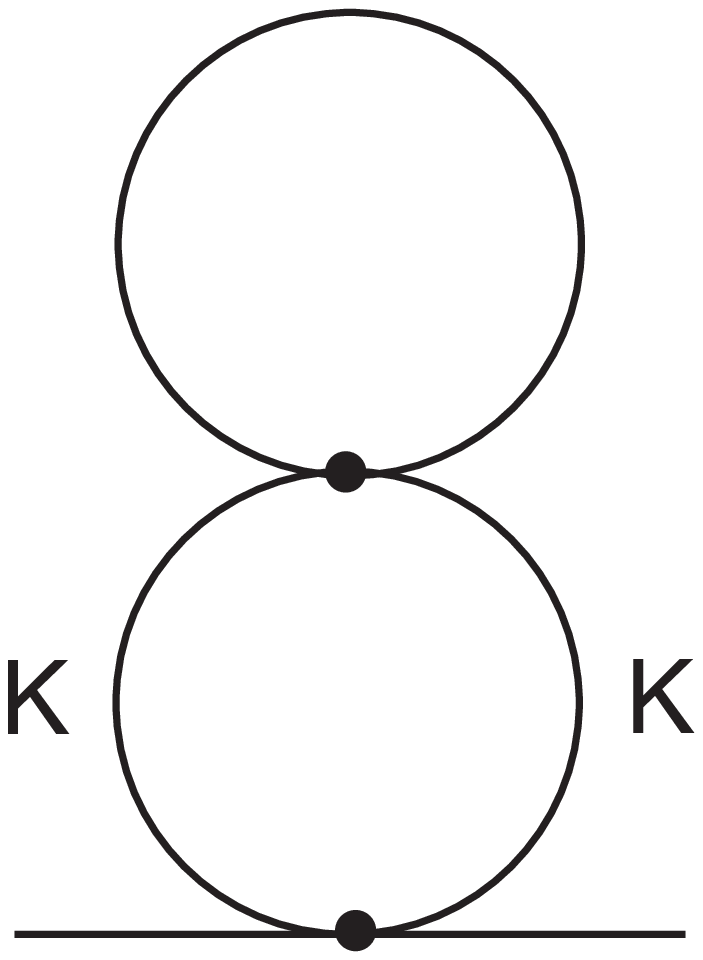,width=4cm}}
\centerline{\it Fig.9}

\end{figure}

Unfortunately there is a serious problem with the naive RTF, presented
above. Singularities, so-called pinch singularities, coming from 
$\delta $-functions are possible in diagrams with two or more propagators,
as e.g. in Fig.9. The lower loop in this diagram is proportional to
$\int d^4K \Delta^2(K)\sim \int d^4K\> [\delta(K^2-m^2)]^2\sim 
\delta(0)=\infty $. 

A possible solution of this problem is the so-called ``doubling of 
degrees of freedom''. According to the KMS-condition the time
of the fields goes from $t=0$ to $t=-i\beta$. This contour can be deformed
in order to include the real time axis by going first from $t=0$ to $t=\infty$
above the real time axis and then back to $t=-i\beta$ below the real time
axis. This deformation corresponds to two different kind of fields, one
existing above and one below the real time axis. Then the
propagator, which contains two fields, is given by a
$2\times 2$ matrix (for details see e.g. Ref.[2])
\beq
\Delta (K)=\left (\begin{array}{cc} \frac{1}{K^2-m^2+i\epsilon} & 0\\
                              0 & \frac{-1}{K^2-m^2-i\epsilon}\\
            \end{array} \right ) 
-2\pi i\, \delta (K^2-m^2)\> \left (\begin{array}{cc}
n_B(k_0) & \theta (-k_0)+n_B(k_0)\\
\theta (k_0)+n_B(k_0) & n_B(k_0) \\ \end{array} \right ),
\label{2.23} 
\eeq
where we have used here and in the following  $n_B(k_0)\equiv n_B(|k_0|)$.

The Feynman rules for the propagator read now 
$i\> \Delta_{ij}(K)$ with $i,j\epsilon \{1,2\}$. 
The vertex is given by $i\> (-1)^j\> g^2\> 4!$. Fields from above and below 
the real time axis are not mixed at a vertex and the vertex of the
``type-2-fields'' has an additional minus sign coming from the
anti-time ordering of the fields below the real time axis.

Again we consider the tadpole diagram as example. Also 
self energies are now given by matrices. The $\Pi_{11}$
component, given by Fig.10, reads
\bed
\Pi_{11}=\frac{i}{2}\> i\> (-1)\> g^2\> 4!\> \int \d4k \> 2\pi \> 
n_B(k_0)\> \delta (K^2-m^2),\nonumber 
\eed
which agrees with (\ref{2.16}). The components
$\Pi_{12}=\Pi_{21}=0$, since all legs of the vertex must have
the same index.  The component $\Pi_{22} = -\Pi_{11}$ because of the
minus sign from the ``type-2-field'' vertex.

\begin{figure}

\centerline{\psfig{figure=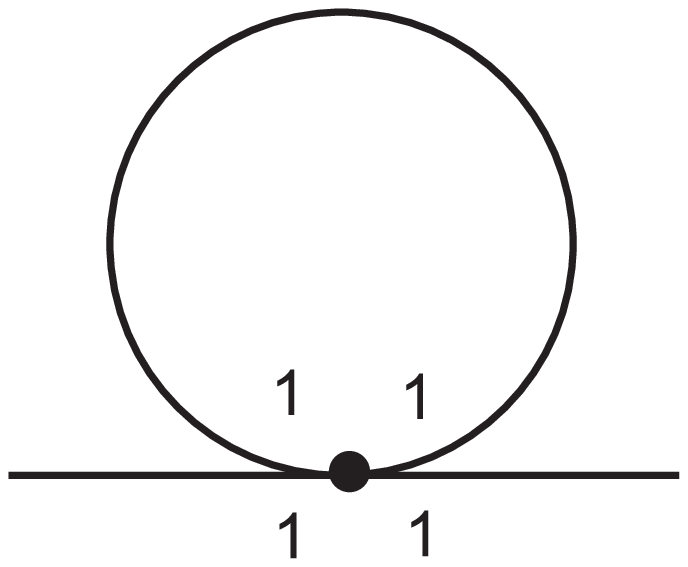,width=4cm}}
\centerline{\it Fig.10}

\end{figure}

It is important to note that the individual components of the propagator
are not independent:
\bea
&& \Delta_{11}-\Delta_{12}-\Delta_{21}+\Delta_{22}\nonumber \\
&& =\frac{1}{K^2-m^2+i\epsilon}-2\pi i\, \delta (K^2-m^2)\, n_B(k_0)+2\pi i\, \delta (K^2-m^2)\, [\theta (-k_0)+n_B(k_0)]\nonumber \\
&& +2\pi i\, \delta (K^2-m^2)\, [\theta (k_0)+n_B(k_0)]-\frac{1}{K^2-m^2-i\epsilon}-2\pi i\, \delta (K^2-m^2)\, n_B(k_0)\nonumber \\
&& =\frac{1}{K^2-m^2+i\epsilon}-\frac{1}{K^2-m^2-i\epsilon}+2\pi i\, \delta (K^2-m^2)\nonumber \\
&& =\frac{-2i\epsilon}{(K^2-m^2)^2+\epsilon^2}+2\pi i\, \delta (K^2-m^2)\nonumber \\
&& =2i\> {\rm Im}\, \frac{1}{K^2-m^2+i\epsilon}+2\pi i\, \delta (K^2-m^2)=0,
\label{2.24}
\eea
where we have used for the last step the important relation
\beq
\delta (K^2-m^2)=-\frac{1}{\pi }\> {\rm Im}\, \frac{1}{K^2-m^2+i\epsilon}. 
\label{2.25}
\eeq
Note that in contrast to ITF factors $i \epsilon $ are important in RTF.

Now we consider the dangerous 2-loop diagrams, for which 
there are 2 contributions to $\Pi_{11}$ as shown in Fig.11.
The integral for the lower loop contains
$\Delta_{11}(K)\Delta_{11}(K)-\Delta_{12}(K)\Delta_{21}(K)$.
(Note the minus sign from the ``type-2-field'' vertex, which implies
that the rules of matrix multiplication do not hold.) Explicitly
this expression can be written as
\bea
&&\left [\frac{1}{K^2-m^2+i\epsilon}-2\pi i\, \delta (K^2-m^2)\, n_B(k_0)\right ]^2-(-2\pi i)\, 
\delta (K^2-m^2)\, [\theta (-k_0)+n_B(k_0)]\nonumber \\
&& \times (-2\pi i)\, \delta (K^2-m^2)\, [\theta (k_0)+n_B(k_0)]\nonumber \\
&&=\left (\frac{1}{K^2-m^2+i\epsilon}\right )^2-\frac{1}{K^2-m^2+i\epsilon}\> 4\pi i\, 
\delta (K^2-m^2)\, n_B(k_0)-4\pi^2\, \delta^2 (K^2-m^2)\, n_B^2(k_0)\nonumber \\
&&+4\pi^2\, \delta^2 (K^2-m^2) [n_B(k_0)+n_B^2(k_0)]\nonumber \\
&& {\buildrel (\ref{2.25})\over =} \left (\frac{1}{K^2-m^2+i\epsilon}\right )^2-\frac{1}{K^2-m^2+i\epsilon}\> 4\pi i\, 
\delta (K^2-m^2)\, n_B(k_0)\nonumber \\
&&-\frac{1}{2\pi i}\> \left (\frac{1}{K^2-m^2+i\epsilon}-\frac{1}{K^2-m^2-i\epsilon}\right )\>
4\pi^2\, \delta (K^2-m^2)\, n_B(k_0)\nonumber \\
&&=\left (\frac{1}{K^2-m^2+i\epsilon}\right )^2-\left (\frac{1}{K^2-m^2+i\epsilon}+\frac{1}{K^2-m^2-i\epsilon}\right )\>
2\pi i\, \delta (K^2-m^2)\, n_B(k_0)\nonumber \\
&& {\buildrel (\ref{2.25})\over =} {\Delta_0}^2(K)+n_B(k_0)\> [{\Delta_0}^2(K)-{\Delta_0^*}^2(K)],\nonumber
\eea
where $\Delta_0=1/(K^2-m^2+i\epsilon )$. One observes that pinch singularities
coming from products of $\delta$-functions are absent due to a cancellation 
between the to diagrams in Fig.11. Similar arguments hold for higher order 
diagrams.

\begin{figure}

\centerline{\psfig{figure=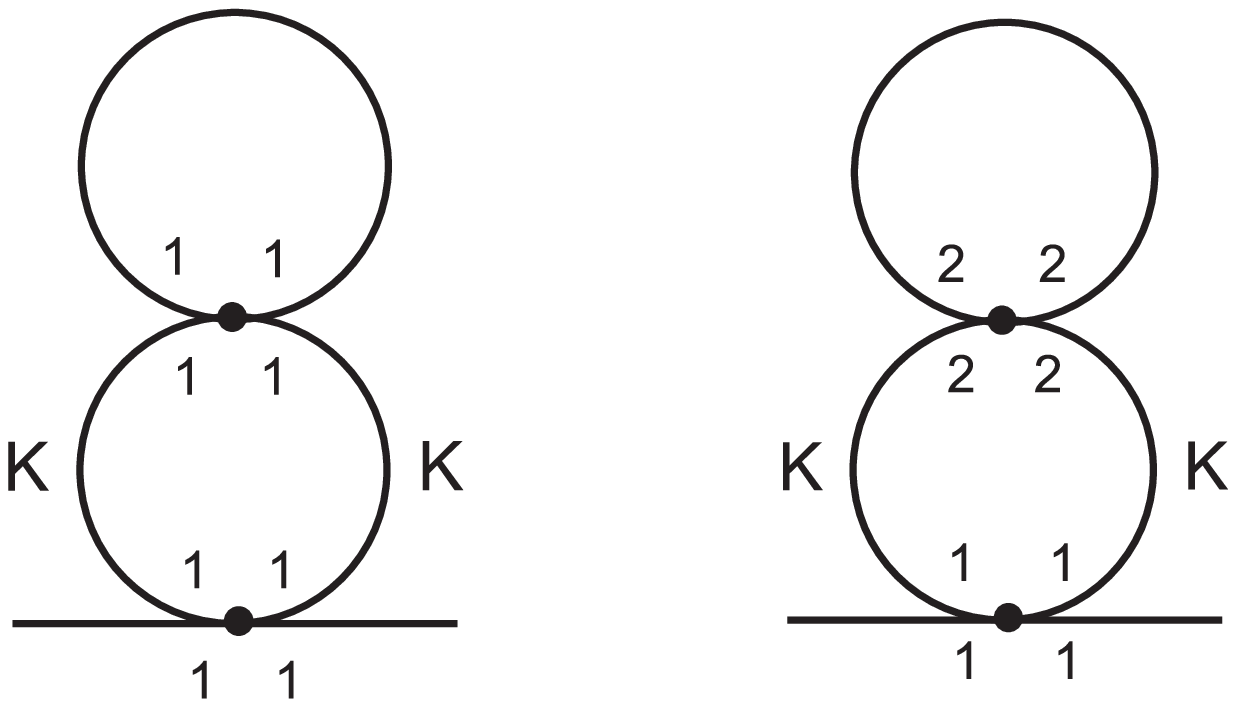,width=6cm}}
\centerline{\it Fig.11}

\end{figure}

Now we will introduce a very convenient representation of the RTF,
the {\it Keldysh representation}. It is constructed from linear 
combinations of the components of the RTF Green functions. The
new components of the propagator are defined as
\bea
&& \mbox{Retarded propagator:}\; \Delta_R\equiv \Delta_{11}-\Delta_{12},\nonumber \\
&& \mbox{Advanced propagator:}\; \Delta_A\equiv \Delta_{11}-\Delta_{21}, \label{2.26} \\
&& \mbox{Symmetric propagator:}\; \Delta_S\equiv \Delta_{11}+\Delta_{22}.
\nonumber
\eea
The component $\Delta_S$ is often denoted as $\Delta_F$ in the literature.
There are only 3 components, which is sufficient  because of (\ref{2.24}).

The inverse relations read
\bea
\Delta_{11} &=& \frac{1}{2}\> (\Delta_S+\Delta_A+\Delta_R),\nonumber \\
\Delta_{12} &=& \frac{1}{2}\> (\Delta_S+\Delta_A-\Delta_R),\nonumber \\
\Delta_{21} &=& \frac{1}{2}\> (\Delta_S-\Delta_A+\Delta_R),\nonumber \\
\Delta_{22} &=& \frac{1}{2}\> (\Delta_S-\Delta_A-\Delta_R).
\label{2.27}
\eea

Explicitly the retarded propagator is given by
\bea
\Delta_R&=&\frac{1}{K^2-m^2+i\epsilon}-2\pi i\, \delta (K^2-m^2)\, n_B(k_0)
+2\pi i\, \delta (K^2-m^2)\, [\theta (-k_0)+n_B(k_0)]\nonumber \\
&=&\frac{1}{K^2-m^2+i\epsilon}-\theta(-k_0)\> \left [\frac{1}{K^2-m^2+i\epsilon} 
-\frac{1}{K^2-m^2-i\epsilon}\right ]\nonumber \\
&=&\left \{ \begin{array}{c} (K^2-m^2+i\epsilon)^{-1}\; \; \; \; \; k_0>0\\
(K^2-m^2-i\epsilon)^{-1}\; \; \; \; \; k_0<0 \end{array} \right. \nonumber \\
&=&\frac{1}{K^2-m^2+i\, \mbox{sgn}(k_0) \epsilon}.
\label{2.28}
\eea
It has both poles above the real axis, as shown in Fig.12. Analogously we get
for the advanced propagator
\beq
\Delta_A=\frac{1}{K^2-m^2-i\, \mbox{sgn}(k_0) \epsilon},
\label{2.29}
\eeq
which has both poles below the real axis.
The symmetric propagator follows as
\bea
\Delta_S&=&\frac{1}{K^2-m^2+i\epsilon}-\frac{1}{K^2-m^2-i\epsilon}
-4\pi i\, \delta (K^2-m^2)\, n_B(k_0)\nonumber \\
&=&-2\pi i\, \delta (K^2-m^2)\, [1+2n_B(k_0)].
\label{2.30}
\eea

\begin{figure}

\centerline{\psfig{figure=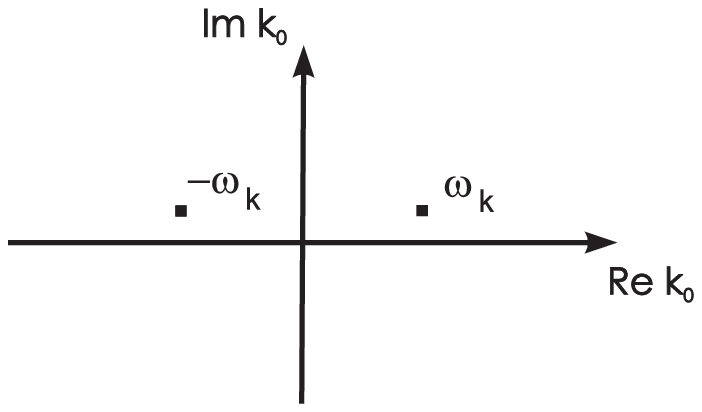,width=6cm}}
\centerline{\it Fig.12}

\end{figure}

From these explicit expressions, we see that only the symmetric
propagator contains a distribution function. This fact allows for a fast 
calculation of the thermal contributions of diagrams.

Self energies are related to the bare and full propagators ($\Delta$, 
$\Delta^*$) via the Dyson-Schwinger equation (see below),
$\Pi = \Delta^{-1}-{\Delta^*}^{-1}$, from which together with (\ref{2.24})
follows
\beq
\Pi_{11}+\Pi_{12}+\Pi_{21}+\Pi_{22}=0.\label{2.31}
\eeq

Analogously to the propagator we define
\bea
\Pi_R&=&\Pi_{11}+\Pi_{12},\nonumber \\
\Pi_A&=&\Pi_{11}+\Pi_{21},\nonumber \\
\Pi_S&=&\Pi_{11}+\Pi_{22},
\label{2.32}
\eea
where the different sign in $\Pi_{R,A}$ is a consequence of (\ref{2.31}).

At this point we will give two important relations for the propagators and 
self energies in the Keldysh representation, which we will use later on:
\beq
\Delta_S(K)=[1+2n_B(k_0)]\> \mbox{sgn}(k_0)\> [\Delta_R(K)-\Delta_A(K)],
\label{2.33}
\eeq
\beq
\Pi_S(K)=[1+2n_B(k_0)]\> \mbox{sgn}(k_0)\> [\Pi_R(K)-\Pi_A(K)].
\label{2.34}
\eeq
The proof in the case of the bare propagator is simple (see problem \#4)
using explicit expressions for the different components (\ref{2.28}) to
(\ref{2.30}) and the relation (\ref{2.25}). However, it can be shown
that (\ref{2.33}) holds also for the full propagator in equilibrium. Out of
equilibrium, on the other hand, modifications are necessary,
as we will discuss in chapter 5.

Finally we will consider again the scalar tadpole as an example for the use
of the Keldysh representation
\bea
\Pi_{R} &=& \Pi_{11} =\frac{i}{2}\> i\> (-1)\> g^2\> 4!\> \int \d4k \> i\, \Delta_{11}(K)\nonumber \\
&=& i\> 12\> g^2\> \int \d4k \> \frac{1}{2}\> (\Delta_S+\Delta_A+\Delta_R)\nonumber \\
&=& 12\> g^2\> \int \d4k \> 2\pi \> n_B(k_0)\> \delta (K^2-m^2),\nonumber 
\eea
which agrees with (\ref{2.16}). The last step follows from the fact that the 
$T=0$-contribution vanishes after renormalization
and only $\Delta_S$ contains a $T>0$-contribution.

\newpage

\section{Hard Thermal Loop Resummation (HTL)}

The HTL resummation technique has been developed in the
late 80's and in the beginning of the 90's by Braaten and Pisarski
({\it Braaten, Pisarski, Nucl. Phys. B337 (1990) 569}) in order to cure 
serious problems of gauge theories at finite temperature
using perturbation theory. For if one uses only bare propagators 
(and vertices), IR singularities and gauge dependent results
have been encountered. A famous example is the damping rate of
a plasma wave in the QGP, which turned out to be different in different gauges.

Braaten and Pisarski suggested the following solution.
Instead of using bare propagators (and vertices) effective propagators,
constructed by resumming certain diagrams, the so-called HTL
self energies, should be adopted. In this way an improved perturbation theory
has been invented.

\subsection{HTL Self Energies}

First we have to isolate the diagrams which should be resummed into
effective propagators. The starting point is the separation of scales
in the weak coupling limit. In a plasma of massless particles, we have
the momentum scale $T$ (hard) and $gT$ (soft) assuming $g\ll 1 $.
HTL self energies are now given by 1-loop diagrams, where the 
external momenta are soft and the loop momenta hard.
As an important example we will consider the photon self energy
or polarization tensor in QED shown in Fig.13. At $T=0$ it reads,
using standard Feynman rules  
\beq
\Pi^{\mu \nu}(P)=-i\> e^2\> \int \d4k\> {\rm tr}\, [\gamma^\mu
\, S(Q)\, \gamma^\nu \, S(K)],
\label{3.1}
\eeq
where $S$ is the electron propagator. 

\begin{figure}[b]

\centerline{\psfig{figure=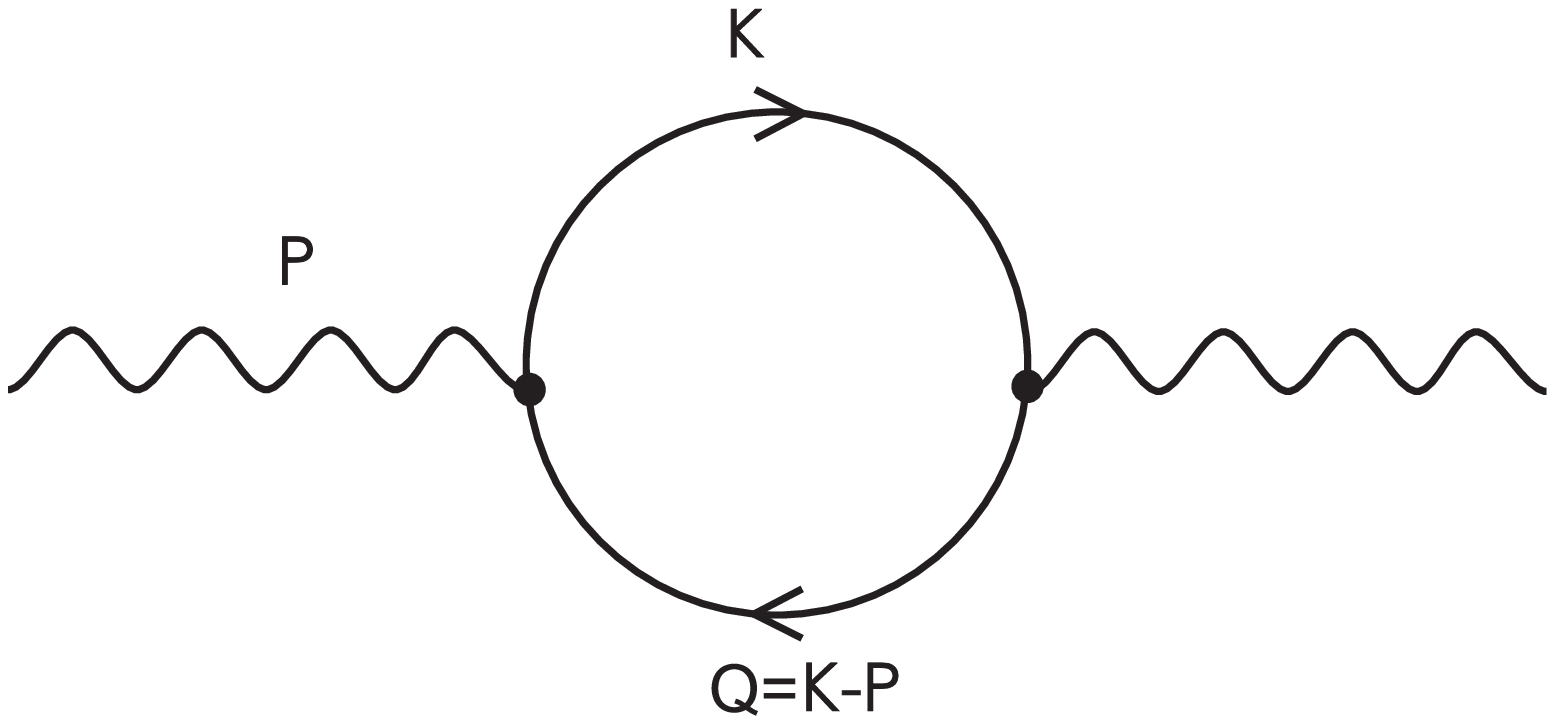,width=10cm}}
\centerline{\it Fig.13}

\end{figure}

For $T>0$ we will adopt the RTF. Then we find according to 
Fig.14 for the retarded self energy
\bea
&&\Pi_R^{\mu \nu}(P)=\Pi_{11}^{\mu \nu}(P)+\Pi_{12}^{\mu \nu}(P)
\nonumber \\
&&=-i\> e^2\> \int \d4k\> \left \{ {\rm tr}\,[\gamma^\mu
\, S_{11}(Q)\, \gamma^\nu \, S_{11}(K)]-{\rm tr}\,[\gamma^\mu
\, S_{21}(Q)\, \gamma^\nu \, S_{12}(K)]\right \}.\nonumber 
\eea
First we will restrict ourselves to the 
longitudinal component $\Pi_R^L\equiv \Pi_R^{00}$.
Furthermore we neglect the bare electron mass, assuming $T\gg m_e$, 
which might be realized for example in a Supernova plasma ($T\simeq 10$
MeV).
We write $S_{ij}(K)\equiv K\sla \> \tilde \Delta_{ij}(K)$,
where $\tilde \Delta_{ij}$ follows from $\Delta_{ij}$ by replacing
$n_B$ by $-n_F$ (compare (\ref{2.21}) and (\ref{2.22})).

\begin{figure}

\centerline{\psfig{figure=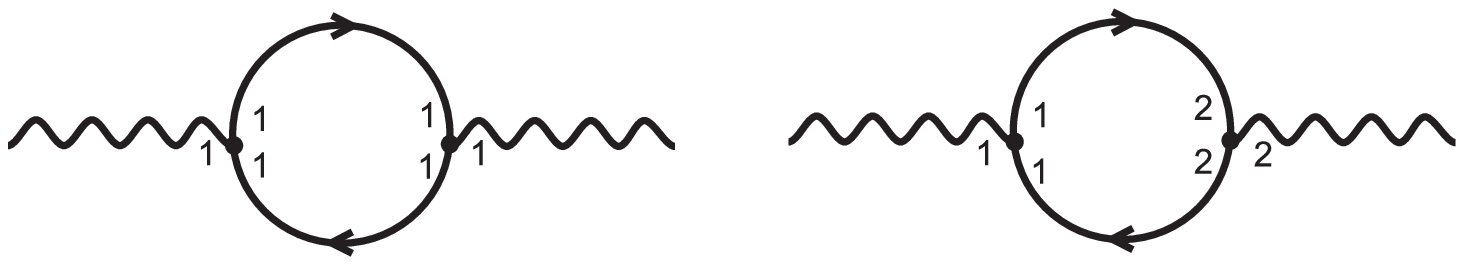,width=10cm}}
\centerline{\it Fig.14}

\end{figure}

The trace over the $\gamma$-matrices gives
\bea
&& {\rm tr}\, [\gamma^\mu\, Q\sla \, \gamma^\nu \, K\sla]
=4\> [Q^\mu \, K^\nu + K^\mu \, Q^\nu -g^{\mu \nu}\, (Q\cdot K)]
\nonumber \\
&& {\buildrel {\mu =\nu =0}\over =} 4\, (2q_0k_0-Q\cdot K)
=4(q_0k_0+{\bf q}\cdot {\bf k}),\nonumber
\eea
from which we obtain
\bed
\Pi_R^L(P)=-4i\> e^2\> \int \d4k\> 
(q_0k_0+{\bf q}\cdot {\bf k})\> \left[\tilde \Delta_{11}(Q)
\, \tilde \Delta_{11}(K)-\tilde \Delta_{21}(Q)\, 
\tilde \Delta_{12}(K)\right ].\nonumber
\eed
The term in the square brackets of this expression 
within the Keldysh representation reads
\bea
&& [ ... ] {\buildrel (\ref{2.27})\over =}\frac{1}{4}\>
[(\tilde \Delta_S(Q)+\tilde \Delta_A(Q)+\tilde \Delta_R(Q))\>  
(\tilde \Delta_S(K)+\tilde \Delta_A(K)+\tilde \Delta_R(K))
\nonumber \\
&&\hspace*{1cm}-(\tilde \Delta_S(Q)-\tilde \Delta_A(Q)
+\tilde \Delta_R(Q))\> (\tilde \Delta_S(K)+\tilde \Delta_A(K)
-\tilde \Delta_R(K))]\nonumber \\
&&=\frac{1}{2}\> [\tilde \Delta_S(Q)\, \tilde \Delta_R(K)+
\tilde \Delta_A(Q)\, \tilde \Delta_S(K)+\tilde \Delta_A(Q)\,  
\tilde \Delta_A(K)+\tilde \Delta_R(Q)\, \tilde \Delta_R(K)].
\nonumber
\eea
Note that there are no terms $\sim \tilde \Delta_S(Q)\, 
\tilde \Delta_S(K)\sim \delta (Q^2)\, \delta (K^2)$,
which lead to singularities for $P=0$.
Also terms $\tilde \Delta_R(K)\, \tilde \Delta_A(K)$,
which cannot be integrated, since there is 
no possibility to close the $k_0$-contour (pinch singularity), 
are absent.

The $k_0$-integration over $\sim \tilde \Delta_A(Q)\, 
\tilde \Delta_A(K)$ and $\sim \tilde \Delta_R(Q)\, 
\tilde \Delta_R(K)$ reduces to zero, as one can close the contour 
always in a half plane that does not contain a pole. Also these
terms do not contribute to the $T>0$-part of the self energy. 
Hence we are left with
\bed
\Pi_R^{L}(P)=-2i\> e^2\> \int \d4k\> 
(q_0k_0+{\bf q}\cdot {\bf k})\> \left[\tilde \Delta_{S}(Q)
\, \tilde \Delta_{R}(K)+\tilde \Delta_{A}(Q)\, 
\tilde \Delta_{S}(K)\right ]\nonumber
\eed
Replacing $K$ by $-Q$ in the first term and using $\Delta_{R}(-Q)
=\Delta_{A}(Q)$ this expression can be simplified further on,
\bea
\Pi_R^{L}(P) &=& -4i\> e^2\> \int \d4k\> 
(q_0k_0+{\bf q}\cdot {\bf k})\> \tilde \Delta_{A}(Q)\, 
\tilde \Delta_{S}(K)\nonumber \\
& {\buildrel{(\ref{2.29}),(\ref{2.30})}\over =}& -8\pi\> e^2\>\int \d4k\> 
(q_0k_0+{\bf q}\cdot {\bf k})\> [1-2n_F(k_0)]\> \delta(K^2)\>
\frac{1}{Q^2-i\rm{sgn}(q_0)\epsilon}.\nonumber
\eea

So far this expression is exact. Now we will consider the HTL-approximation,
i.e., $P\al eT$ and  $K\ag T$. First we discuss the $T=0$-part, corresponding to the
1 in the square brackets in the above formula. At $T=0$ the only scale
is the external momentum $P$. Therefore ${\Pi_R^L}^{T=0}\sim e^2\, P^2$ 
holds. As we will see below this term is of higher order ${\cal O}(e^4)$
for soft momenta $P^2\sim e^2T^2$ compared to the finite temperature part.

The $T>0$-contribution yields after integrating over $k_0$
\bea
\Pi_R^{L}(P) &=& 16\pi\> e^2\>\int \d4k\> 
\frac{n_F(k_0)}{2k}\, [\delta(k_0-k)+\delta(k_0+k)]\>\nonumber \\
&& \frac{(k_0-p_0)k_0+{({\bf k}-{\bf p})}\cdot {\bf k} }
{(k_0-p_0)^2-({\bf k}-{\bf p})^2-i\rm{sgn}(k_0-p_0)\epsilon}\nonumber \\
&=& \frac{e^2}{2\pi^3}\>\int d^3k\> 
\frac{n_F(k)}{k}\> \Biggl [\frac{(k-p_0)k+({\bf k}-{\bf p})\cdot {\bf k}}
{(k-p_0)^2-({\bf k}-{\bf p})^2-i\rm{sgn}(k-p_0)\epsilon}\nonumber \\
&& + \frac{(k+p_0)k+({\bf k}-{\bf p})\cdot {\bf k}}
{(k+p_0)^2-({\bf k}-{\bf p})^2-i\rm{sgn}(-k-p_0)\epsilon}\Biggr ].
\label{3.2}
\eea
Assuming the HTL-approximation, $|p_0|$, $p \ll k$, we will
expand the integrand for small $p_0/k$ and $p/k$:
\bea
[...]&\simeq &\frac{2k^2-p_0k-{\bf p}\cdot {\bf k}}
{-2kp_0+2{\bf k}\cdot{\bf p}+P^2-i\epsilon}+\frac{2k^2+p_0k-{\bf p}\cdot {\bf k}}
{2kp_0+2{\bf k}\cdot {\bf p}+P^2+i\epsilon}\nonumber \\
&\simeq & \frac{k}{-p_0+p\eta -i\epsilon}+\frac{k}{p_0+p\eta +i\epsilon}
+\frac{1}{2}\> \frac{-p_0-p\eta}{-p_0+p\eta -i\epsilon}\nonumber \\
&+& \frac{1}{2}\> \frac{p_0-p\eta}{p_0+p\eta +i\epsilon}
-\frac{1}{2}\> \frac{P^2}{(-p_0+p\eta -i\epsilon)^2}
-\frac{1}{2}\> \frac{P^2}{(p_0+p\eta +i\epsilon)^2}
+O\left(\frac{p}{k}\right ).\nonumber
\eea
The integration over $\eta={\bf p}\cdot {\bf k}/(pk)$ goes from -1 to 1.
The sum of the first two terms is odd under $\eta \rightarrow -\eta$.
Hence the first two terms vanish after integrating over $\eta$.
Then we end up with the final result
\bea
\Pi_R^{L}(P) &=& \frac{e^2}{2\pi^2}\>\int_0^\infty dk\> 
k\> n_F(k)\> \int_{-1}^1 d\eta \Biggl [\frac{-p_0-p\eta}{-p_0+p\eta -i\epsilon}+
\frac{p_0-p\eta}{p_0+p\eta +i\epsilon}\nonumber \\
&-&\frac{P^2}{(-p_0+p\eta -i\epsilon)^2}
-\frac{P^2}{(p_0+p\eta +i\epsilon)^2}\Biggr ]\nonumber \\
&=& -3\, m_\gamma ^2\left (1-\frac{p_0}{2p}\ln \frac{p_0+p+i\epsilon}{p_0-p+i\epsilon}
\right ),
\label{3.3}
\eea
where we introduced the effective thermal photon ``mass''
$m_\gamma^2=e^2T^2/9$, coming from the $k$-integration
over the distribution function. The lower limit of this integration
is zero, although we assumed that $k\gg p$. However, the error
introduced in this way is of higher order, as can be seen
in the following way. Adopting a lower limit $eT\ll k^*\ll T$,
we may replace $n_F(k)$ by its zero momentum limit $1/2$.
Then $k^*$ is the only scale of the soft loop momentum part
of the self energy. Hence it is of order $e^2{k^*}^2\ll e^2T^2$ and
can be neglected compared to the HTL contribution. Also note that
$\Pi^{T>0} \sim e^2 \gg \Pi^{T=0} \sim e^4$, justifying the neglect
of the vacuum part.

Analogously we find for the transverse part of the self energy
\bea
\Pi_R^{T}(P) &\equiv& \frac{1}{2}\> \left (\delta_{ij}-\frac{p_ip_j}{p^2}\right )\> \Pi_{ij}(P)\qquad
(i,\> j\; \epsilon \; \{1,2,3\})\nonumber \\
&=& \frac{3}{2}\, m_\gamma ^2\> \frac{p_0^2}{p^2}\> \left [1-\left (1-\frac{p^2}{p_0^2}
\right )\>
\frac{p_0}{2p}\> \ln \frac{p_0+p+i\epsilon}{p_0-p+i\epsilon} \right ]. 
\label{3.4}
\eea
At $T=0$ there is only one independent component of the polarization tensor
due to Lorentz and gauge invariance: 
\bed
\Pi_{\mu \nu}^{T=0}=\left (g_{\mu \nu}-\frac{P_\mu P_\nu}{P^2} \right )\> \Pi (P).
\nonumber
\eed
At $T>0$ Lorentz invariance is broken, since we have chosen the heat bath as the rest frame.
But transversality, $P^\mu \Pi_{\mu\nu}=0$, still holds as a consequence of gauge invariance.
This leads to two independent components, e.g. $\Pi^{L,T}$ (see e.g. Ref.[1]).

The advanced self energy follows from the retarded one simply by the replacement
\beq
\Pi_A^{L,T}=\Pi_R^{L,T}(i\epsilon \rightarrow -i\epsilon).
\label{3.5}
\eeq 
Finally, the symmetric self energy is given by (see problem \# 5)
\bea
\Pi_S^{L}(P) &=& -\frac{4ie^2}{\pi p}\> \theta (p^2-p_0^2)
\>\int_0^\infty dk\> k^2\> n_F(k)\> [1-n_F(k)]\nonumber \\
&=& -6\pi i\, m_\gamma ^2\> \frac{T}{p}\> \theta (p^2-p_0^2),\label{3.6}\\
\Pi_S^{T}(P)&=& -3\pi i\, m_\gamma ^2\> \frac{T}{p}\> \left (1-\frac{p_0^2}{p^2}\right )\> 
\theta (p^2-p_0^2).\label{3.7}
\eea
Note that it is of lower order, ${\cal O}(eT^2)$, than the retarded and advanced self energies for 
soft momenta $p\sim eT$. Furthermore it is purely imaginary.

Now a couple of remarks are in order:

1. Using the ITF one obtains the same results. In this case one has to continue analytically
the discrete imaginary energy $p_0=2\pi i nT$ to real continuous values.

2. The HTL approximation is equivalent to the high temperature approximation, 
$T\gg p$, $|p_0|$, which has first been studied by {\it Klimov} and {\it Weldon} in 1982,
and to the semiclassical approximation discussed already in 1960 by {\it Silin}.
 
3. In contrast to the tadpole $\Pi^{L,T}$ are momentum dependent.

4. For $p_0^2<p^2$ the self energy has an imaginary part
\beq
\ln \frac{p_0+p\pm i\epsilon}{p_0-p\pm i\epsilon}=\ln \left |\frac{p_0+p}{p_0-p}\right |
\mp i\pi \> \theta (p^2-p_0^2), 
\label{3.8}
\eeq
describing the collisionless energy transfer from the plasma modes (see below)
to the thermal particles of the plasma (Landau damping).

\sm

5. In the static limit, $p_0\rightarrow 0$, the longitudinal self energy reduces to
\beq
\Pi^{L}_{R,A}(p_0\rightarrow 0, p)=-3m_\gamma^2,
\label{3.9}
\eeq
which leads to Debye screening of the electric interaction due to the presence
of charges in the plasma. However, the transverse part reduces to
\beq
\Pi^{T}_{R,A}(p_0\rightarrow 0, p)=0,
\label{3.10}
\eeq
i.e., there is no static magnetic screening in the plasma.

\begin{figure}

\centerline{\psfig{figure=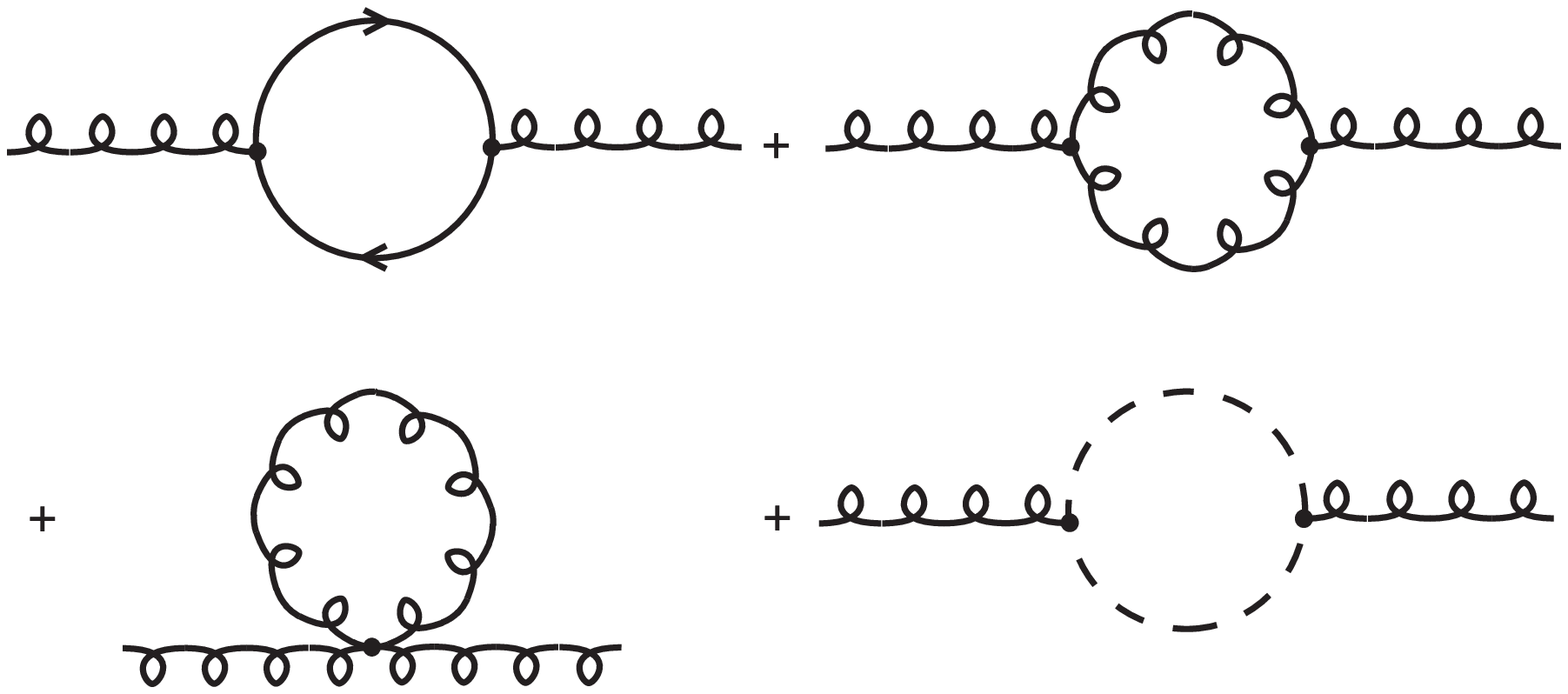,width=12cm}}
\centerline{\it Fig.15}

\end{figure}

6. Considering QCD instead of QED, the gluon self energy is given by Fig.15.
Surprisingly (\ref{3.3}) to (\ref{3.7}) still hold within the HTL approximation if
we simply replace 
\beq
m_\gamma^2 \rightarrow m_g^2=\frac{g^2T^2}{3}\> \left (1+\frac{N_f}{6}\right ),
\label{3.11}
\eeq
where $N_f$ is the number of active flavors in the QGP. Since the 1-loop QED polarization
tensor is gauge invariant, the same holds for the QCD one, which has the same form.

\begin{figure}[b]

\centerline{\psfig{figure=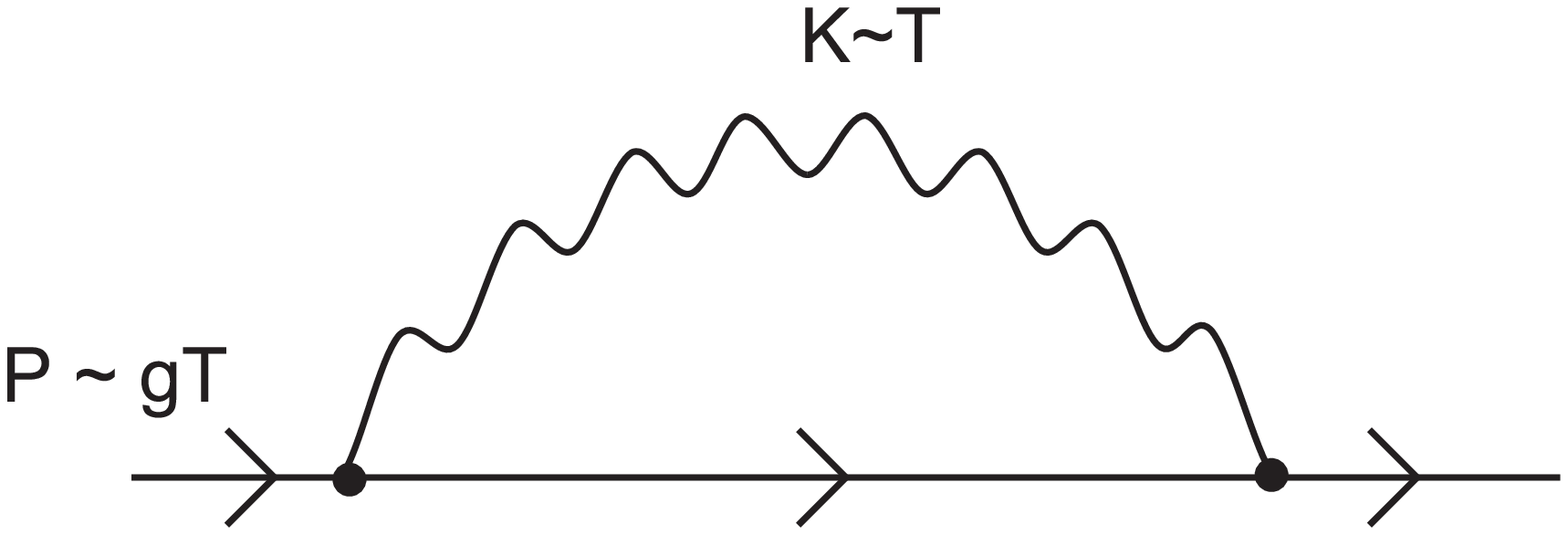,width=6cm}}
\centerline{\it Fig.16}

\end{figure}

7. The fermion self energy can also be derived within the HTL approximation (see Fig.16).
Starting from the most general ansatz for massless fermions at $T>0$,
\beq
\Sigma_R(P)=-a(p_0,p) P\sla - b(p_0,p)\gamma_0
\label{3.12}
\eeq
with
\bea
a(p_0,p) & = & \frac {1}{4p^2}\> \left [{\rm tr}(P\sla \> \Sigma_R )
- p_0\> {\rm tr}(\gamma _0 \> \Sigma_R )\right ],\nonumber\\
b(p_0,p) & = & \frac {1}{4p^2}\> \left [P^2\> {\rm tr}(\gamma _0\> \Sigma_R )
- p_0\> {\rm tr} (P\sla \> \Sigma_R )\right ],
\label{3.13}
\eea
one finds in the HTL approximation
\bea
{\rm tr}(P\sla \> \Sigma_R) & = & 4\> m_F^2\; ,\nonumber \\
{\rm tr}(\gamma _0\> \Sigma_R) & = & 2\> m_F^2\> \frac {1}{p}\> \ln \frac
{p_0+p+i\epsilon}{p_0-p+i\epsilon}
\label{3.14}
\eea
with the effective fermion mass $m_F^2=e^2T^2/8$ (QED) and $m_F^2=g^2T^2/6$ (QCD).
Again one can show that the fermion self energy, (\ref{3.12}) to (\ref{3.14}),
is gauge independent.

\subsection{Effective propagators and dispersion relations}

\subsubsection{$\phi^4$-theory}

The pole of the bare propagator, $\Delta (K)=1/(K^2-m^2)$,
describes the dispersion relation of a non-interacting, scalar particle:
$K^2-m^2=k_0^2-\omk ^2=0$, i.e. $k_0=\omk =\sqrt{k^2+m^2}$.
(Here we use the notation $\Delta (K)=\Delta_R (K)$ and omit $i\epsilon $.)

We can construct an effective propagator by resumming the self energy using the 
Dyson-Schwinger equation as shown in Fig.17. This diagrammatic equation reads
\bea
i\Delta^*&=&i\Delta+i\Delta(-i\Pi_1)i\Delta^*\nonumber \\
\Delta^*&=&\Delta+\Delta \Pi_1 \Delta +\Delta \Pi_1 \Delta \Pi_1 \Delta +...\nonumber \\
&=& \Delta \sum_{n=0}^\infty (\Pi_1\Delta )^n\nonumber \\
& {\buildrel {g\ll 1}\over =}& \Delta \frac{1}{1-\Pi_1 \Delta}=\frac{1}{\Delta^{-1}-\Pi_1}
=\frac{1}{K^2-m^2-\Pi_1}.
\label{3.15}
\eea
The pole of the effective propagator determines the dispersion relation of an interacting 
collective mode with the effective mass $M=\sqrt{m^2+\Pi_1}$. (Effective masses,
generated by the interaction with a medium, have been introduced in various problems
in physics, e.g. the effective mass of electrons in crystals.)
The dispersion relation of the collective scalar particle is simply given by the one of a
massive particle, $\omega (k) =\sqrt{k^2+M^2}$ (see dashed line in Fig.20).

\begin{figure}

\centerline{\psfig{figure=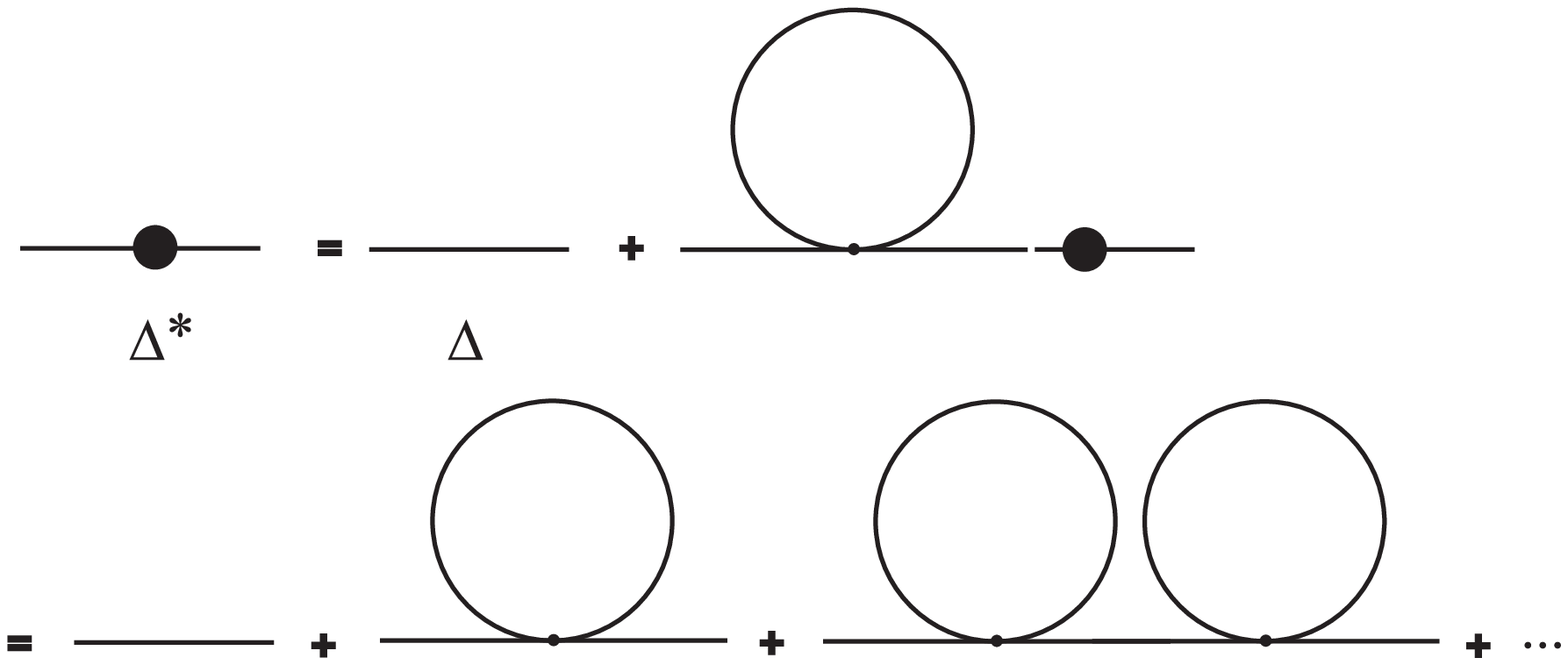,width=12cm}}
\centerline{\it Fig.17}

\end{figure}

\subsubsection{QED}

In gauge theories we have to fix the gauge in order to determine the dispersion relations
for a gauge boson from its propagator. At finite temperature it is convenient to choose  
the Coulomb gauge ${\bf \nabla}\cdot {\bf A}=0$. Since Lorentz invariance is broken,
the choice of a non-covariant gauge is no problem. The bare propagator is then given by
\bea
D_{00}(K) &\equiv & D_L(K)=\frac{1}{k^2},\qquad D_{0i}=D_{i0}=0,\nonumber \\
D_{ij}(K) &\equiv & \left (\delta_{ij}-\frac{k_ik_j}{k^2}\right )\> D_T(K),\qquad D_T(K)=\frac{1}{K^2},
\label{3.16}
\eea
where we used $D=D_{R,A}$, $\Pi=\Pi_{R,A}$.
The longitudinal propagator $D_L$ cannot be associated with a real photon but with the
Coulomb potential, i.e. electric interactions. The transverse propagator
$D_T$, on the other hand, describes 2 transverse massless photons and the
magnetic interaction. 

\begin{figure}

\centerline{\psfig{figure=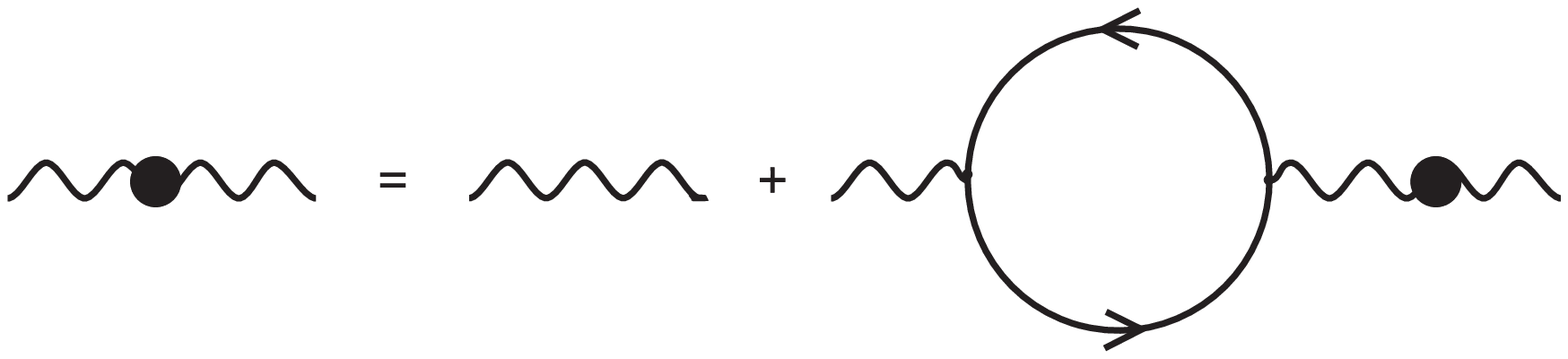,width=12cm}}
\centerline{\it Fig.18}

\end{figure}

The Dyson-Schwinger equation, shown in Fig.18 gives the effective propagator
\bea
D^*_L(K)&=&\frac{1}{k^2-\Pi_L(K)},\nonumber \\
D^*_T(K)&=&\frac{1}{K^2-\Pi_T(K)}.
\label{3.17}
\eea
(See problem \# 6).

Using the HTL-approximation for $\Pi_{L,T}$ we find in the static limit: 
$D^*_L(k_0=0)=1/(k^2-m_D^2)$ with the Debye mass $m_D^2=3m_\gamma^2$,
leading to screening of electric fields by the presence of charges in the plasma.
For the transverse propagator we get in this limit $D^*_T(k_0=0)=1/k^2$, i.e., 
there is no screening of static magnetic fields.

The pole of the effective propagator, ${D^*}_{L,T}(K)^{-1}=0$, determines the
HTL dispersion relations.  Due to the complicated momentum dependence of the
HTL self energies only numerical solutions are possible, which are shown in Fig.19.
Let me make the following remarks:

1. There are two branches. The longitudinal mode  $\omega _L$ is called a plasmon
as in non-relativistic plasma physics. It corresponds to a collective longitudinal 
photon mode that is absent in vacuum.

2. For $k\rightarrow 0$ we have $\omega_{L}(0)=\omega_T(0)=m_\gamma$ which is
called the plasma frequency.

3. For $k\rightarrow \infty$ we find  $\omega_{L,T}(k\rightarrow \infty)= k$, i.e.,
we recover the free dispersion relation, since the self energy $\Pi \sim eT$
can be neglected for large momenta $k\gg eT$. 

4. Because of $\omega_{L,T}>k$ we have  ${\rm Im}\, \Pi_{L,T}=0$, i.e., the HTL dispersion
relations are undamped (see below).

5. The gluon dispersion relation follows from the one of a photon by simply
replacing $m_\gamma$ by $m_g$.

6. In the case of massless fermions the HTL resummed fermion propagator leads also
to two branches as shown in Fig.20. The $\omega_+$ branch corresponds to a  collective 
electron or quark mode, whereas the $\omega_-$ branch, which has a negative ratio of 
helicity to chirality, does not exist in vacuum. Interestingly this so-called 
plasmino has a minimum at finite momentum, leading to interesting consequences for possible
observables of the QGP (see below).

Finally we will discuss the symmetric HTL resummed propagator.
We start from (\ref{2.33}), which also holds for the full propagator in equilibrium,
\bed
D^{*\, L}_S(K)=[1+2n_B(k_0)]\> \mbox{sgn}(k_0)\> [D^{*\, L}_R(K)-D^{*\, L}_A(K)]. \nonumber 
\eed
Combining (\ref{3.5}) and (\ref{3.8}) we see that ${\rm Im}\, \Pi^L_A =-{\rm Im}\, \Pi^L_R$
holds. Therefore we have
\bed
D^{*\, L}_{R,A}(K)=\frac{1}{k^2-{\rm Re}\, \Pi^L_R(K)\pm i{\rm Im}\, \Pi^L_R(K)},
\nonumber
\eed
if ${\rm Im}\, \Pi^L_R\neq 0$ (i.e., for $k^2>k_0^2$ in HTL-approximation).
From this we find
\bed
D^{*\, L}_{R}(K)-D^{*\, L}_{A}(K)=\frac{2i{\rm Im}\, \Pi^L_R}
{(k^2-{\rm Re}\, \Pi^L_R(K))^2+({\rm Im}\, \Pi^L_R(K))^2}=2i\> {\rm Im}\, D_R^{*\, L}(K).
\nonumber
\eed
Now we introduce the spectral function
$\rho_L=-{\rm Im}\, D_R^{*\, L}/\pi $.
Then we can write
\beq
D_S^{*\, L}(K)=-2\pi i\> [1+2n_B(k_0)]\> \mbox{sgn}(k_0)\> \rho_L(K).
\label{3.18}
\eeq
The spectral function is of the Breit-Wigner form
\bed
\rho_L(K)=-\frac{1}{\pi}\> \frac{{\rm Im}\, \Pi^L_R}
{(k^2-{\rm Re}\, \Pi^L_R(K))^2+({\rm Im}\, \Pi^L_R(K))^2}
\nonumber 
\eed
describing quasiparticles with finite width. Note, however, that the quasiparticle ``mass''
${\rm Re}\, \Pi$ and ``width'' ${\rm Im}\, \Pi$ depend on momentum and energy. 
In the HTL approximation the self energy has an imaginary part only below
the light cone ($k^2>k_0^2$). Then the spectral function can be decomposed
into a pole contribution and a cut contribution
\bea
\rho_L(K)&=&\rho_L^{\rm pole}(K)+\rho_L^{\rm cut}(K),
\nonumber \\
\rho_L^{\rm pole}(K)&=&\mbox{sgn}(k_0)\> \delta(k^2-{\rm Re}\, \Pi_L^R(K)),
\nonumber \\
\rho_L^{\rm cut}(K)&=&-\frac{1}{\pi}\> \frac{{\rm Im}\, \Pi^L_R(K)}
{(k^2-{\rm Re}\, \Pi^L_R(K))^2+({\rm Im}\, \Pi^L_R(K))^2}\> \theta(k^2-k_0^2),
\nonumber
\eea
where one has to use the HTL results for ${\rm Re}\, \Pi^L_R(K)$ and 
${\rm Im}\, \Pi^L_R(K)$.

\newpage

\begin{figure}[t]

\centerline{\psfig{figure=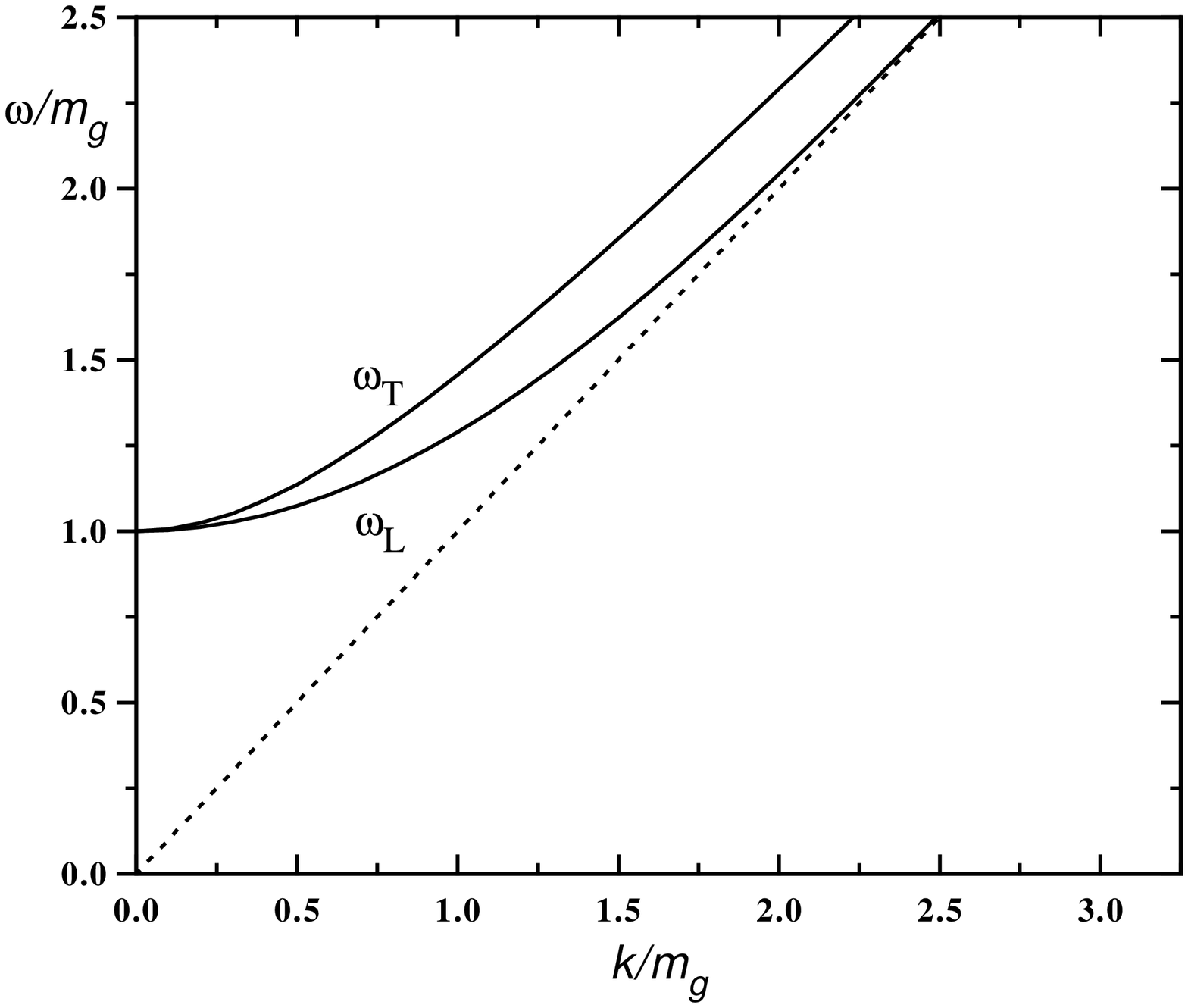,width=12cm}}
\vspace*{1cm}
\centerline{\it Fig.19}

\end{figure}

\vspace*{-10cm}

\begin{figure}

\vspace*{-11.5cm}
\hspace*{4.4cm}
\centerline{\psfig{figure=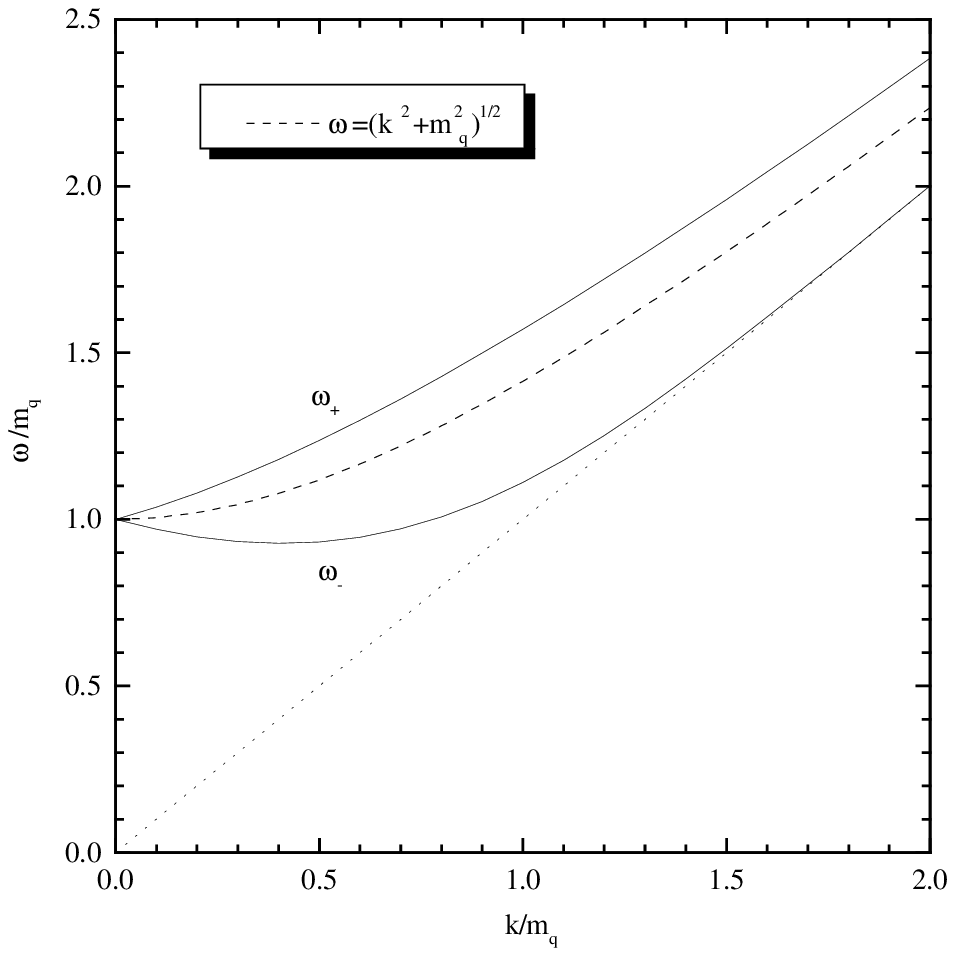,width=10cm}}
%\vspace*{cm}

\end{figure}

\vspace*{16.5cm}

\centerline{\it Fig.20}

\newpage

\subsection{HTL resummation technique}

As we said before, naive perturbation theory at $T>0$ suffers from
a problem, namely infrared singularities and gauge dependent results.
The reason for this undesirable behaviour is that the naive perturbative expansion 
is incomplete at $T>0$. Infinitely many higher order diagrams with more and more loops
can contribute to lower order in the coupling constant. These diagrams can be taken into 
account  by resummation.

\subsubsection{Massless $\phi^4$-theory}

The lowest order tadpole shown in Fig.7 has been calculated already above, 
$\Pi_1=g^2T^2=M^2$. To next order of the scalar self energy we find
a diagram as in Fig.9. This diagram exhibits a logarithmic
infrared singularity because of $\Pi_2\sim g^4\int d^4K/K^4$.
Naively we expect that the self energy can be written as $\Pi=\Pi_1+\Pi_2+...=g^2T^2+O(g^4)$.

\begin{figure}

\centerline{\psfig{figure=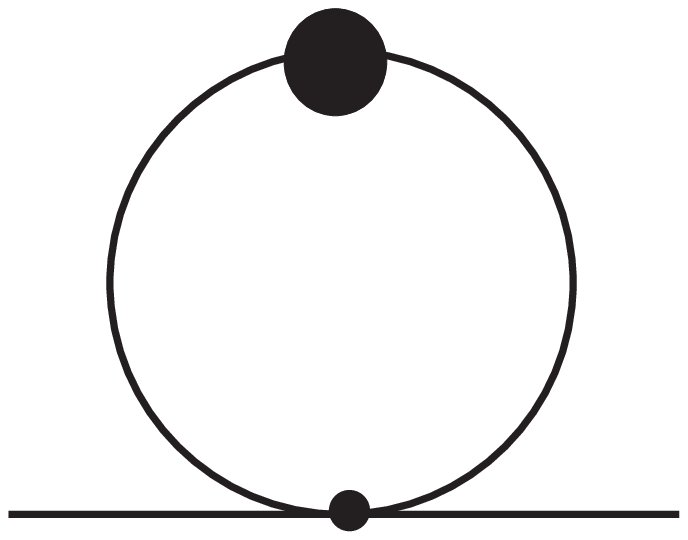,width=4cm}}
\centerline{\it Fig.21}

\end{figure}

Now we consider the tadpole diagram $\Pi^*$
in Fig.21, where we have an effective propagator,
$i\Delta^*=i/(K^2-M^2)$, in the loop instead of a bare one. Since the effective
propagator follows from a resummation of the bare tadpole, $\Pi^*$ is given by Fig.22,
i.e., it is a sum of all daisy diagrams. In order to calculate $\Pi^*$ we simply have to replace
the bare mass in $\Pi_1$ by the effective mass $M$ and obtain
\bed
\Pi^*=12\> g^2\> \int \dk \> [1+2n_B(\Omega_K)],\qquad \Omega_k\equiv \sqrt{k^2+M^2}.
\nonumber
\eed
The $k$-integral cannot be done analytically, but an
expansion for small $g$ is possible (see problem \# 8):
\beq
\Pi^* = g^2T^2\> [1-\frac{3}{\pi}\> g+O(g^2)].
\label{3.19}
\eeq

\begin{figure}[b]

\centerline{\psfig{figure=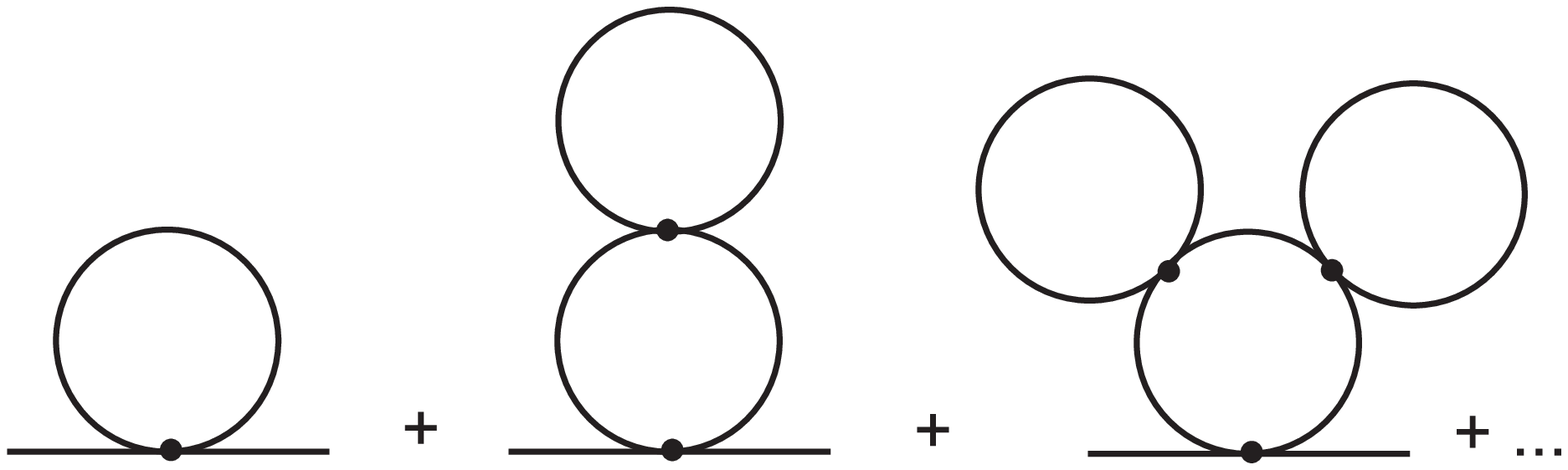,width=10cm}}
\centerline{\it Fig.22}

\end{figure}

This surprising result deserves two comments:

1. $\Pi^*$ is infrared finite, although it contains (infinitely many) infrared divergent diagrams.

2. The correction to $\Pi_1$ is of order $g$ not $g^2$, i.e., it is
a non-perturbative correction, since it is not a power of the perturbative
expansion parameter $g^2$.

These observations suggest the following recipe:

1. Isolate terms $\sim g^2T^2$. Here: $\Pi _1$.

2. Construct the effective propagator by resummation: Here: $\Delta^*$.

3. Use $\Delta^*(K)$ as in perturbation theory, if $k_0$ {\it and} $k$ are soft ($\sim gT$),
because then all terms in the denominator of the effective propagator are of the same order,
$\Delta^* =1/(g^2T^2)$. If $k_0$ {\it or} $k$ are hard ($\sim T$), however,
it is sufficient to use the bare propagator, which is then of order 
$\Delta =1/K^2\sim 1/T^2$.

In $\Pi^*$ we have integrated over $k$ and $k_0$ from 0 to $\infty $.
The soft momentum range is important for the correction to $\Pi_1$.
Therefore it is necessary to use $\Delta^*$.

\subsubsection{Gauge theories}

Here we do not want to discuss the HTL resummation technique for gauge theories in detail,
but will only take over the arguments from the scalar theory. Our findings above
suggest the following strategy:

\begin{figure}

\centerline{\psfig{figure=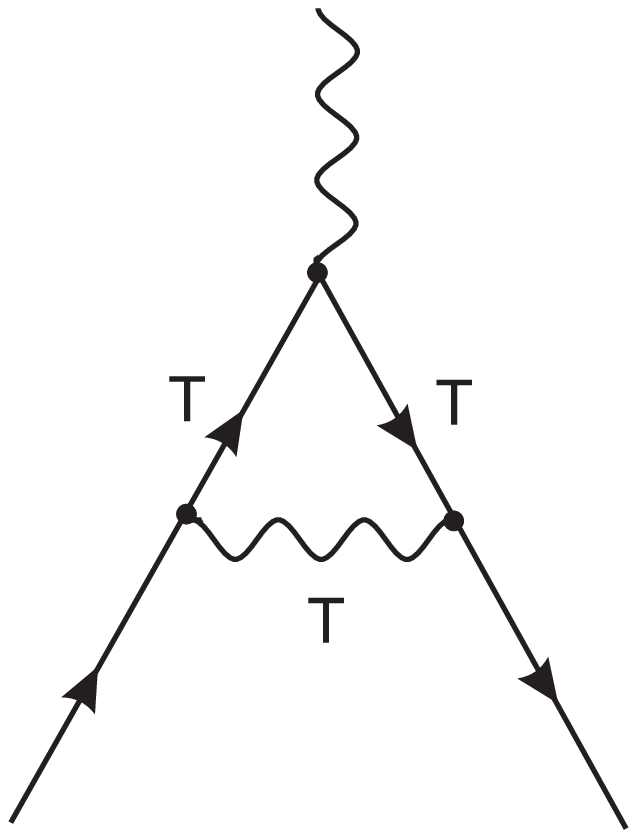,width=4cm}}
\centerline{\it Fig.23}

\end{figure}

{\it 1. step: Isolation of self energies $\sim g^2T^2$}

These are the HTL self energies $\Pi_{L,T}$ and $\Sigma $, which we discussed already in detail.
In contrast to the scalar theory there is a new aspect: due to Ward identities self energies are
related to vertices, e.g.
\bed
ie\> [\Sigma(P_1)-\Sigma(P_2)]=(P_1+P_2)_\mu \Gamma^\mu(P_1,P_2).
\nonumber
\eed
Therefore we have to consider also HTL vertex corrections,
as shown in Fig.23, where all internal lines are hard $\sim T$.

\begin{figure}

\centerline{\psfig{figure=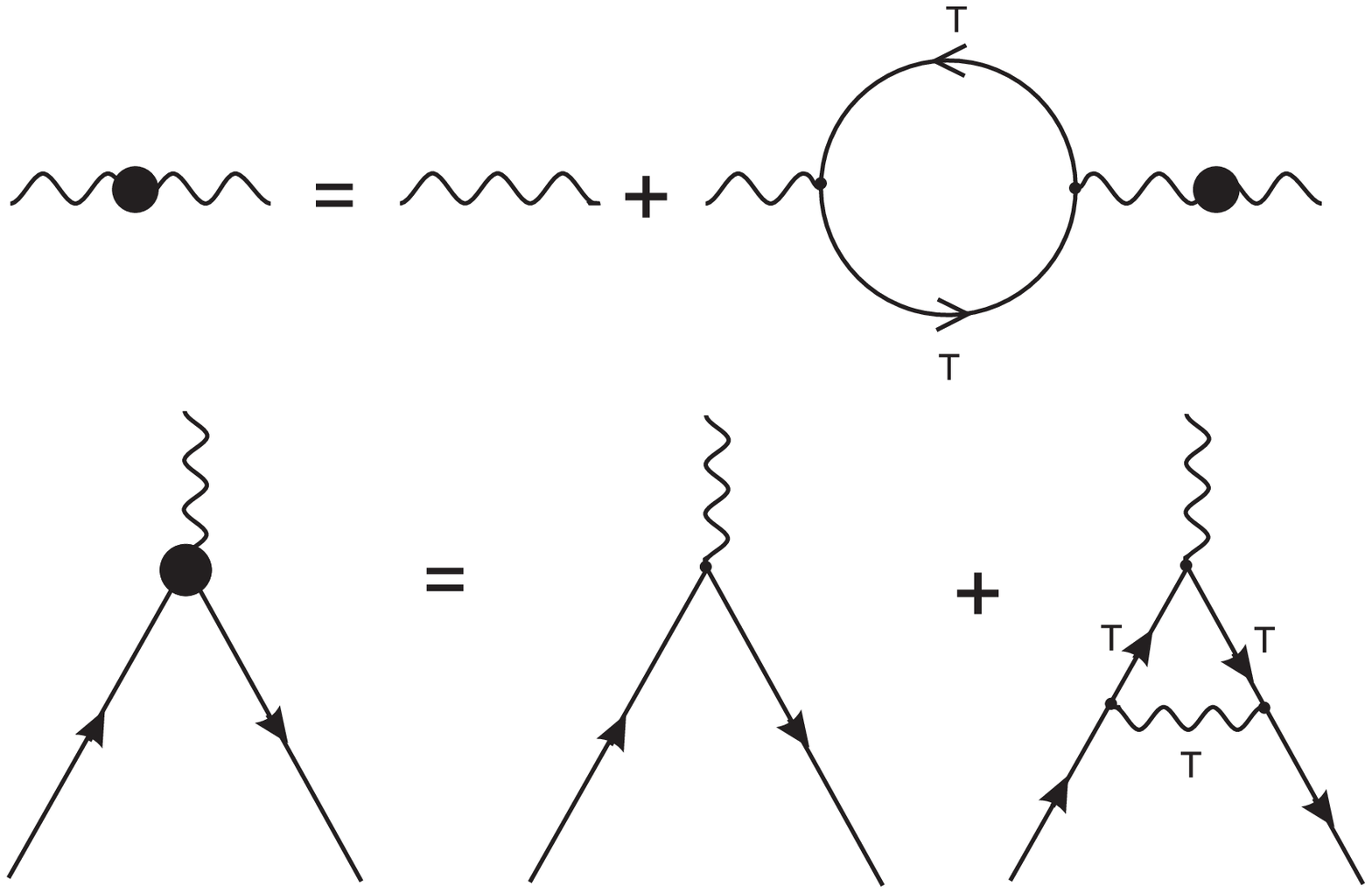,width=10cm}}
\centerline{\it Fig.24}

\end{figure}

{\it 2. step: Effective propagators and vertices}

Effective propagators are constructed as above making use of the Dyson-Schwinger equations.
The HTL vertex following simply by adding the HTL correction to the bare vertex.
Examples are shown in Fig.24.

\newpage

{\it 3. step: Effective perturbation theory}

We use effective propagators and vertices if {\it all} legs are soft;
otherwise bare propagators and vertices are sufficient.
In this way, contributions of the same (or even lower) order in $g$ are included,
gauge independent results are obtained, and
screening effects are included leading to an
improved IR behaviour.

Summarizing, a large progress compared to naive perturbation theory
has been achieved, although there are still problems, as we will see below.

\newpage

\section{Applications}

\subsection{Muon damping rate}

The simplest application of the HTL resummation technique for gauge theories
is the interaction or damping rate $\gamma $ of a heavy fermion with mass $M$
in a relativistic plasma, e.g. muons in an $e^+$-$e^-$-plasma at $m_e\ll T\ll 
M_\mu $. The interaction rate is related to the mean free path by
$\lambda =1/\gamma $. 

The damping or interaction rate describes the damping of a particle with the
time evolution $\exp(-i\omega t)$ (plane wave). In general the frequency
$\omega $ has a real and an imaginary part: $\omega = {\rm Re}\, \omega
+i\> {\rm Im}\, \omega $. Defining $\gamma \equiv -{\rm Im}\, \omega$
we have $\exp(-i\omega t)=\exp(-i{\rm Re}\,\omega t)\> \exp(-\gamma t)$. 
As an example we consider a scalar field. Its dispersion relation
is given by $\omega^2-k^2-\Pi(\omega , k)=0$, from which we find
\beq
({\rm Re}\, \omega -i\gamma)^2-k^2-{\rm Re}\, \Pi({\rm Re}\, \omega 
-i\gamma ,k) -i {\rm Im}\, \Pi ({\rm Re}\, \omega -i\gamma ,k)=0.
\nonumber
\eeq
In the case of no overdamping, $\gamma \ll {\rm Re}\, \omega$, we get
\beq
{\rm Re}^2\omega-k^2-{\rm Re}\, \Pi ({\rm Re}\, \omega,k)=0,
\nonumber
\eeq
from which the dispersion relation ${\rm Re}\, \omega =\omega (k)$
follows, and
\beq
-2i\> {\rm Re}\, \omega \> \gamma =i\> {\rm Im}\, \Pi ({\rm Re}\, \omega, k),
\nonumber
\eeq
which leads to
\beq
\gamma = -\frac{1}{2\omega (k)}\> {\rm Im}\, \Pi (\omega (k),k).
\nonumber
\eeq

\begin{figure}[b]

\centerline{\psfig{figure=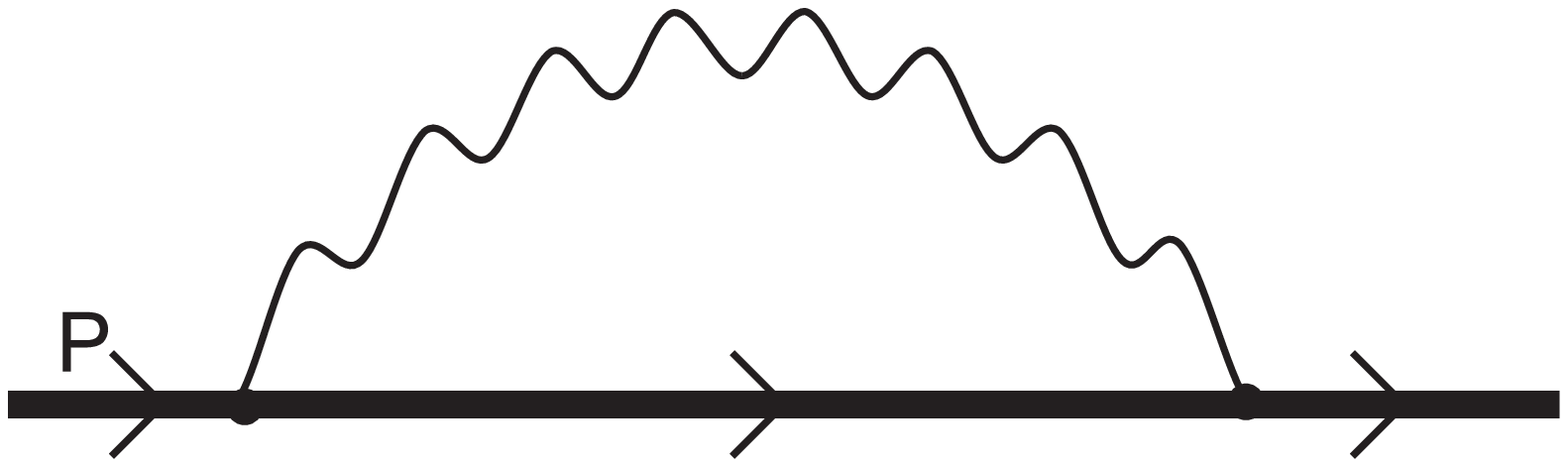,width=6cm}}
\centerline{\it Fig.25}

\end{figure}

In the case of a fermion with mass $M$ the interaction rate is given by
\beq
\Gamma (E) = -\frac {1}{2E}\> [1-n_F(E)]\> {\rm tr}\, [(P\sla +M)\>
{\rm Im}\, \Sigma_R (E, {\bf p})],
\label{4.1}
\eeq
where $E^2=p^2+M^2$ and $\Sigma_R$ is the retarded self energy of the massive
fermion.

In naive perturbation theory the self energy to lowest order is given
by Fig.25, which has no imaginary part on-shell,
${\rm Im}\, \Sigma_R(E,{\bf p})=0$. (A bold line denotes a massive
fermion.) For according to cutting rules the
imaginary part of $\Sigma$ can be related to the matrix element of Fig.26,
which describes the emission or absorption of a photon from a bare muon.
However, this process is forbidden by energy-momentum conservation.

\begin{figure}

\centerline{\psfig{figure=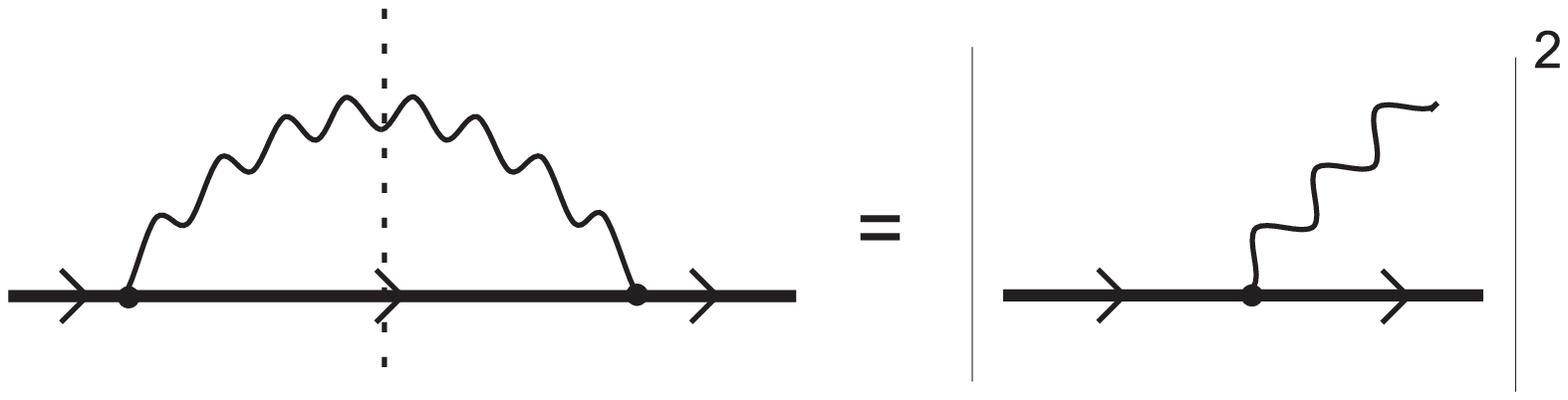,width=12cm}}
\centerline{\it Fig.26}

\end{figure}

To next order we have to consider the 2-loop diagram of Fig.27. 
Using cutting rules it corresponds $\mu e^\pm \rightarrow \mu e^\pm$ 
scattering, which leads to an energy loss of a muon in the 
$e^+$-$e^-$-plasma. From naive power counting we expect
$\Gamma (E)\sim \alpha^2$. However, due to the exchange of a massless
photon the damping rate turns out to be quadratically IR divergent.

\begin{figure}

\centerline{\psfig{figure=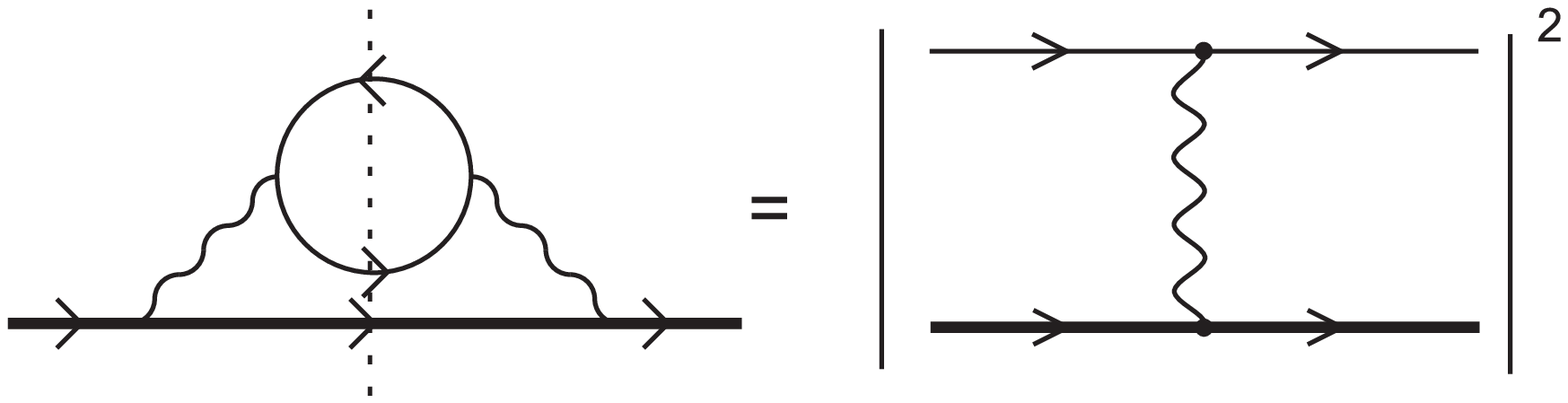,width=12cm}}
\centerline{\it Fig.27}

\end{figure}

Therefore we have to adopt the HTL method, which leads to the diagram of 
Fig.28. Due to the non-vanishing imaginary part of the photon self energy
contained in the HTL photon propagator of the exchanged photon this
diagram has a finite imaginary part and corresponds to scattering via
the exchange of a soft photon (virtual Landau damping). Note that the HTL 
photon self energy contains hard (thermal) electrons. Furthermore the 
HTL resummed photon propagator contains Debye screening. Note also that
we do not need an effective muon-photon vertex or an effective muon propagator
as the momenta $P$, $P'$ are always hard due to the large muon mass.

\begin{figure}

\centerline{\psfig{figure=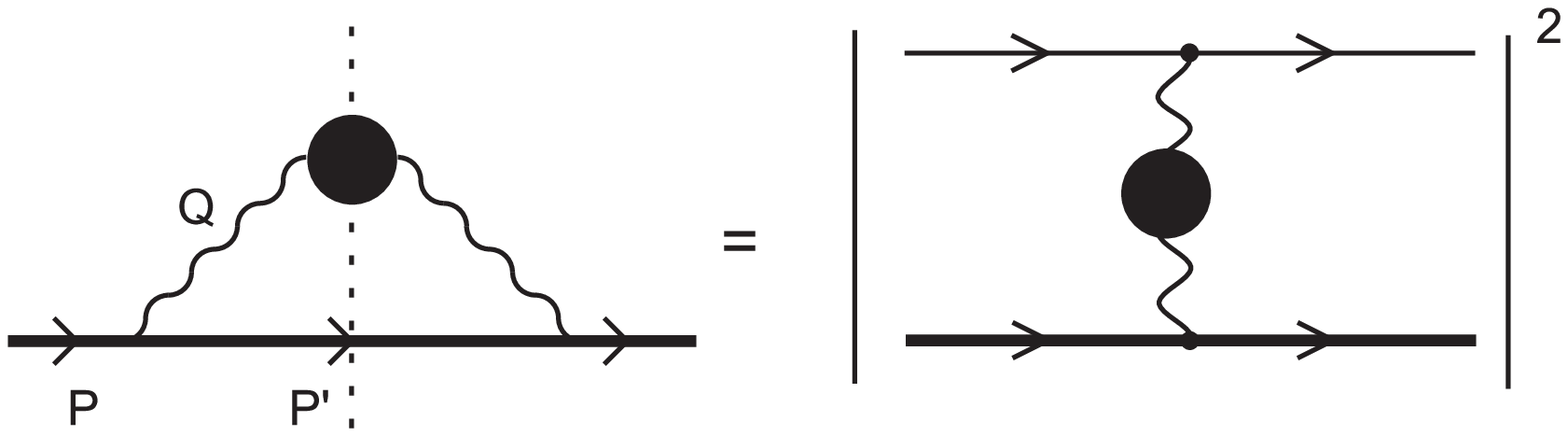,width=12cm}}
\centerline{\it Fig.28}

\end{figure}

Using standard Feynman rules we find at $T=0$
\beq
\Sigma ^* (P) = i\> e^2\> \int \frac {d^4Q}{(2\pi )^4}\> D_{\mu \nu }
^* (Q)\> \gamma ^\mu\> S(P')\> \gamma ^\nu .
\label{4.2}
\eeq

At finite temperature, using the RTF, we find
\bea
&&\Sigma_R^* (P)=\Sigma_{11}^* (P)+\Sigma_{12}^* (P)\nonumber \\
&&=ie^2\> \int \frac {d^4Q}{(2\pi )^4}\> [D_{\mu \nu }^{*\, 11} (Q)
\gamma ^\mu\> S^{11}(P')\> \gamma ^\nu -D_{\mu \nu }^{*\, 12} (Q)
\gamma ^\mu\> S^{12}(P')\> \gamma ^\nu ].
\nonumber
\eea
Defining $S(P')=({P\sla}'+M)\> \tilde \Delta(P')$  and using
\bed
D_{\mu \nu }^{*} \gamma ^\mu\>({P\sla}'+M) \> \gamma ^\nu
{\buildrel (\ref{3.16}) \over =} D_L^*\gamma^0({P\sla}'+M)\gamma^0
+D_T^*\left (\delta_{ij}-\frac{q_iq_j}{q^2}\right ) \gamma^i
({P\sla}'+M)\gamma^j \nonumber
\eed
and
\bea
&& {\rm tr}\, [(P\sla +M)\gamma^0({P\sla}'+M)\gamma^0]=4(2p_0p_0'-P\cdot
P'+M^2)\nonumber \\
&& \left (\delta_{ij}-\frac{q_iq_j}{q^2}\right )\> {\rm tr}\, [(P\sla +M)
\gamma^i({P\sla}'+M)\gamma^j]=8[p_0^2-p_0q_0-({\bf p}\cdot \hat{\bf q})^2
+{\bf p}\cdot{\bf q}-M^2]\nonumber
\eea
we find
\bea
& \! & {\rm tr}\, [(P\sla +M)\> \Sigma_R^* (P)]  =\nonumber \\
& \! & 4i e^2\> \int \frac {d^4Q}{(2\pi )^4} \bigl \{\tilde \Delta_{11}(P')\>
\bigl [ D_{11}^{L\, *} (Q)\> (p_0^2+p^2-p_0q_0- {\bf p}\cdot {\bf q}
+M^2) \nonumber \\
& \! & +2\> D_{11}^{T\, *} (Q)\> 
(p_0^2-p_0q_0+{\bf p}\cdot {\bf q}-({\bf p}\cdot
\hat {\bf q})^2-M^2)\bigr ] - \bigl [ (11)\rightarrow (12)\bigr ]\bigr \}.
\nonumber
\eea
Using the Keldysh representation we obtain
\bed
\tilde \Delta_{11}\, D_{11}^*-\tilde \Delta_{12}\, D_{12}^*
{\buildrel (\ref{2.27}) \over =} \frac{1}{2}\> [\tilde \Delta_R\, D_S^*
+\tilde \Delta_R\, D_A^*+\tilde \Delta_S\, D_R^*
+\tilde \Delta_A\, D_R^*].\nonumber
\eed

The leading contribution in the coupling constant $e$
comes from $n_B(k_0\sim eT)=1/[\exp(|k_0|/T)-1]\simeq 
T/|k_0|\sim 1/e$, which appears only in the first term.
Therefore to lowest order we have
\bea
&&\hspace*{-0.5cm}{\rm tr}\, [(P\sla +M)\> \Sigma_R^* (P)] {\buildrel (\ref{3.18}) \over =} 
4\pi e^2\> \int \frac {d^4Q}{(2\pi )^4} \> [1+2n_B(q_0)]\> \mbox{sgn}(q_0)\nonumber \\
&&\hspace*{-0.5cm}[(p_0^2+p^2-p_0q_0- {\bf p}\cdot {\bf q}+M^2)\> \rho_L(Q)+2(p_0^2-p_0q_0
+{\bf p}\cdot {\bf q}-({\bf p}\cdot \hat {\bf q})^2-M^2)\> \rho_T(Q)]\> \tilde \Delta_R(P').
\nonumber
\eea
The term
${\rm Im}\, \Sigma$ comes from ${\rm Im}\, \tilde \Delta_R(P')=-\pi \,
\mbox{sgn}(p'_0)\, \delta({P'}^2-M^2)$.
Hence we find 
\bea
& \! & \Gamma (E)=\frac{2\pi^2e^2}{E}\> [1-n_F(E)]\> \int \frac {d^4Q}{(2\pi )^4}\> [1+2n_B(q_0)]\> \mbox{sgn}(q_0)\nonumber \\
& \! & [(2E^2-Eq_0- {\bf p}\cdot {\bf q})\> \rho_L(Q)+2(p^2-Eq_0
+{\bf p}\cdot {\bf q}-({\bf p}\cdot \hat {\bf q})^2)\> \rho_T(Q)]\>\nonumber \\
& \! &\mbox{sgn}(E-q_0)\> \frac{1}{E'}\> [\delta(E-q_0-E')+\delta(E-q_0+E')].
\label{4.3}
\eea

Now we want to make the following 
approximations:

1. $E\gg T$ from which $n_F(E)\simeq 0$ follows.

2. The HTL approximation for the photon propagator: $q$, $q_0\ll T\ll E$. This 
leads to 

\bu \hspace*{0.2cm}$[1+2n_B(q_0)]\mbox{sgn}(q_0)\simeq 2T/q_0$,

\bu \hspace*{0.2cm} $\mbox{sgn}(E-q_0)=+$,

\bu \hspace*{0.2cm} the following simplification in the argument of the 
first $\delta$-function:

\bed
E'=\sqrt{p'^2+M^2}=\sqrt{({\bf p}-{\bf q})^2+M^2}\simeq \sqrt{E^2-2{\bf p}\cdot{\bf q}}
\nonumber
\eed

\bed
\simeq E\left (1-\frac{{\bf p}\cdot{\bf q}}{E^2}\right )=E-{\bf v}\cdot{\bf q}, \qquad 
{\bf v}=\frac{{\bf p}}{E},\nonumber 
\eed

\bu \hspace*{0.2cm} $\delta(E-q_0+E')=0$,

\bu \hspace*{0.2cm} $1/E'\simeq 1/E$.

Using these approximations we find
\bed
\Gamma (E)=4\pi^2e^2T\> \int \frac {d^4Q}{(2\pi )^4}\> \frac{1}{q_0}
\> [\rho _L(Q)+(v^2-({\bf v}\cdot \hat{\bf q}^2))\rho_T(Q)]\>
\delta (q_0-{\bf v}\cdot{\bf q}).\nonumber
\eed
Introducing the angle $\eta$ 
\bed
{\bf v}\cdot{\bf q}=vq\eta, \qquad \eta \equiv \frac{{\bf p}\cdot{\bf q}}{pq},
\qquad v=\frac{p}{E}
\nonumber
\eed
we can use the $\delta$-function for integrating over this angle, using
\bed
\delta(q_0-{\bf v}\cdot{\bf q})=\frac{1}{vq}\> \delta 
(\eta-\frac{q_0}{vq}),
\nonumber 
\eed
which gives
\beq
\Gamma (E) = \frac  {e^2T}{2\pi v}\> \int _0^\infty dq \> q \int _{-vq}
^{vq} \frac{dq_0}{q_0} \> \left [\rho _L(q_0,q)
+\left (v^2-\frac {q_0^2}{q^2}\right )\> \rho _T(q_0 ,q)
\right ].
\label{4.4}
\eeq
Note that we integrate from $0$ to $\infty$, although the momentum $q$ of the
photon is soft. This is possible since the intergrand drops like $1/q^3$, 
i.e., the integral is dominated by soft momenta regardless of the upper limit
as long as it is much larger than $eT$. Hence the hard integration range 
contributes only to higher orders that have been neglected anyway.

From the intergation range of $q_0$ we read off that $q_0<q$. Hence only
the cut contributions (Landau damping) of the spectral functions
$\rho_{L,T}$ are needed.

The intergation over $q$ in (\ref{4.4}) can be done analytically, 
the one over $q_0$ only numerically. However, under the
simplifying assumption of a non-relativistic muon, i.e., 
$M\gg p\Rightarrow v\ll 1 \Rightarrow q_0\ll q$, we can use the
quasistatic approximation for the spectral functions
\bea
\rho _L ^{cut} (q_0 ,q) & = & \frac{3m_\gamma ^2q_0}{2q}\> |D_L^*(q_0,q)|^2
\simeq \frac {3m_\gamma ^2q_0}{2q}\> \frac{1}{(q^2+3m_\gamma ^2)^2},
\nonumber \\
\rho _T ^{cut} (q_0,q) & = & \frac {3m_\gamma ^2\omega (q^2-q_0^2)}
{4q^3}\> |D_T (q_0,q)|^2\simeq \frac {3m_\gamma ^2q_0 q}{4}\> \frac
{1}{q^6+(3\pi m_\gamma ^2q_0/4)^2}.
\nonumber
\eea
Here we can easily identify the Debye screening, i.e. the screening
of static electric fields associated with the Debye mass $m_D^2=3m_\gamma^2$.
Also we see that there is no static ($q_0=0$) magnetic screening.
Using this approximation all integrations can be done exactly yielding
\bea
\Gamma _L(v\ll 1) &=& \frac {e^2T}{4\pi},\nonumber \\
\Gamma _T(v\ll 1) &=& \frac {e^2T}{2\pi }\> v\> \int \frac{dq}{q}.
\nonumber
\eea
As expected the longitudinal part, corresponding to the
exchange of an electric photon, is finite due to Debye screening, 
whereas the transverse part (exchange of magnetic photons) is
still IR divergent. Since $q$ is restricted to soft momenta we adopt an  
UV-cutoff in the above expression for $\Gamma_T$ of the order 
$m_\gamma \sim eT$. (The UV-behaviour $1/q$ instead of $1/q^3$ in $\Gamma_T$
is a consequence of the quasistatic approximation.) 
The IR-cutoff cannot be calculated within the 
HTL method. However, one can show that the damping itself provides an 
IR-cutoff of order $e^2T$. Then we obtain within
the leading log approximation the final result
\beq
\Gamma _T(v\ll 1) = \frac {e^2T}{2\pi }\> v\> \ln \frac {1}{e}.
\label{4.17}
\eeq

Again let us make a few remarks here:

1. We have seen that in the case of the damping rate the HTL method 
reduces the quadratic IR divergence, found in naive perurtbation
theory, to a logarithmic one.

2. Surprisingly $\Gamma \sim e^2$ and not of order $e^4$ as expected from
naive perturbation theory. In other words, the HTL resummation leads to 
a lower order result (anomalous large damping) due to the strong
IR sensitivty of the damping rate.

3. In the case of a heavy quark in the QGP we simply have to replace
$e^2$ in (\ref{4.17}) by $4g^2/3$. 

4. In QCD there is an alternative IR-cutoff, namely a magnetic
screening mass of the order $m_{\rm magn}=g^2T$, as suggested by
non-perturbative arguments such as lattice QCD.

\subsection{Other quantities}

There is a number of other quantities of the QGP which have been computed
using the HTL method. Here we want to give just a
qualitative overview. For details we refer the reader
to Refs.[2,3].

\subsubsection{Energy loss, thermalization times, viscosity}

\begin{figure}

\centerline{\psfig{figure=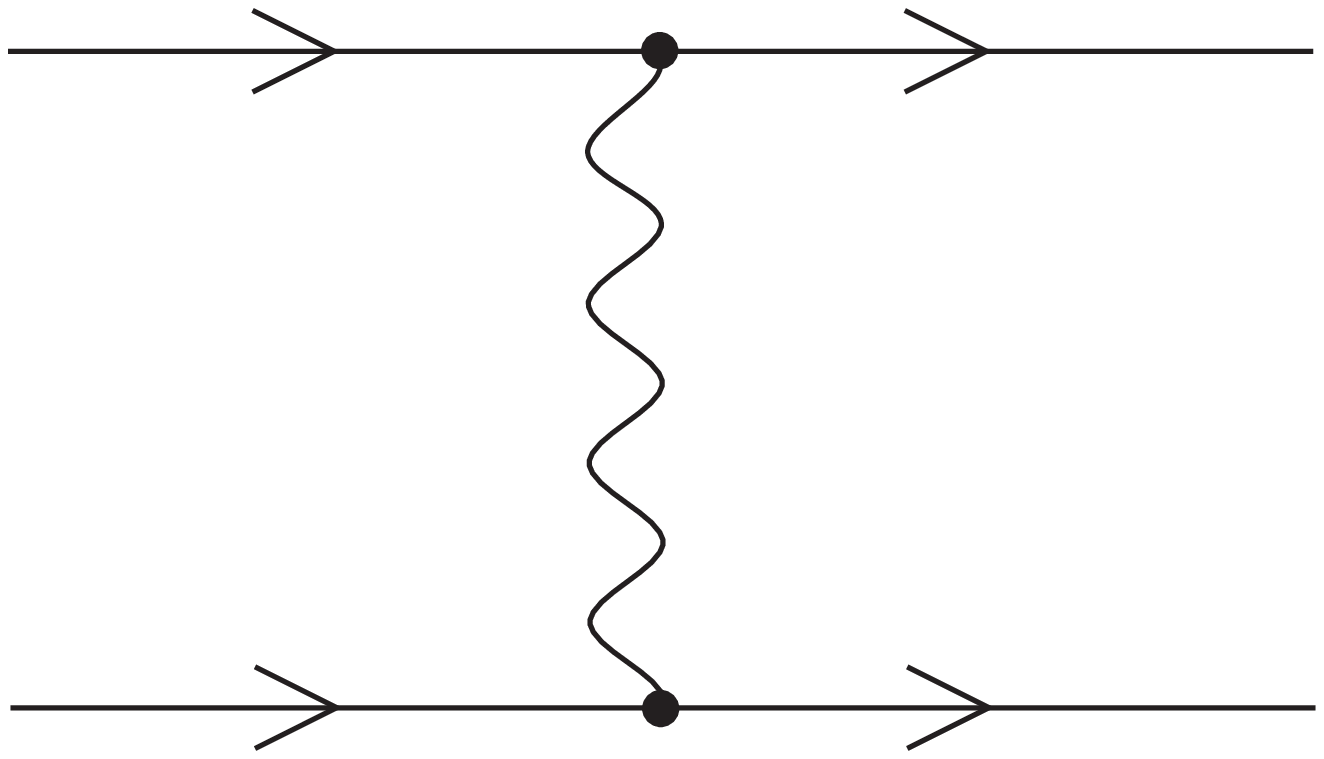,width=4cm}\hspace*{2cm}
\psfig{figure=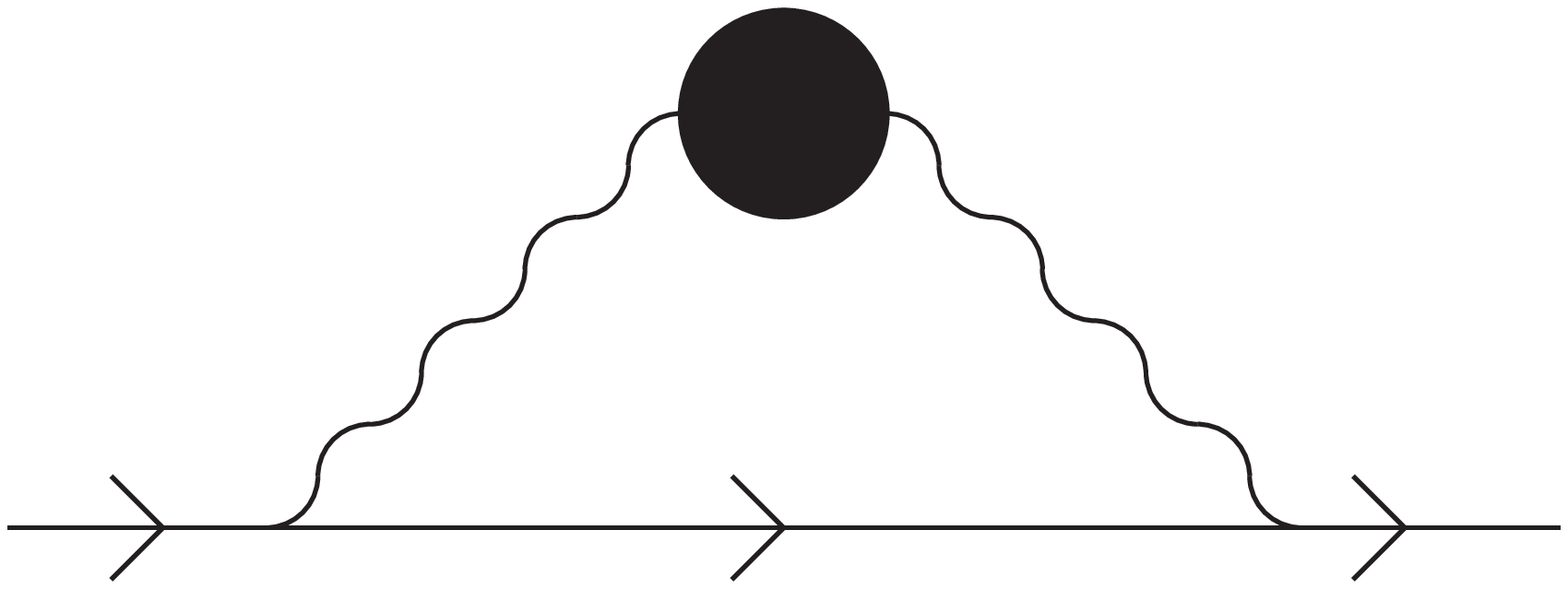,width=5cm}}
\centerline{\it Fig.29\hspace*{5.5cm}Fig.30}

\end{figure}

The energy loss of energetic quarks and gluons ($E\gg T$) in a QGP
is related to jet quenching, which might serve as a signature for the QGP
formation. Energetic quarks from initial hard collisions with a large
transverse momentum have to propagate through the fireball. Depending on the
phase the energy loss in the fireball might be different leading to a 
different energy distribution of jets. The energy loss per unit length
is defined as the average of the energy transfer per collision $\Delta E$
divided by the
mean free path $\lambda =v/\Gamma$. Hence it can be written as
\bed
\frac{dE}{dx}=\frac{1}{v}\> \int d\Gamma \Delta E\nonumber
\eed
Due to the factor $\Delta E$ this expression is only logarithmically 
IR divergent in naive perturbation theory, i.e.
\bed
\frac{dE}{dx}\sim \int \frac{dq}{q}.
\nonumber
\eed
The energy loss 
can be calculated to leading order by introducing a
separation scale $gT\ll q^*\ll T$ ($g\ll 1$!). For
$q>q^*$ it is sufficient to use a bare gluon propagator and we find
from Fig.29
\bed
\left (\frac{dE}{dx}\right )_{\rm hard}\sim \int_{q^*}^E dq ... 
\sim \ln\frac{ET}{{q^*}^2}.
\nonumber 
\eed
For $q<q^*$ we need a HTL gluon propagator as in Fig.30 leading to
\bed
\left (\frac{dE}{dx}\right )_{\rm soft}\sim \int_0^{q^*} dq ... \sim \ln\frac{{q^*}^2}{m_g^2}.
\nonumber 
\eed
Summing up the soft and the hard contributions we find
\bed
\frac{dE}{dx}=\left (\frac{dE}{dx}\right )_{\rm soft}+\left (\frac{dE}{dx}\right )_{\rm hard}
\sim \ln\frac{ET}{m_g},
\nonumber 
\eed
where the arbitrary separation scale $q^*$ has dropped out.

\begin{figure}

\centerline{\psfig{figure=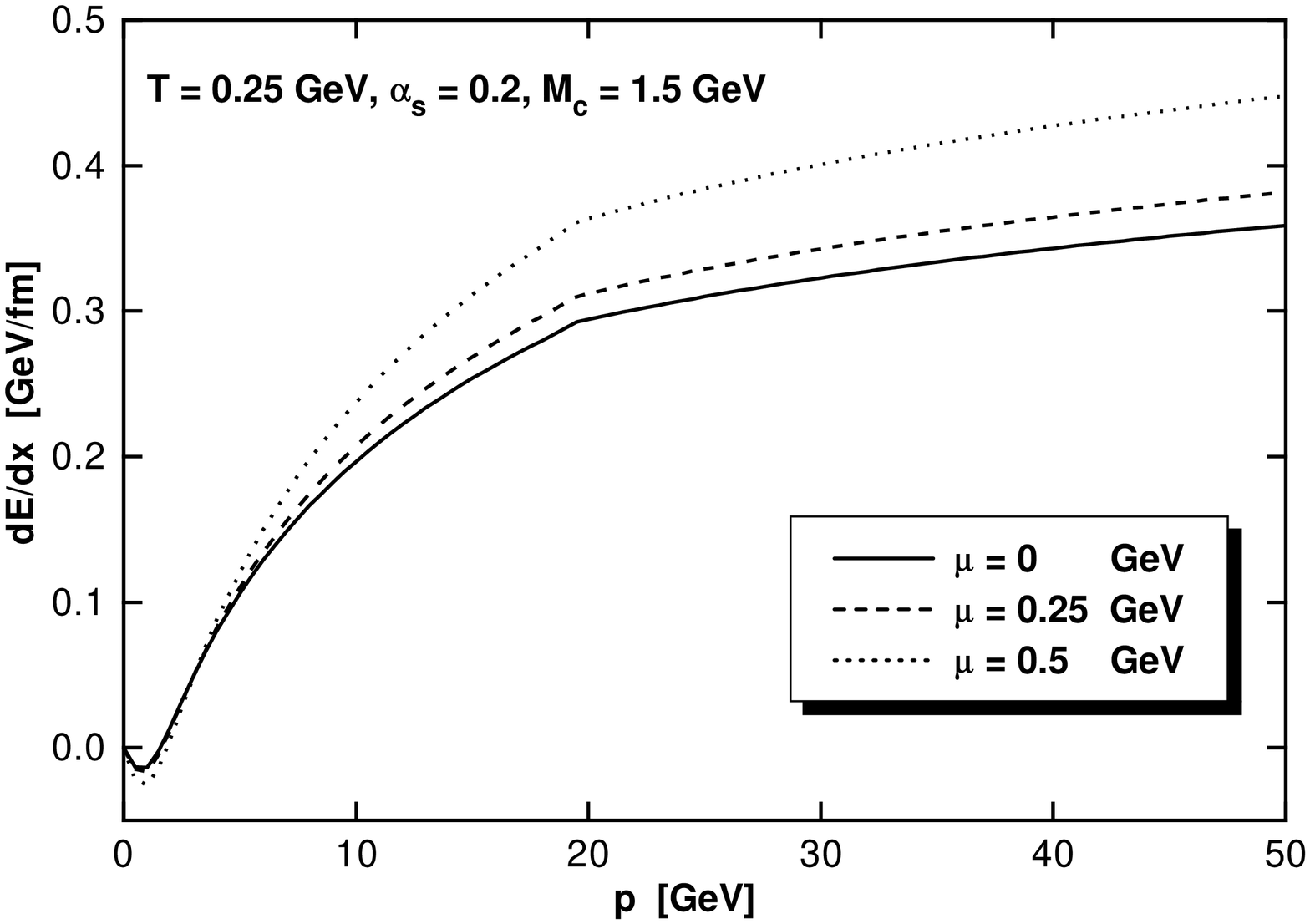,width=8cm}\psfig{figure=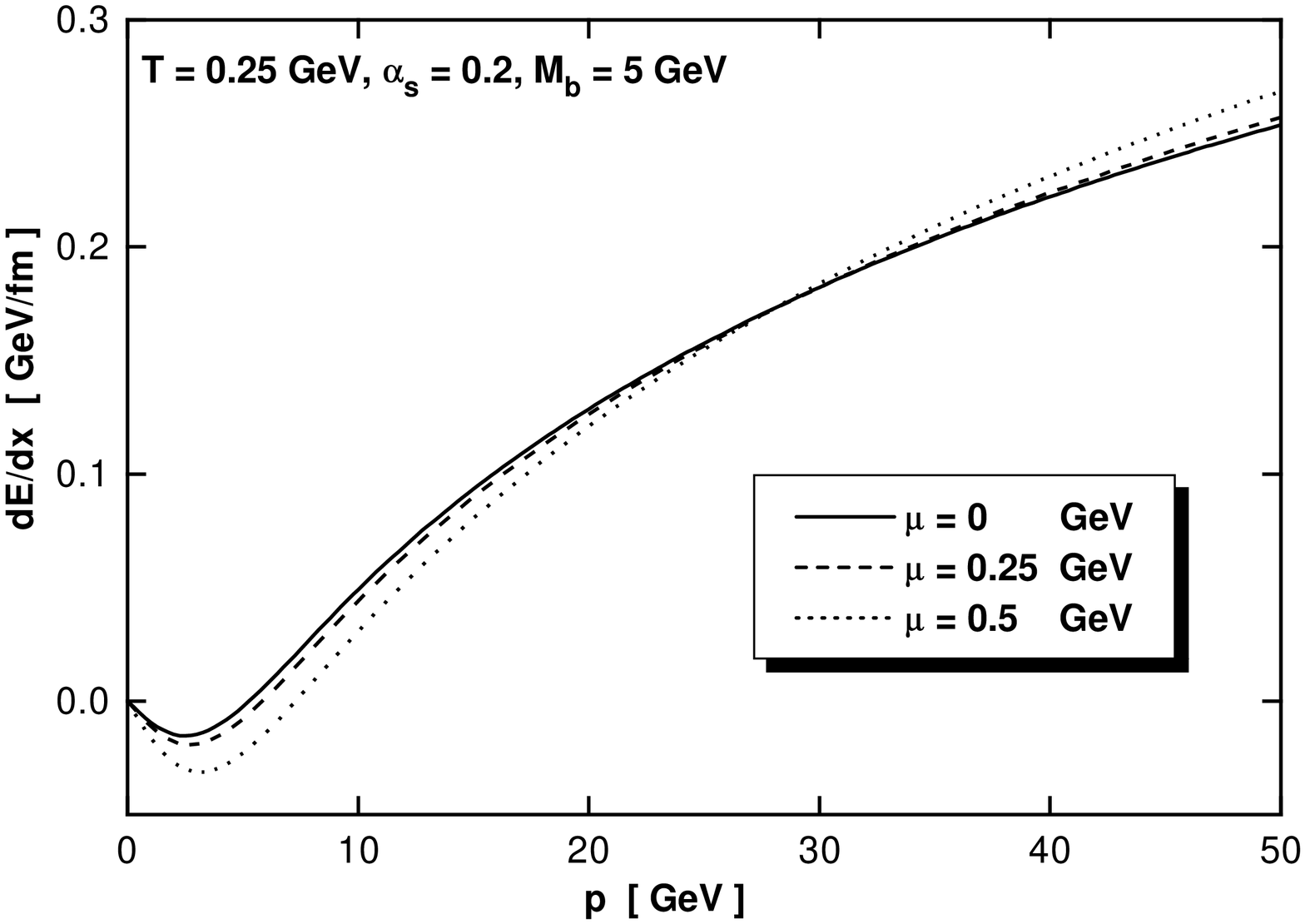,width=8cm}}
\centerline{\it Fig.31}

\end{figure}

The complete calculation yields the generalization of the famous
Bethe-Bloch formula to the case of energetic quarks in a QGP:
\bed
\frac{dE}{dx}\simeq 
\frac{16\pi}{9}\> \alpha_s^2\> T^2\> \ln \frac{9E}{16\pi \alpha_s T},
\nonumber
\eed
shown in Fig.31 for an energetic charm and bottom quark.

Note that $dE/dx\sim \alpha_s^2$ as in naive perturbation theory, 
because there is only a logarithmic IR
singularity in naive perturbation theory in contrast to $\Gamma $.
Besides this collisional energy loss due to elastic scattering, there
is also a radiative energy loss due to gluon bremsstrahlung which turns out 
to be dominating in realistic situations.

The thermalization time and the viscosity of the QGP are also closely related
to the damping rate and can be calculated in a similar way as the energy loss,
yielding for example for the thermalization or momentum relaxation time
\bed
\tau^{-1} \sim \alpha_s\> T\> \ln \frac{0.2}{\alpha_s}.
\nonumber 
\eed
Since the energy in the logarithm - compare with the Bethe-Bloch formula - above
is now replaced by a thermal energy the extrapolation from $\alpha_s \ll 1$ 
to $\alpha_s >0.2$ breaks down.

\subsubsection{Photon and dilepton production}

\begin{figure}

\centerline{\psfig{figure=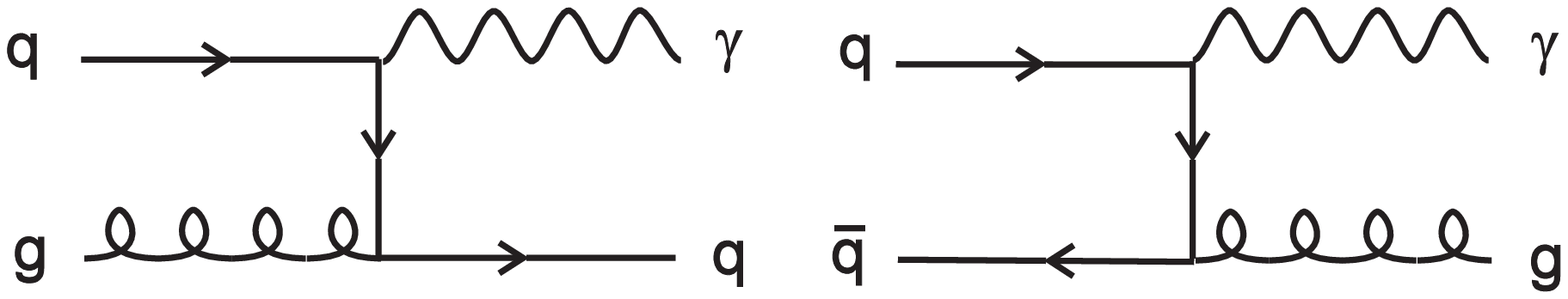,width=10cm}}
\centerline{\it Fig.32}

\end{figure}

The thermal emission of real and virtual photons from the QGP 
has been proposed as another promising signature for the QGP formation in
relativistic heavy ion collisions. Let us first consider
hard real photons with an energy $E\gg T$. To lowest order naive 
perturbation theory these photons are produced by the diagrams 
of Fig.32. The photon production rate follows from the imaginary part of
the two-loop photon self energy, which is related to the diagrams above
again by cutting rules. In the case of a bare quark (massless) propagator we 
encounter a logarithmic IR singularity. Using the HTL method, we
have to consider to lowest order the diagrams of Fig.33, where one HTL quark 
propagator appears. Due to $E\gg T$ it is sufficient to consider only
one effective quark propagator. 

\begin{figure}[b]

\centerline{\psfig{figure=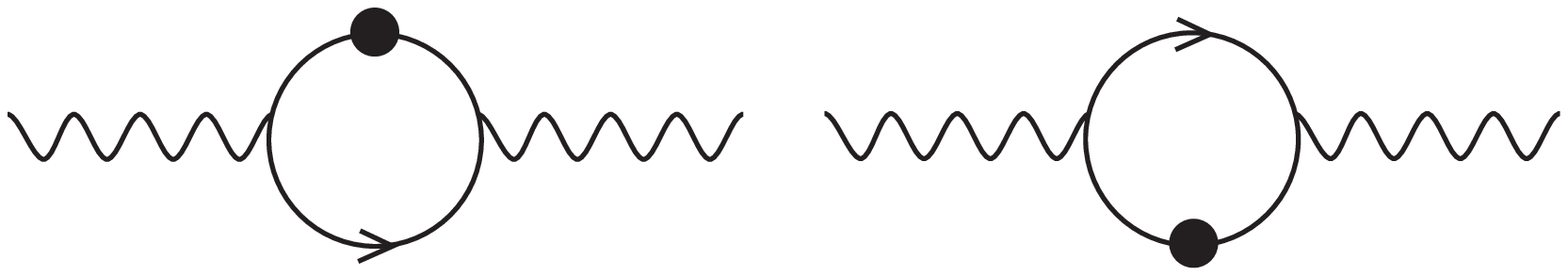,width=10cm}}
\centerline{\it Fig.33}

\end{figure}

After some more or less tedious calculations one finds for the
photon production rate
\begin{equation}
E\frac{dR}{d^3p}=\frac{5\alpha \alpha _s}{18\pi^2}\> e^{-E/T}\> \left (T^2+
\frac{\mu ^2}{\pi ^2}\right )\> \ln \frac{0.13E}{\alpha _sT},
\nonumber
\end{equation}
which is shown in Fig.34.

Soft dileptons, i.e.,
lepton pairs from the decay of soft virtual photons ($E$, $p\sim gT$),
can also be treated within the HTL resummation technique. Since the external
photon momentum of the photon self energy is soft now, two
effective quark propagators and also effective quark-photon vertices 
are necessary as in Fig.35. According to cutting rules this diagrams
contains as physical process the annihilation of collective quarks and 
antiquarks. The non-trivial dispersion relation of the in-medium quarks in
the HTL approximation leads to so-called
Van Hove singularities in the dilepton production rate as in Fig.36,
where a quantity proportional to ${\rm Im}\, {\Pi_\mu}^\mu$ is depicted.

\begin{figure}

\centerline{\psfig{figure=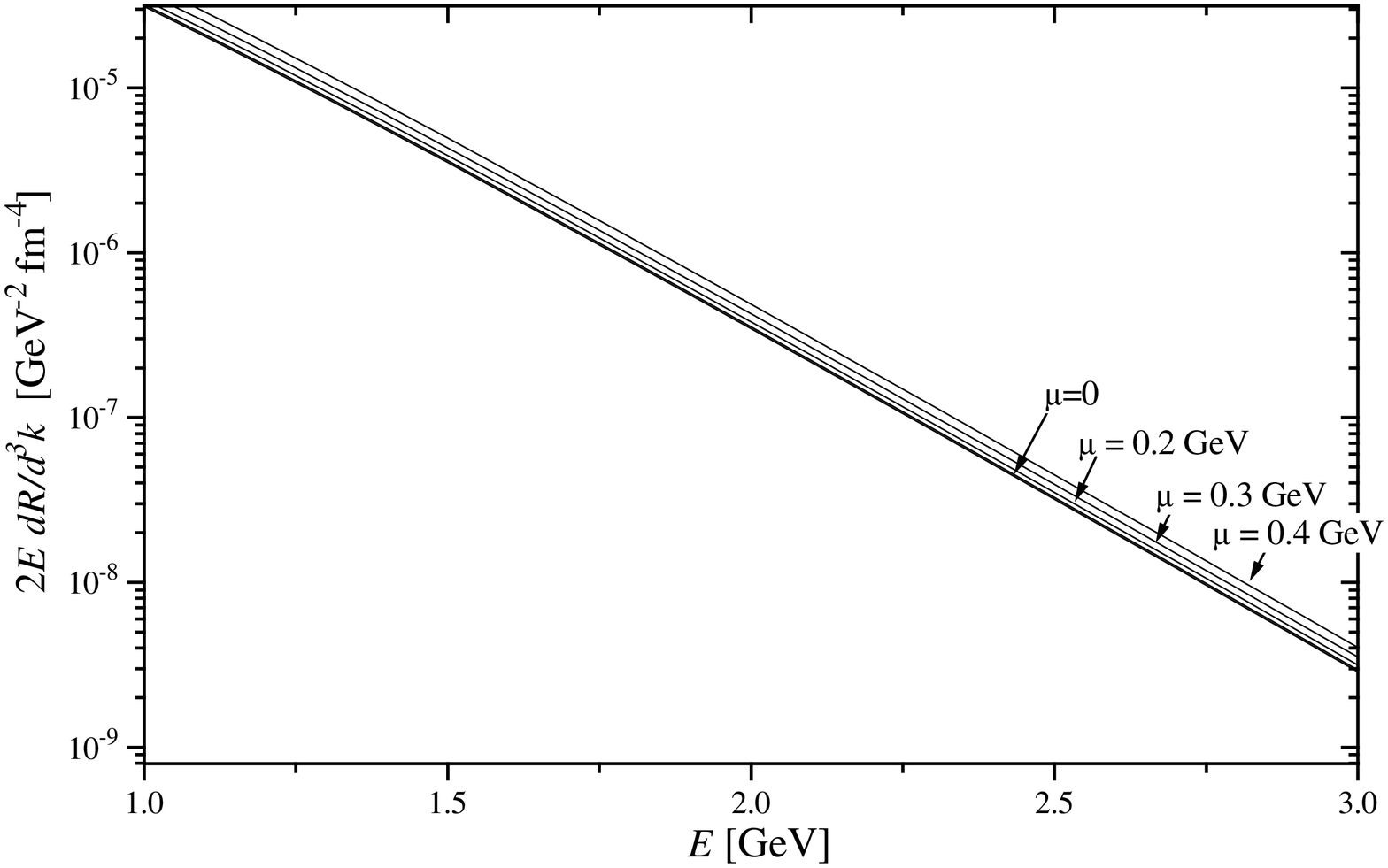,width=10cm}}
\centerline{\it Fig.34}

\end{figure}

Recently it has been  shown
that there are additional contributions to the same order from higher 
order diagrams (e.g. bremsstrahlung) if the invariant photon mass of
is of order $g^2T$. Consequently the photon and
ultrasoft dilepton production rates are non-perturbative and cannot be 
calculated to leading order even using the HTL improved perturbation theory.

Let us shortly summarize the achievements of the
HTL resummation technique:

1. Using HTL resummed propagators and vertices for soft momenta,
contributions of same (or even lower) order at $T>0$ 
as using bare propagators and vertices can be computed.

2. At the same time medium effects (Landau damping, Debye screening) 
are included leading, e.g. to an improved IR-behaviour.

3. Most important it provides gauge independent results for 
physical quantities in contrast to naive perturbation theory.

\begin{figure}[b]

\centerline{\psfig{figure=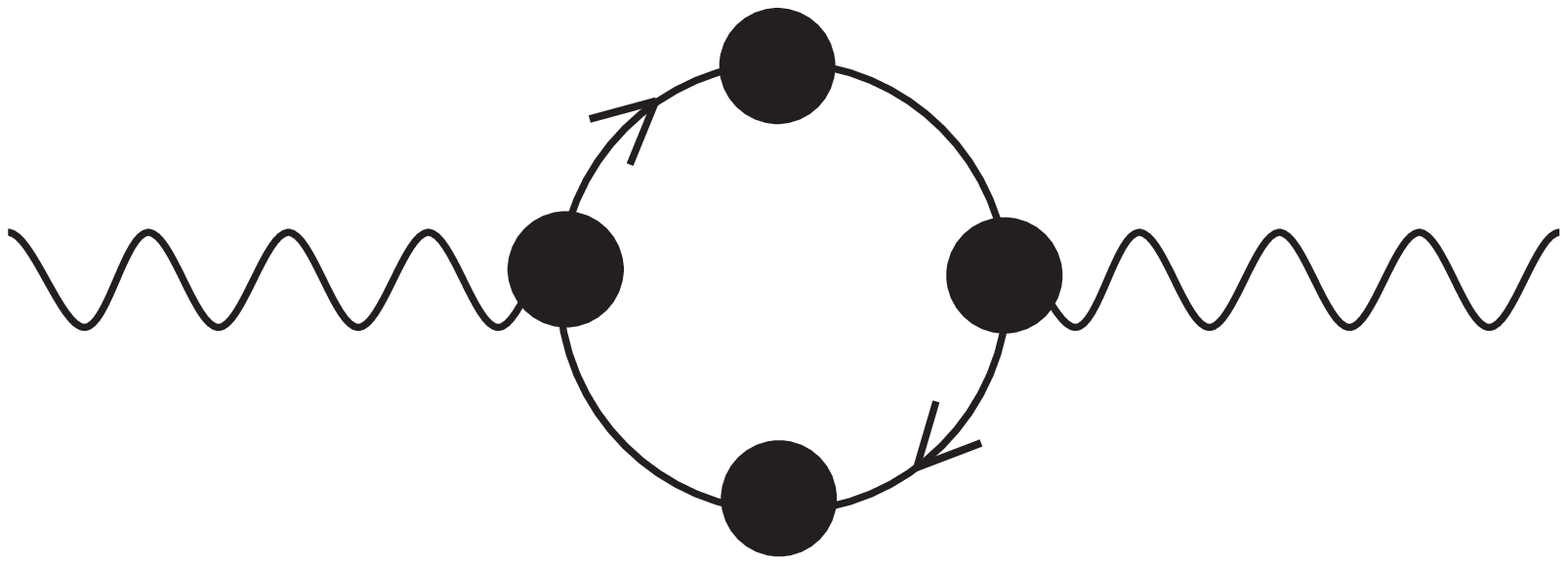,width=8cm}}
\centerline{\it Fig.35}

\end{figure}

\begin{figure}

\centerline{\psfig{figure=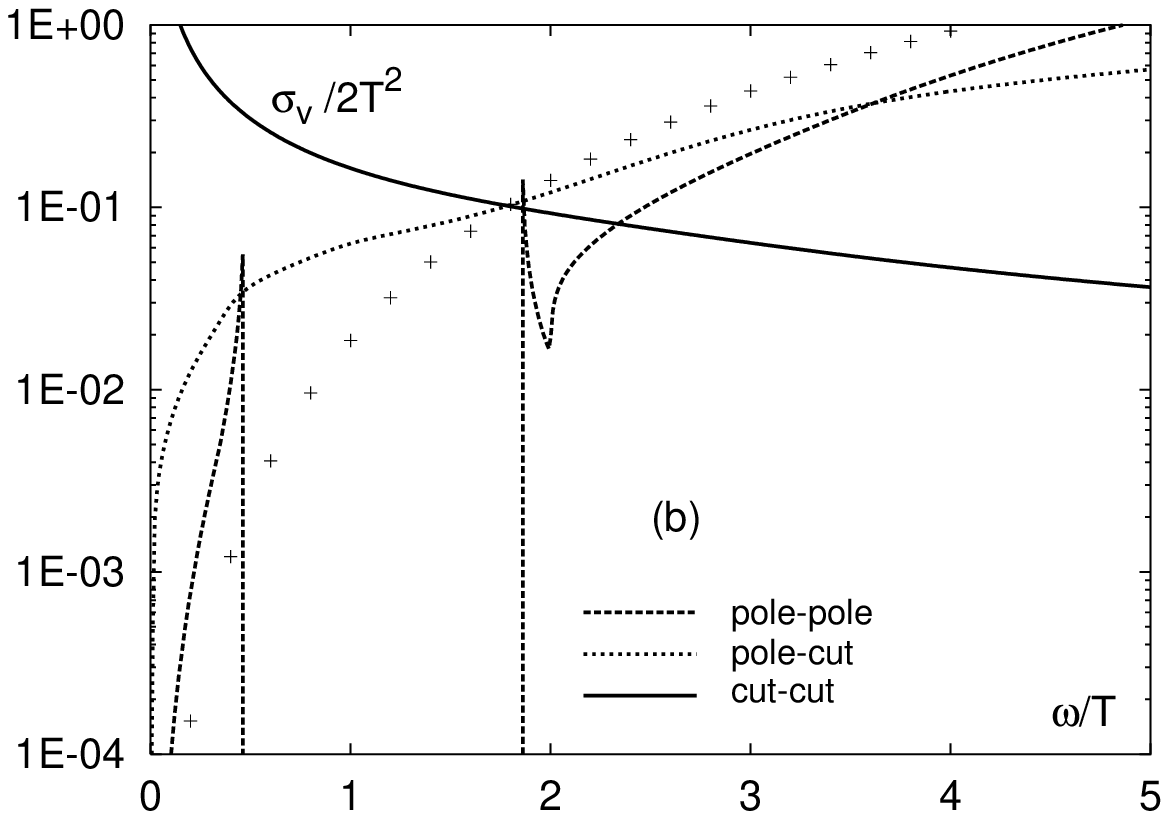,width=10cm}}
\centerline{\it Fig.36}

\end{figure}

However, there are still unsolved problems:

1. The assumption $g\ll 1$ is in contrast to realistic situations. 
In a QGP, expected to be produced in relativistic heavy ion collisions,
we expect $\alpha_s=4\pi/g^2\simeq 0.2 - 0.5$, i.e. $g=1.5 - 2.5$.
The extrapolation of HTL results to such large couplings is questionable 
and in the case of thermal energies $E\simeq T$ impossible (see e.g.
the thermalization time).

2. There is no static magnetic screening in the HTL approximation.
Hence IR singularities from the magnetic sector are not removed completely
(see e.g. damping rates).

3. Some quantities cannot be calculated to leading order
within HTL improved perturbation theory due to a strong IR sensitivity. For these
quantities a non-perturbative resummation beyond HTL seems to be
required (see e.g. the photon production rate).

4. In relativistic heavy ion collisions the QGP fireball will be out of 
equilibrium at least in the early stage. Therefore an extension of the
HTL method to non-equilibrium situations is desirable.

\newpage

\section{Non-Equilibrium}

If a partonic fireball is created in relativistic heavy ion collisions,
it will be out of equilibrium at the beginning. Only by secondary collisions
a thermal and chemical equilibrated QGP might be reached. Therefore we 
have to study non-equilibrium situations to describe the early
stages of the fireball.

Transport models, based on the Boltzmann equation for quarks and gluons,
predict a fast thermalization ($\tau_{\rm thermal}<1$ fm/c),
but a slow (or no) chemical equilibration. Hence we have to
consider non-equilibrium distributions for quarks and gluons. A simple ansatz
for theses distributions in thermal equilibrium, but out of chemical
equilibrium is the following: $f_{F,B}=\lambda (t)\> n_{F,B}(k_0)$.
Here we simply multiply the equilibrium distributions by a time-dependent 
factor (``fugacity''), $\lambda (t)$, which describes the
deviation from chemical equilibrium. At the beginning there will be less
quarks and gluons than in equilibrium, i.e. $0<\lambda (t)<1$.

Here we will discuss the generalization of the HTL method for quasistatic 
non-equilibrium situations, where the equilibration is slow compared to 
the processes under consideration\footnote{The ideas presented in this 
section are based on Ref.[4].}. We are not aiming at a 
description of the equilibration process, which is intrinsically
non-perturbative. The starting point for this program is the RTF, since
the ITF applies only to equilibrium. The basic idea is simply to replace 
the equilibrium  distribution functions in (\ref{2.23}) and (\ref{2.30})
by non-equilibrium ones. Then it is easy to see that (\ref{3.3}) to 
(\ref{3.5}), i.e. the HTL results for the retarded and advanced
polarisation tensor, still hold with
\beq
m_\gamma \rightarrow \tilde m_\gamma=\frac{4e^2}{3\pi^2}\>
\int_0^\infty dk\> k\>  f_F(k) {\buildrel {f_F\rightarrow n_F}
\over \longrightarrow}\frac{e^2T^2}{9},
\label{5.1}
\eeq
where we assumed a locally isotropic momentum distribution $f_F=f_F(k,x)$
depending on $k$ and the space-time coordinate $x$.
Note that in deriving (\ref{3.3}) to (\ref{3.5}) no assumption about the 
existence of a temperature has been made. The HTL method relies only on 
the separation of scales. Instead of the temperature $T$ one might use
the average momentum  $\langle k\rangle$.

Instead of (\ref{3.6}) (symmetric HTL polarization tensor) we get
\bea
\Pi _S^L(P)&=&-\frac{4ie^2}{\pi p}\theta (p^2-p_0^2)\, \int _0^\infty
dk\, k^2\, f_F(k)\, [1-f_F(k)] \nonumber \\
&=& 2iA\, \frac{{\rm Im}\, \Pi _R^L(P)}{p_0}\qquad \mbox{if}\; \; p_0^2<p^2 
\label{5.2}
\eea
with 
\beq
A=\frac{\int _0^\infty dk\, k^2\, f_F(k)\, [1-f_F(k)]}{\int _0^\infty
dk\, k\, f_F(k)}.
\label{5.3}
\eeq
In equilibrium $A$ reduces to $2T$. 
\me
Eq. (\ref{3.17}) (retarded and advanced HTL photon propagator) holds also
using the replacement $m_\gamma \rightarrow \tilde m_\gamma$.

The most interesting quantity is the symmetric HTL propagator. We cannot
use (\ref{2.33}) here anymore as we did for deriving (\ref{3.18}). 
Rather we have to go 
back to the Dyson-Schwinger equation for ${D^*}_S^L$ (see problem \#9). It reads
\bed
{D^*}_S^L={D^*}_{11}^L+{D^*}_{22}^L=D_{11}^L+\sum _{i,j=1}^2 D_{1i}^L\Pi _{ij}^L
{D^*}_{j1}^L+D_{22}^L+\sum _{i,j=1}^2 D_{2i}^L\Pi _{ij}^L{D^*}_{j2}^L.
\nonumber
\eed
Using (\ref{2.27}) and the inverse relations of (\ref{2.32}) we find after some
manipulations
\beq
{D^*}_{S}^L=D_{S}^L+D_{R}^L\Pi _R^L{D^*}_{F}^L+D_S^L\Pi _{A}^L
{D^*}_{A}^L+D_{R}^L\Pi _{S}^L{D^*}_{A}^L.
\label{5.4}
\eeq
It can be shown that this equation is solved by 
\bea
&& {D^*}_{S}^L(P)= [1+2f_B(p_0)]\, \mbox{sgn}(p_0)\, [{D^*}_{R}^L(P)-{D^*}_{A}^L(P)]
\nonumber \\
&& +\{\Pi _S^L(P)-[1+2f_B(p_0)]\, \mbox{sgn}(p_0)\, [\Pi _R^L(P)-\Pi _A^L(P)]\}
\, {D^*}_{R}^L(P)\, {D^*}_A^L(P).
\label{5.5}
\eea    

Here we encounter a term $\sim {D^*}_R(P){D^*}_A(P)$. In the case of bare
propagators such a term produces a pinch singularity, 
i.e. terms containing $\delta^2(P)$.
In equilibrium, where (\ref{2.34}) holds, the term in the curly brackets
vanishes, $\{...\}=0$. This is a consequence of detailed balance. Hence
there is no pinch singularity in equilibrium.

Now we study the dangerous pinch term, i.e. the second term of (\ref{5.5}).
The following equations hold in general:
\beq
{D^*}^L_{R,A}=(p^2-{\rm Re}\, \Pi_R^L\mp i\, {\rm Im}\, \Pi_R^L)^{-1}
\label{5.6}
\eeq
and
\beq
\Pi _R^L-\Pi _A^L =2i\, {\rm Im}\, \Pi _R^L.
\label{5.7}
\eeq
From (\ref{5.6}) we obtain
\beq
{D^*}^L_{R}\> {D^*}^L_{A}=\frac{1}{(p^2-{\rm Re}\, \Pi_R^L)^2+({\rm Im}\, 
\Pi_R^L)^2}=\frac{{D^*}^L_{R}-{D^*}^L_{A}}{2i\, {\rm Im}\, \Pi _R^L},
\label{5.8}
\eeq
if ${\rm Im}\, \Pi _R^L\neq 0$, i.e. for $p_0^2<p^2$ within the HTL approximation.
Inserting (\ref{5.7}) and (\ref{5.8}) in (\ref{5.5}) yields
\beq
{D^*}^L_{S}=\frac{\Pi_S^L}{2i\, {\rm Im}\, \Pi _R^L}\> 
({D^*}^L_{R}-{D^*}^L_{A}).
\label{5.9}
\eeq
Hence there is no pinch singularity if ${\rm Im}\, \Pi _R^L\neq 0$.

The physical interpretation is the following: in the case 
${\rm Im}\, \Pi _R^L\neq 0$ the particles have a finite width (damping).
Consequently the $\delta$ function is replaced by a 
function $\rho (P)$  of the Breit-Wigner form, i.e. a smeared 
$\delta $-function. Using (\ref{5.2}) and the definition of the
spectral function (see section 3.2.2) one finds
\beq
{D^*}^L_{S}(P)=-2\pi i\> \frac{A}{p_0}\> \tilde \rho_L(P)
\label{5.10}
\eeq
with $\tilde \rho_L(P)$ follows from $\rho_L(P)$ using the replacement
$m_\gamma \rightarrow \tilde m_\gamma$.
In equilibrium, where we have $A\rightarrow 2T$, (\ref{5.10}) reduces to
(\ref{3.18}) in the HTL approximation $p_0 \ll T$, for which $[1+2n_B(p_0)]\, 
\mbox{sgn}(p_0)\rightarrow 2T/p_0$. 

\begin{figure}

\centerline{\psfig{figure=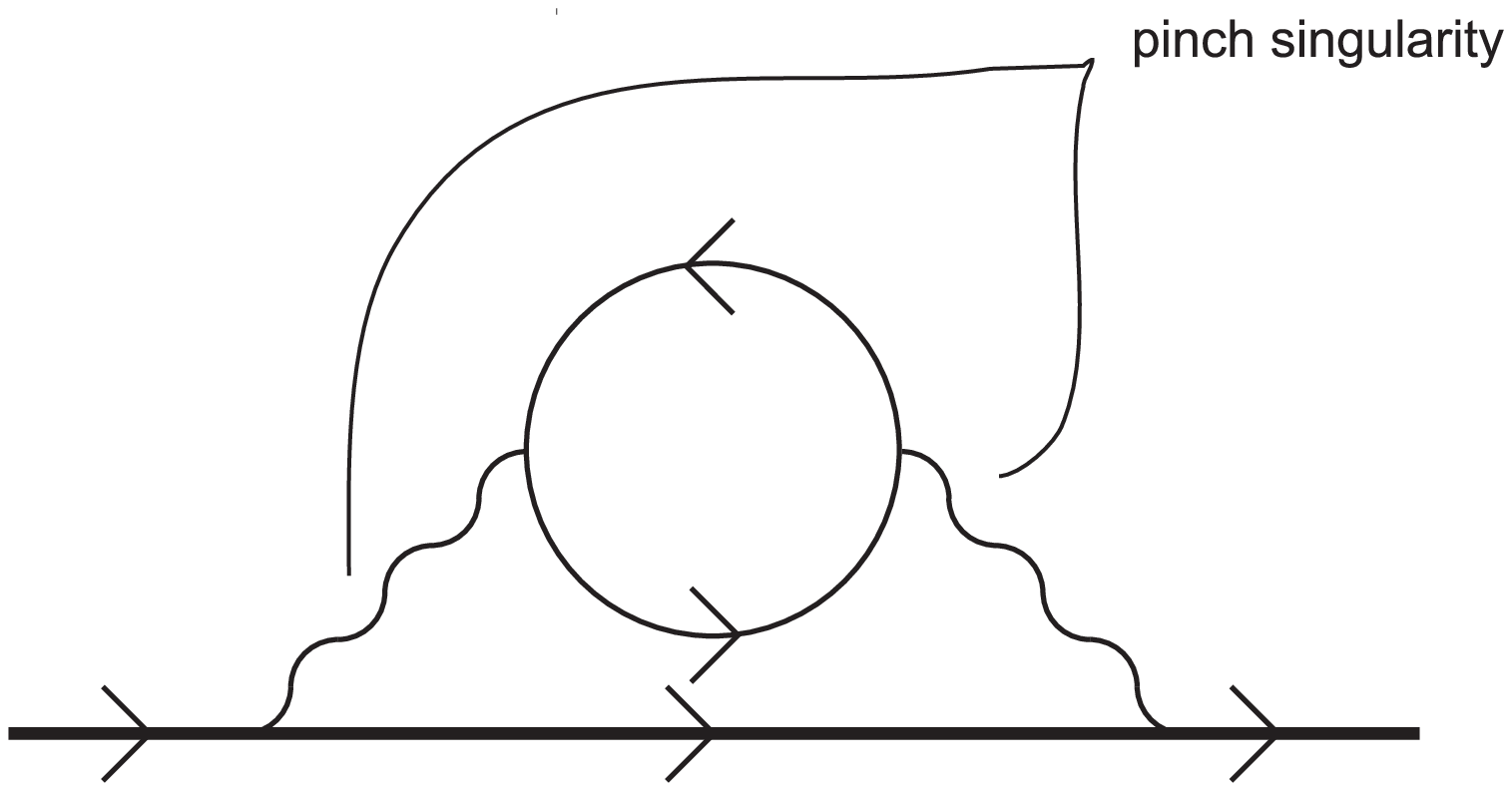,width=6cm}}
\centerline{\it Fig.37}

\end{figure}

As an application we discuss again the damping rate of a muon in a 
QED plasma. The leading order contribution in naive perturbation theory
suffers from a pinch singularity in non-equilibrium since there are two photon
propagators of the same momentum (see Fig.37). In the HTL approximation, 
however, the 2-loop diagram of Fig.38 is replaced by a 1-loop diagram
containing an effective gluon propagator, as in section 4.1. The photon
momentum is restricted to the $Q^2<0$ (see (\ref{4.4})), where ${\rm Im}\, 
\Pi_R^L\neq 0$. Hence the HTL method also removes the pinch singularity 
found in naive perturbation theory. Eq. (\ref{4.4}) reads now
\bea
\Gamma_{\rm neq} (E) &=& \frac  {e^2A}{4\pi v}\> \int _0^\infty dq \> q \int _{-vq}
^{vq} \frac{dq_0}{q_0} \> \left [\tilde \rho _L(q_0,q)
+\left (v^2-\frac {q_0^2}{q^2}\right )\> \tilde \rho _T(q_0 ,q)
\right ]\nonumber \\
&=&\frac{A}{2T}\> \Gamma_{\rm eq}(E).
\label{5.11}
\eea
This result does not depend on $\tilde m_\gamma $ since it
drops out after performing the integrations in (\ref{5.11}).

\begin{figure}[b]

\centerline{\psfig{figure=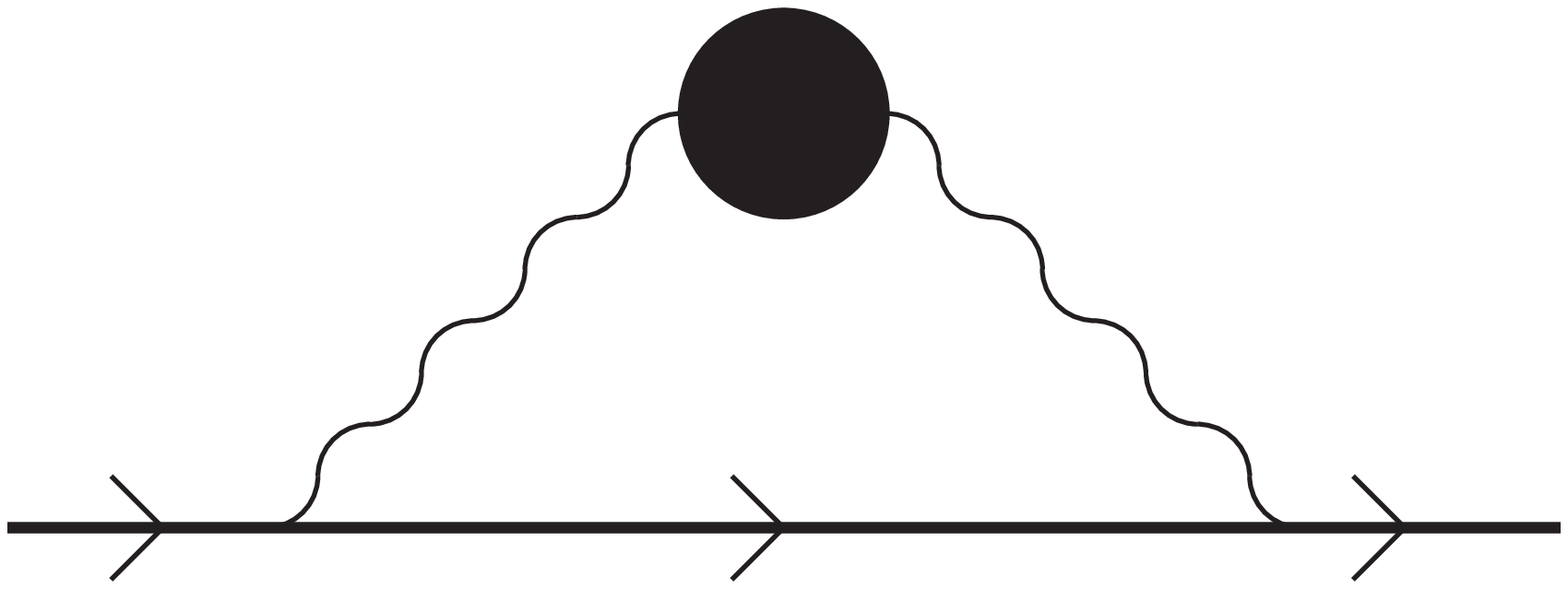,width=6cm}}
\centerline{\it Fig.38}

\end{figure}

Summarizing, we have shown, how the HTL technique can be generalized to 
quasistatic non-equilibrium situations. The symmetric HTL propagator 
contains no pinch singularity below the light cone due to Landau damping. 
In general, we can say that pinch singularities come from using bare 
propagators in perturbation theory. Resummed propagators for quasiparticles 
with a finite width
(Breit-Wigner) do not lead to pinch singularities. 
In equilibrium pinch singularities are absent
also for non-interacting particles due to detailed balance.

\newpage

\section{Problems}

\begin{enumerate}

\item
Calculate the particle and energy density of an ideal quark-gluon plasma
containing up-, down-quarks, and gluons. Assume that the quarks are massless
and that there are as many antiquarks as quarks. Use the result to estimate 
the critical particle and energy density for the deconfinement transition 
at $T_c=170$ MeV.

\medskip

Hint: Use 
$$\zeta(n)=\frac{1}{(n-1)!} \int_0^\infty dx \frac{x^{n-1}}{e^{x}-1}=
\frac{1}{(1-2^{1-n})(n-1)!} \int_0^\infty dx \frac{x^{n-1}}{e^{x}+1}$$ 
with $\zeta(3)=1.202$ and $\zeta(4)=\pi^4/90$.
\bigskip

\item
Show that 
\bea
I\equiv T\sum_{n=-\infty}^\infty f(k_0=2\pi inT) &=&\frac{1}{2\pi i}
\biggl \{\int_{-i\infty}^{i\infty}dk_0 \frac{f(k_0)+f(-k_0)}{2}\nonumber \\
&+&\int_{-i\infty+\epsilon}^{i\infty+\epsilon}dk_0[f(k_0)+f(-k_0)]n_B(k_0)
\biggr \},\nonumber
\eea
if $f(k_0)$ has no poles on the imaginary axis.

\medskip

Hint: Show first
$$I=\frac{1}{4\pi i}\oint_C dk_0 f(k_0) \coth(\beta k_0/2),$$
where $C$ are the contours around the poles of $\coth(\beta k_0/2)$.
\bigskip

\item
Show that the propagator in the Keldysh representation 
$$\Delta_K=\left (\begin{array}{cc}
0&\Delta_A\\ \Delta_R&\Delta_S \end{array}\right )$$
follows from the RTF propagator
$$\Delta=\left (\begin{array}{cc}
\Delta_{11}&\Delta_{12}\\ \Delta_{21}&\Delta_{22} \end{array}\right )$$
by applying an orthogonal transformation
$$Q=\frac{1}{\sqrt{2}}\left (\begin{array}{cc}
1&-1\\ 1&1\end{array}\right ).$$
\bigskip

\item
Show that the symmetric propagator can be written as
$$\Delta_S(K)=[1+2n_B(k_0)]\, {\rm sgn}(k_0)\, [\Delta_R(K)-\Delta_A(K)].$$
\bigskip

\item
Derive the symmetric longitudinal photon self energy in the HTL approximation
and show that it fulfils the relation
$$\Pi^L_S(K)=[1+2n_B(k_0)]\, {\rm sgn}(k_0)\, [\Pi^L_R(K)-\Pi^L_A(K)].$$

\medskip

Hint: Use 
$$n_F(k)[1-n_F(k)]=-T\frac{d}{dk}n_F(k).$$
\bigskip

\item
Show that the Dyson-Schwinger equation for the full photon propagator in 
Coulomb gauge leads to
$$D_L^*(K)=\frac{1}{k^2-\Pi_L},$$
$$D_T^*(K)=\frac{1}{K^2-\Pi_T}.$$
\bigskip

\item
Calculate the photon dispersion relation in the HTL approximation 
for small $k\ll \omega$ up to order $k^2$. 

\medskip

Hint: $\ln{\frac{1+x}{1-x}}=2(x+\frac{x^3}{3}+\frac{x^5}{5}+ ...).$
\bigskip

\item
Show that the tadpole diagram in massless $\phi^4$-theory containing the
effective propagator with mass $M=gT$ (sum of daisy diagrams) is given 
for small $g\ll 1$ by
$$\Pi^*=g^2T^2\left [1-\frac{3}{\pi} g+ O(g^2)\right ].$$

\medskip

Hint: Show that the UV divergent part is of $O(g^4)$ after regularizing it 
by subtracting the 1-loop tadpole at $T=0$ and introduce a separation scale 
$gT\ll k^*\ll T$ for evaluating the momentum integral in the small $g$ limit.
\bigskip

\item
Show that the Dyson-Schwinger equation for the full symmetric propagator 
reads
$$\Delta_S^*=\Delta_S+\Delta _R \Pi_R \Delta_S^*+\Delta_S \Pi_A \Delta_A^*
+\Delta_R \Pi_S \Delta_A^*$$
and that it is solved by the ansatz
\begin{eqnarray}
\Delta_S(K)&=&[1+2n_B(k_0)]{\rm sgn}(k_0)[\Delta_R^*(K)-\Delta_A^*(K)]\nonumber
\\ &+&\{\Pi_S(K)-[1+2n_B(k_0)]{\rm sgn}(k_0)[\Pi_R(K)-\Pi_A(K)]\}\Delta_R^*(K)
\Delta_A^*(K).\nonumber
\end{eqnarray}

\medskip

Hint: Use $\Delta_{R,A} \Pi_{R,A} \Delta_{R,A}^*=\Delta_{R,A}^*-\Delta_{R,A}$.

\end{enumerate}

\newpage

\centerline{{\bf Literature}}

\medskip

\noindent
1. J.I. Kapusta, {\it Finite-Temperature Field Theory} (Cambridge
University Press, Cambridge, 1989).

\medskip

\noindent
2. M. Le Bellac, {\it Thermal Field Theory} (Cambridge University Press,
Cambridge 1996).

\medskip

\noindent
3. M.H. Thoma, {\it Applications of High-Temperature Field Theory to 
Heavy-Ion Collisions}, in {\it Quark-Gluon Plasma 2}, ed. R.C. Hwa
(World Scientific, Singapore, 1995) p.51 (hep-ph/9503400).

\medskip

\noindent
4. M.E. Carrington, H. Defu, and M.H. Thoma, {\it Equilibrium and 
Non-Equilibrium Hard Thermal Loop Resummation in the Real Time Formalism},
Eur. Phys. J. C7 (1999) 347 (hep-ph/9708363).

\end{document}